%% file: paperJTv2.tex
\documentclass[12pt]{article}
\pdfoutput=1
\usepackage{amsmath,amsfonts,amssymb,amsthm}
\usepackage{nicefrac}
\usepackage{cite}

\usepackage{times}

\usepackage{xcolor}

\usepackage{geometry}
\usepackage{hyperref}

\newcommand{\poubelle}[1]{}

\input{victorsmacros}

\numberwithin{equation}{section}

\title{New boundary conditions for AdS$_2$}
\author{}

\hypersetup{
    colorlinks=true,
    linkcolor=blue,
    filecolor=magenta,      
    urlcolor=cyan,
    citecolor = blue,
}

\begin{document}

\hfuzz=1.5pt
%%%%%%%%%%%%%%%%%%%%%%%%%%%%%%%%%%%%%%%%%%%%%%%%%%%%%%%%%%%%%%%%%%%%%%

\def\mytitle{New boundary conditions for AdS$_2$}

\pagestyle{myheadings} \markboth{\textsc{\small Godet, Marteau}}{%
  \textsc{\small New boundary conditions for AdS$_2$}} \addtolength{\headsep}{4pt}

\begin{flushright}\small
\end{flushright}

\begin{centering}

  \vspace{1cm}

  \textbf{\LARGE{\textbf{\mytitle}}}

%\vspace{1cm}

%{\huge Notes}

  \vspace{1.5cm}

  {\large Victor Godet$^{a}$ and Charles Marteau$^{b}$}

\vspace{1.5cm}

\begin{minipage}{1.0\textwidth}\small \it 
\begin{center}
$^a$ Institute for Theoretical Physics, University of Amsterdam, 1090 GL Amsterdam, Netherlands\\
e-mail: victor.godet.h@gmail.com\\
\vspace{0.3cm}
$^b$ Centre de Physique Théorique, CNRS, Institut Polytechnique de Paris, France\\
e-mail: marteau.charles.75@gmail.com
\end{center}
\end{minipage}

\end{centering}

\vspace{3cm}

\begin{center}
  \begin{minipage}{.9\textwidth}
    \textsc{Abstract}. We describe new boundary conditions for AdS$_2$ in Jackiw-Teitelboim gravity. The asymptotic symmetry group is enhanced to  $\r{Diff}(S^1)\ltimes C^\infty(S^1)$ whose breaking to  $\r{SL}(2,\R)\times \r{U}(1)$ controls the near-AdS$_2$ dynamics. The action reduces to a boundary term which is a generalization of the Schwarzian theory and can be interpreted as the coadjoint action of the warped Virasoro group. This theory reproduces the low-energy effective action of the complex SYK model.  We compute the Euclidean path integral and derive its relation to the random matrix ensemble of Saad, Shenker and Stanford. We study the flat space version of this action, and show that the corresponding path integral also gives an ensemble average, but of a much simpler nature.  We explore some applications to near-extremal black holes.
  \end{minipage}
\end{center}

\vfill

\thispagestyle{empty}
\newpage

\begingroup
\hypersetup{linkcolor=black}
\tableofcontents
\endgroup

\section{Introduction}

 AdS$_2$  plays a special role in quantum gravity because it stands as the lowest dimensional realization of the AdS/CFT correspondence \cite{Maldacena:1997re}. It also appears universally near the horizon of near-extremal black holes  \cite{Ferrara:1995ih,FerraraKallosh1996,FerraraKallosh1996a,Astefanesei:2006dd,Kunduri:2007vf,Kunduri:2008tk,Kunduri:2013ana}, which suggests that AdS$_2$  holography will play an important role in understanding quantum black holes \cite{ Strominger:1997eq,Maldacena:1998bw,Strominger:1998yg,Sen:2008yk,Sen2009b}. Despite this, the AdS$_2$/CFT$_1$ correspondence is considered to be dynamically trivial since it was shown that AdS$_2$ does not support finite energy excitations \cite{Maldacena:1998uz}. More recently, after the discovery of the SYK model, it was understood that to make AdS$_2$ holography dynamical, one has to go to the so-called near-AdS$_2$ regime. This is achieved by breaking the conformal symmetry and going slightly away from the infrared fixed point \cite{Almheiri:2014cka, Maldacena:2016upp, Jensen:2016pah, Engelsoy:2016xyb}, which is referred to as the nAdS$_2$/nCFT$_1$ correspondence \cite{ Maldacena:2016hyu}.

A canonical realization of this duality is obtained in the Jackiw-Teitelboim (JT) theory of gravity \cite{Jackiw:1984je, Teitelboim:1983ux} where it was shown that the near-AdS$_2$ dynamics is controlled by a boundary action involving the Schwarzian derivative. The same action also governs the low-energy regime of the SYK model \cite{PhysRevLett.70.3339, Kit.KITP.2}, demonstrating a holographic duality between JT gravity and a subsector of the SYK model.

 The appearance of the Schwarzian action is tied to the pattern of spontaneous and explicit symmetry breaking $\r{Diff}(S^1)\to\r{ SL}(2,\R)$ that controls the near-AdS$_2$ physics. In \cite{Stanford:2017thb}, it was understood that the Schwarzian action can be seen as a Hamiltonian generating a $\r{U}(1)$ symmetry on a coadjoint orbit of the Virasoro group. This allowed the authors to compute its partition function, which, thanks to the Duistermaat-Heckman theorem \cite{Duistermaat:1982vw}, is one-loop exact.

Even though JT gravity is usually thought as an effective theory, arising in the low-energy description of more complicated systems, such as higher-dimensional black holes, it can also be studied as a UV complete theory in its own right. One of the motivation to do this was to use the simplicity of this theory to probe some features of the spectral form factor, which is a diagnosis of the discreteness of the black hole spectrum \cite{Maldacena:2001kr, Papadodimas:2015xma, Cotler:2016fpe,Saad:2018bqo}. This was considered in \cite{Saad:2019lba} where the full Euclidean path integral of JT gravity is computed. The authors showed that the gravitational theory is not holographically dual to a single quantum mechanical theory but rather to a statistical  ensemble of theories. More precisely, this ensemble corresponds to a double-scaled matrix integral whose leading density of eigenvalues matches with the density of states of the Schwarzian theory. This leads to the more general suggestion that gravitational path integrals should be interpreted holographically in term of ensemble averages. This story has been generalized in various ways \cite{Stanford:2019vob, Iliesiu:2019lfc, Kapec:2019ecr}.  

JT gravity has also played a central role in recent developments on the information paradox and the black hole interior. After introducing a coupling between an evaporating AdS$_2$ black hole and an external bath, the entropy of the bath was shown to follow the Page curve \cite{Penington:2019npb, Almheiri:2019psf}, using a semi-classical version of the RT/HRT/EW formula for entanglement entropy \cite{Ryu:2006bv, Hubeny:2007xt,Engelhardt:2014gca,Faulkner:2013ana}. This led to the island prescription \cite{Almheiri:2019hni, Almheiri:2019yqk} which was proven using replica wormholes \cite{Penington:2019kki, Almheiri:2019qdq} appearing as saddle-points in the replicated Euclidean path integral \cite{Lewkowycz:2013nqa, Faulkner:2013ana, Dong:2017xht}. It was noticed that these new geometries could only make sense if some kind of average was taking place.

To specify a proper classical solution space and identify the asymptotic  symmetries of JT gravity, one needs to gauge-fix the metric. This is usually done by writing the metric in Fefferman-Graham gauge and imposing a Dirichlet boundary condition, which leads to the Schwarzian action together with its $\r{SL}(2,\mathbb{R})$ symmetry. However other gauge choices and boundary conditions can be considered \cite{Grumiller:2017qao,Gonzalez:2018enk}, leading to new boundary actions and new symmetry groups.

In this paper, we study JT gravity in Bondi gauge. The latter leads to an enhancement of the asymptotic symmetry group from the usual $\r{Diff}(S^1)$ to a warped version of the latter, \emph{i.e.} with an additional local $\r{U}(1)$ symmetry. The dynamics is controlled by a generalization of the Schwarzian action, which can also be understood as the generator of a Hamiltonian symmetry on a coadjoint orbit of the warped Virasoro group. As a result, the near-AdS$_2$ physics is controlled by the pattern of symmetry breaking to $\r{SL}(2,\R)\times \r{U}(1)$. Our boundary action matches with the low-energy effective action of the complex SYK model. This shows that our version of JT gravity is holographically dual to a subsector of the complex SYK model. We  compute the full Euclidean path integral, including sums over topologies, and show that it leads to a simple refinement of the random matrix ensemble of Saad, Shenker and Stanford. We also find a connection to  warped CFTs which suggests a route towards a better understanding of  Kerr/CFT. Finally, we study the $\widehat{\r{CGHS}}$ model as a flat space analog of JT gravity. We  compute the Euclidean path integral and show that this theory is also dual to an ensemble average, albeit a much simpler one.

\subsection{Summary of results}

The usual boundary condition for JT gravity fixes the boundary metric in Fefferman-Graham gauge, and corresponds to a choice of AdS$_2$ metric of the form
\begin{equation}\label{intro:FGAdS2}
ds^2=r^2\le(1-{s(t)\ov 2r^2} \ri)^2dt^2+\frac{dr^2}{r^2}~,
\end{equation}
where $t\sim t+\b$ and $s(t)$ is an arbitrary periodic function. The corresponding asymptotic symmetry group is  $\r{Diff}(S^1)$ and acts on $s(t)$. This infinite-dimensional symmetry gets spontaneously broken to $\r{SL}(2,\R)$ when choosing $s$ to be
\be\label{intro:sval}
s(t)={2\pi^2\ov \beta^2}~.
\ee
The near-AdS$_2$ dynamics, captured by the JT dilaton, also breaks explicitly this symmetry. It is controlled by the boundary action 
\be\label{intro:SchwAction}
I[f] = -\g \int_{S^1} dt   \le( \frac{2\pi^2}{\beta^2} f'(t)^2+  \{f(t),t\}\ri)~,
\ee
where $\{f(t),t\}$ is the Schwarzian derivative of $f$. This theory describes Goldstone mode fluctuations parametrized by $f(t) \in \r{Diff}(S^1) $ around the background \eqref{intro:sval}. 
% In this context, the boundary action is given by the renormalized extrinsic curvature term $\Phi(K-1)$.

 In this paper, we consider new boundary conditions for JT gravity, which are naturally formulated in (Euclidean) Bondi gauge 
\be\label{intro:AdSBondi}
ds^2 = 2\le({r^2\ov 2}-i P(\tau)r +T(\tau) \ri) d\tau^2 -2i d\tau dr ~,
\ee
where $\tau\sim\tau+\b$ and $P(\tau)$ and $T(\tau)$ are arbitrary periodic functions. In Sec.\ \ref{sec:boundaryaction}, we obtain the boundary term required to make the corresponding variational problem well-defined. In this case, it is not given by the extrinsic curvature term. We show that the asymptotic symmetry group gets enhanced to the group $\r{Diff}(S^1)\ltimes C^\infty(S^1)$ which gets spontaneously broken to $\r{SL}(2,\R)\times \r{U}(1)$ after we choose the thermal values 
\begin{equation}\label{intro:valPT}
\begin{split}
&P(\tau)=  {\mu\ov \b}, \qq T(\tau)= -\frac{2\pi^2}{\beta^2}-\frac{\mu^2}{2\b^2}~,
 \end{split}
\end{equation}
where $\mu$ will be interpreted as a chemical potential for the $\r{U}(1)$ symmetry. The explicit breaking is controlled by the boundary action
\be\label{intro:bdyAction}
I[f,g]= \gamma\int_{S^1} d\tau\le(T f'^2 -{1\ov 2}g'^2 + P f' g' + { g'  f'' \ov f'}  -g'' \ri)+ \g \frac{\mu}{\beta}\int_{S^1} d\tau\left(P-\frac{f''}{f'}+g'\right)~.
\ee
This corresponds to Goldstone mode fluctuations parameterized by $(f,g)\in \r{Diff}(S^1)\ltimes C^\infty(S^1)$ around the background \eqref{intro:valPT}. In particular, the effective action has a $\r{U}(1)$ symmetry, which corresponds to shifting $g$ by a constant. The gravitational charges are computed in Sec.\ \ref{Gravitational charges} and agree with the Noether charges of the boundary theory. As in the Schwarzian case, our AdS-Bondi boundary action can be reinterpreted as a particle moving on rigid AdS$_2$. 

\ms

In Sec.\ \ref{Relation to the Schwarzian action}, we show that the Schwarzian dynamics is embedded in the Bondi description. We find that imposing an additional boundary condition that fixes  $P$  in the solution space reduces the asymptotic symmetry group to $\r{Diff}(S^1)$. The boundary action \eqref{intro:bdyAction} then becomes the Schwarzian action \eqref{intro:SchwAction} with the relation
\be
{-}s(\tau)=T(\tau)+{1\ov 2}P(\tau)^2 -P'(\tau)~.
\ee
This expression of $s$ in terms of $P$ and $T$ is also  derived by constructing a diffeomorphism that relates the solution space of Bondi gauge to the Fefferman-Graham one.

\ms

In App.\ \ref{app:NewCount}, we consider an alternative way to derive  our boundary action, which is shown to arise from a new counterterm for JT gravity:
\be
I = \k \int \,du \sqrt{-h} \, (\Phi K-n^\mu \p_\mu\Phi) ~.
\ee

\pg{Euclidean path integral.} The computation of the partition function requires an integration over a coadjoint orbit of the warped Virasoro group, which corresponds to the thermal values of $P$ and $T$ given in \eqref{intro:valPT}. This is greatly simplified by the fact that our boundary action generates a $\r{U}(1)$ symmetry on this orbit. As a result of the Duistermaat-Heckman theorem, the partition function is one-loop exact, like in the Schwarzian case. The result is
\be
Z(\b,\mu) \propto \frac{\gamma^2}{\beta^2}\exp\le( {2\pi^2\g \ov \b} - {\g\mu^2\ov 2\b}\ri)~.
\ee
Interpreting $\mu$ as a chemical potential associated to the $\r{U}(1)$ symmetry, the partition function at fixed charge $Q$ takes the form
\be
Z(\b,Q)  \propto  \frac{\gamma^{3/2}}{\beta^{3/2}}\exp\left(\frac{2\pi^2\gamma}{\beta}-\frac{Q^2\beta}{2\gamma}\right)~,
\ee
which follows from a Fourier transform. We see that for $Q=0$, we recover the partition function of the Schwarzian theory \cite{Stanford:2017thb}. The additional information carried by our AdS-Bondi formulation is therefore contained in this additional $\r{U}(1)$ charge. From the partition function, we obtain the leading density of states 
\be
\rho_0(E,Q) \propto\sinh\left(2\pi\sqrt{2\gamma E-Q^2}\right)~.
\ee
Using similar techniques, we can compute the contribution of the trumpet geometry 
\begin{equation}
Z^{\mathrm{trumpet}}(\beta, Q, b) \propto \frac{\gamma^{1/2}}{\beta^{1/2}}\exp\left(-\frac{\gamma b^2}{2\beta}-\frac{Q^2\beta}{2\gamma}\right),
\end{equation}
which also depends on the geodesic length $b$ of the small end. 

This last result is a crucial step to compute the full gravitational path integral with prescribed boundary conditions. This is because any hyperbolic Riemann surface with $n$ asymptotic boundaries can be constructed by gluing $n$ trumpets to a genus $g$ surface with $n$ geodesic boundaries. This property was exploited in \cite{Saad:2019lba} to match the Euclidean path integral of JT gravity with a random matrix ensemble.

Our AdS-Bondi formulation allows us to fix two conditions at each boundary: the renormalized length and the $\r{U}(1)$ charge. In Sec.\ \ref{Sec:EucPath} we show that the full gravitational path integral, including all topologies, can be computed using the following prescription: the insertion of a boundary of length $\beta$ and charge $Q$ corresponds to the insertion of $\Tr\,e^{-\b (H+{Q^2/(2\g))}}$ in the matrix model of Saad, Shenker and Stanford. More precisely we have
\begin{equation}
Z_{\r{grav}}(\{\beta_i\}, \{Q_i\})= 
\langle \mathrm{Tr}\,e^{-\beta_1 (H+Q_1^2/(2\g))}\dots \,\mathrm{Tr}\,e^{-\beta_n (H+Q_n^2/(2\g))}\rangle_{\r{SSS}}~,
\end{equation}
where $H$ is the random matrix while the $Q_i$'s are scalars that shift the ground state energy at each boundary. We see that setting all the $\r{U}(1)$ charges to zero, we recover the prescription of Saad-Shenker and Stanford.

\pg{Relation to complex SYK.} The complex SYK model  \cite{Davison_2017, Bulycheva:2017uqj, Gu:2019jub}  is a version of the SYK model where the Majorana fermions are replaced by complex fermions. It is also maximally chaotic and solvable at large $N$, while being closer to condensed-matter systems, such as strange metals \cite{Guo:2020aog}. As the Majorana SYK model, it can be described at low temperature by an effective action which takes the form 
\be\label{intro:IcSYK}
I_\r{cSYK} = {N K\ov 2} \int_{S^1} d\tau\le(\tilde{g}'+{2\pi i\cE\ov \b}f' \ri)^2 -{N\g_\r{SYK}\ov 4\pi^2} \int_{S^1} d\tau\,\le\{\r{tan}(\tfrac\pi\b f), \tau\ri\}~.
\ee
We show in Sec.\ \ref{Complex SYK} that our boundary action for JT gravity matches with the effective action of complex SYK, after a field redefinition given in \eqref{SYKredefg}. The matching between the complex SYK parameters and the gravitational parameters is given by
\be
K N= \g \l_\r{AdS}^2 \la_0^2,\qq \g_\r{SYK} N = 4\pi^2 \g \l_\r{AdS}^2, \qq \cE = {\mu-\mu_0\ov 2\pi \la_0} ~.
\ee
Here, $K,N,\g_\r{SYK}$ and $\cE$ are parameters of the complex SYK model while $\l_\r{AdS}$, $\g = \k \bar\phi_r$ and $\mu$ are gravitational parameters. The constants $\la_0$ and $\mu_0$ are free parameters that appear in the identification. This matching shows that JT gravity in Bondi gauge is dual to a subsector of the complex SYK model. This is on the same footing as the relationship between JT gravity in Fefferman-Graham gauge and the Majorana SYK model. A similar match between a flat space version of our boundary action and a particular scaling limit of the complex SYK model was achieved in \cite{Afshar:2019axx}.

\pg{Warped symmetry of AdS$_2$.} The AdS-Bondi gauge gives rise to an enhancement of the asymptotic symmetry group to $\r{Diff}(S^1)\ltimes C^\infty(S^1)$. In Sec.\ \ref{Gravitational charges} we show that the corresponding gravitational charges belong to a centerless representation of the asymptotic symmetry algebra. However, the solution space transforms in the coadjoint representation of a centrally extended version of the group whose central charges are
\begin{equation}
c=0,\quad k=-2\quad  \text{and}\quad \lambda=-1.
\end{equation}
The corresponding algebra is the twisted warped Virasoro algebra \cite{Detournay:2012pc}. For $k\neq 0$, the twist parameter $\lambda$ can be removed by a redefinition the generators, leading to a warped Virasoro algebra with central charge $c = -24 \la^2/k$.  In gravity, it is natural to rescale the currents $P$ and $T$ by $\k^{-1} = 8\pi G^{(2)}_N$ which leads to
\be
c= 12 \k,\qq k = -2 \k~.
\ee
For the AdS$_2$ factor in the near horizon region of the extreme Kerr black hole, we have $\k=J$, therefore we obtain $c=12 J$ as in Kerr/CFT \cite{Guica:2008mu}. Both central charges also match the ones derived for the warped symmetry of Kerr in \cite{Aggarwal:2019iay}. This observation indicates that the near-AdS$_2$ analysis might shed light upon the lack of knowledge of the classical phase space in Kerr/CFT \cite{Dias:2009ex, Amsel:2009ev}. Further comments are given in Sec.\ \ref{Sec:warped CFTs}, where we also discuss the connection to warped CFTs, following the known relation between the complex SYK model and warped CFTs \cite{Chaturvedi:2018uov}.

\pg{Deformations of Reissner-Nordström.} In Sec.\ \ref{sec:RN}, we  show that the Bondi gauge for AdS$_2$ captures deformations of an extremal black hole in a finer way than the usual Fefferman-Graham gauge. We illustrate this for the Reissner-Nordström black hole. A deformation of the extremal geometry is generally parametrized by deviations $\d r_+$ and $\d r_-$ for the outer and inner horizons
\be
r_+ = r_0 + \la \d r_+ + O(\la^2), \qq r_- = r_0 - \la \d r_-+ O(\la^2)~,
\ee
where $r_0$ is the extremal horizon. After an appropriate change of coordinate, the near-horizon geometry is obtained by taking the $\lambda\to 0$ limit. In Fefferman-Graham gauge, we obtain the AdS$_2$ metric \eqref{intro:FGAdS2} in Lorentzian signature with 
\be
s(t)= - {1\ov 8}{(\d r_+ +\d r_-)^2}~.
\ee
We see that $s(t)$ is only sensitive to the sum $\d r_+ +\d r_-$. In AdS-Bondi gauge, we obtain the Lorentzian version of the metric \eqref{intro:AdSBondi} with
\bea
P(u) =  {\d r_+-\d r_-\ov 2 M_0},\qq T(u)= {\d r_+ \d r_-\ov 2}~,
\eea
where $u$ is the retarded time. As a result, we see that the Bondi gauge is a finer probe of deformations of the extremal geometry. It can distinguish between the deformations $\d r_+$ and $\d r_-$ independently. This information is ultimately captured in the chemical potential $\mu$ associated with the new $\r{U}(1)$ symmetry.

\pg{Embedding in near-extreme Kerr.} In Sec.\ \ref{Sec:breakingKerr}, we  study the near-extreme Kerr black hole for which JT gravity cannot be obtained by Kaluza-Klein reduction. The linearized perturbation capturing the Schwarzian mode in near-extreme Kerr was described in  \cite{Castro:2019crn}. We repeat this analysis and show that the Bondi near-AdS$_2$ dynamics described in this paper can be embedded in near-extreme Kerr. In particular, the infinite-dimensional warped symmetry algebra of Bondi AdS$_2$ is realized via the 4d diffeomorphism
\be
u \to\mathcal{F}(u),\qq  r\to\frac{1}{\mathcal{F}'}\left(r+\mathcal{G'}(u)\right),\qq
 \phi \to \phi - \mathcal{G}(u),\qq \theta\to \theta~,
\ee
which is shown to preserve a phase space of linearized perturbations described by the ansatz \eqref{deformed geometry}. We  also see that our $\r{U}(1)$ symmetry, which was obtained in the asymptotic symmetry group of Bondi AdS$_2$, corresponds here to rotations around the black hole axis in the four-dimensional geometry. The corresponding $\r{U}(1)$ charge is then simply the change in angular momentum due to the perturbation.

\pg{Flat holography in two dimensions.} A flat space version of our boundary action was derived in \cite{Afshar:2019axx} from a modified version of the CGHS model, dubbed $\widehat{\text{CGHS}}$, which we study in Sec.\ \ref{sec:CGHSandboundaryaction}. This gives rise to a flat space analog of JT gravity.\footnote{Flat JT gravity, which was first considered in \cite{Dubovsky:2017cnj} to derive the exact gravitational S-matrix in two dimensions, is not the right theory to consider to study the Euclidean path integral, for reasons that are explained in Sec.\ \ref{sec:Flat}.} We show that the boundary action is equivalent to a particle moving on a rigid 2d Minkowski spacetime. We describe the thermal solution, which is a 2d analog of the Schwarzschild black hole, as depicted in Fig.\ \ref{FlatBparticles}. The corresponding symmetry breaking pattern is 
\begin{equation}
\r{Diff}(S^1)\ltimes C^\infty(S^1)\to \r{ISO}(2)\times \r{U}(1),
\end{equation}
where the residual symmetry corresponds to a warped version of the 2d Poincaré group. We compute the corresponding gravitational charges and show that they realize a representation of the symmetry algebra.

We compute the partition function which produces a linear density of states 
\be
\rho(E) = 2\pi \g^2 E~.
\ee
We also compute the contribution of the cylinder to the gravitational path integral. This bulk geometry is the only regular flat surface which connects two asymptotic boundaries and is depicted in Fig.\ \ref{Fig:Cylinder}. The result is
\be\label{intro:Zcylinder}
Z^\text{cyl}(\b_1,\b_2)  = {4\pi^2\g\ov \b_1+\b_2}~.
\ee
This non-vanishing answer implies that the $\widehat{\text{CGHS}}$ model should be holographically dual to an ensemble average. Let us introduce the notation $\ln Z(\b_1)\dots Z(\b_n)\rn$ to represent the Euclidean path integral with $n$ asymptotic circles of lengths $\b_1,\dots,\b_n$. The fact that the cylinder does not vanish implies that
\be
\ln Z(\b_1)Z(\b_2) \rn \neq \ln Z(\b_1) \rn\ln Z(\b_2)\rn~.
\ee
This indeed shows that the path integral should be interpreted as an ensemble average. The answer \eqref{intro:Zcylinder} is not the universal answer for double-scaled matrix ensembles so the dual of the $\wh{\text{CGHS}}$ model has to be something different. The only regular flat surfaces with asymptotic boundaries are the plane (or disk) and the cylinder. Therefore the path integral with an arbitrary number of boundaries is completely determined using Wick contractions involving the cylinder and the disk. This implies that the corresponding third-quantized theory is a Gaussian theory. Thus, the $\wh{\text{CGHS}}$ model constitutes an interesting example of a theory where the full Euclidean path integral can be done, while not being completely trivial and giving rise to an ensemble average.

The asymptotically flat 2d black hole shown in Fig.\ \ref{FlatBparticles} seems to be a interesting setup to study black hole evaporation and the information paradox. In contrast with the AdS setups that were studied recently, it does not require the introduction of a coupling with an external bath, because radiation can escape to null infinity. Another observation is that the simplicity of flat Riemann surfaces might constrain the existence of replica wormholes. For recent discussions on the information paradox in asymptotically flat spacetime, we refer to \cite{Laddha:2020kvp, Anegawa:2020ezn,Gautason:2020tmk, 1793416,Krishnan:2020oun}.

\section{JT gravity in Bondi gauge}\label{sec:BondiJT}

We consider near-AdS$_2$ gravity using Bondi gauge instead of the usual Fefferman-Graham (FG) one. Bondi gauge has been extensively studied in three and four dimensions  (see \cite{Barnich:2010eb} for a good review on both cases).  We will show that in this gauge, JT gravity has an enhanced asymptotic symmetry group $\r{Diff}(S^1)\ltimes C^\infty(S^1)$ which gets broken to $\r{SL}(2,\R)\times \r{U}(1)$. We will derive the boundary action and compute the gravitational charges. We will also show that this action can be interpreted as the worldline action of a boundary particle.

% In particular we will recover the Schwarzian boundary action in Bondi gauge. 

% Another advantage of this gauge is that it is written in a coordinate system that allows for a smooth and non-trivial flat space limit, it is therefore a nice setting for taking the flat limit of the AdS$_2$ solution space.

\subsection{Boundary action for Bondi AdS}
\label{sec:boundaryaction}

We consider the Euclidean action for Jackiw–Teitelboim gravity in two dimensions
\begin{equation}
I_\r{JT}[\Phi,g]=\frac{\kappa}{2}\int d^2x \sqrt{g}\,\Phi\left(R+2\right) + I_\p~,
\label{JT action}
\end{equation} 
where $\k= (8\pi G^{(2)}_N)^{-1}$ and the AdS radius has been rescaled to one. Since we are formulating the theory in another gauge, we need to derive the boundary term $I_\p$ by studying the variational problem. The variation of the action at first order is given by
\begin{equation}
\delta I_{\mathrm{JT}}=\int d^2x\,\left( E_\Phi\delta \Phi+ E_g^{\mu\nu}\delta g_{\mu\nu}+ \partial_\mu \Theta^\mu\right) + \d I_\p,
\label{deltaS}
\end{equation}
where $\Theta^\mu$ is the remaining term when all the integrations by part have been made. This is the term which carries information about the boundary action. The equations of motion are
\bea
E_\Phi\=\frac{\kappa}{2}\sqrt{g}\left(R+2\right), \\
E_g^{\mu\nu}\=\frac{\kappa}{2}\sqrt{g}\left(\nabla^\mu\nabla^\nu \Phi-g^{\mu\nu}\nabla_\rho\nabla^\rho \Phi+ g^{\mu\nu}\Phi\right).
\eea
Since we are considering a theory of gravity, a proper analysis of the phase space is needed in order to capture the physical degrees of freedom and to extract the symmetries that are not pure gauge.  
% In the next section, we will show how the usual description in FG gauge is embedded into the Bondi gauge formulation. 
We will consider Bondi gauge, which consists in imposing two gauge-fixing conditions on the metric
\begin{equation}
g_{rr}=0\quad \text{and}\quad g_{ru}=-1.
\end{equation}
We start by describing it in Lorentzian signature but we will soon Wick-rotate to Euclidean signature, in which most of our study takes place. The metric takes the simple form 
\begin{equation}
ds^2=2V(u,r)du^2-2dudr,
\label{Bondi}
\end{equation}
where the $r$ coordinate is null and $u$ is a retarded time. The scalar curvature is $R=2\partial_r^2V$. From there, we obtain the most general metric with constant negative scalar curvature in Bondi gauge
\begin{equation}
R=-2\quad\Leftrightarrow\quad V=-\frac{r^2}{2}+P(u)r+T(u),
\label{Vonshell}
\end{equation}
where $P$ and $T$ are any function of the retarded time. The asymptotic Killing vectors, \emph{i.e.} the vector fields which preserve the form of \eqref{Bondi} on-shell, are 
\begin{equation}
\xi=\varepsilon(u)\partial_u-(\varepsilon' r -\eta(u))\partial_r,
\end{equation}
for any function $\varepsilon$ and $\eta$ of the retarded time.  The corresponding variations of $P$ and $T$ are
\begin{equation}
\begin{split}
\delta_\xi P &=\varepsilon P'+\varepsilon'P+\varepsilon''-\eta,\\
\delta_\xi T &=\varepsilon T'+2\varepsilon' T+\eta P-\eta'.\\
\end{split}
\label{TransfoPandT}
\end{equation}
The set of all the vector fields $\xi_{(\varepsilon,\eta)}$ forms an infinite-dimensional Lie algebra whose exponentiation gives the asymptotic symmetry group. The corresponding algebra is called BMS$_2$ and also corresponds to the symmetries of the flat version of Bondi gauge \cite{Afshar:2019axx}. The transformations of $P$ and $T$ will be interpreted later in terms of a coadjoint representation of the asymptotic symmetry Lie algebra.

Having properly specified the phase space for the metric, we will use it to compute the boundary term. The metric \eqref{Bondi} with $V$ given by \eqref{Vonshell} is automatically solution of the equation $E_\Phi$, while the $uu$-component of $E_g^{\mu\nu}$ gives
\begin{equation}
\Phi=\varphi_1(u)r+\varphi_0(u).
\end{equation}
These two new functions also transform under the spacetime symmetry $\xi$
\begin{equation}
\begin{split}
&\delta_\xi \varphi_0 =\varepsilon  \varphi_0'+\eta \varphi_1 ,\\
&\delta_\xi \varphi_1 =\varepsilon  \varphi_1'-\varepsilon'  \varphi_1.\\
\end{split}
\label{TransfoPhi}
\end{equation}
The two other components of $E_g^{\mu\nu}$ give two evolution equations for $\varphi_0$ and $\varphi_1$
\begin{equation}
\begin{split}
&\varphi_1'+P\varphi_1+\varphi_0=0,\\
&\varphi_0''-P\varphi_0'+\varphi_1T'+2T\varphi_1'=0.
\end{split}
\label{Dilaton}
\end{equation}
From now when we say on-shell we mean that these two equations are satisfied (and their linearized version for linear perturbations). The total solution space is parametrized by four functions of the retarded time $(P,T,\varphi_0, \varphi_1)$ whose equations of motion are given by \eqref{Dilaton}.

In the following, we would like to consider the Euclidean version of JT gravity. We will therefore perform a Wick rotation by defining
\begin{equation}
\tau=iu\quad \text{with}\quad \tau\sim \tau+\beta.
\end{equation}
The Euclidean time lies on a circle of length $\beta$. We  also make the following replacements 
\begin{equation}
\begin{split}
&P(u)\to iP(\tau), \quad T(u)\to -T(\tau),\\
&\varepsilon(u)\to -i\varepsilon(\tau), \quad \eta(u)\to i\eta(\tau),\\
&\varphi_0(u)\to -\varphi_0(\tau),\quad \varphi_1(u)\to i\varphi_1(\tau) .
\end{split}
\label{Dictionary}
\end{equation}
All the new functions are periodic in $\tau$. These replacements are chosen so that the equations of motion \eqref{Dilaton} and the expressions for the field variations \eqref{TransfoPandT} and \eqref{TransfoPhi} are unchanged after the Wick rotation. From now on, we will only consider  the Euclidean theory.

This being specified we deduce that the term $\partial_\tau\Theta^\tau$ does not contribute in \eqref{deltaS}. For the moment we will also suppose that there is only one boundary for the AdS$_2$ spacetime, situated at $r=\infty$, so that the third term in $\delta I_{\mathrm{JT}}$ becomes a boundary term
\begin{equation}
B\equiv\int d\tau\, \Theta^r(\tau,r=\infty)=-i\kappa \int d\tau \,(\varphi_0 \delta P-\varphi_1 \delta T).
\end{equation}
For the variational problem to be well posed, we need $B$ to be canceled by the variation of the boundary term in the action: 
\begin{equation}
\delta I_\partial+B=0,
\end{equation}
when we perturb around a solution. This will ensure that solutions to the equations of motions really correspond to extrema of $I_{\r{JT}}$. Consider the following boundary action 
\begin{equation}
I_\partial=-i\kappa\int d\tau \left(\varphi_1 T- \varphi_0 P + \varphi_0' -\frac{\varphi_1'}{\varphi_1}\varphi_0-\frac{\varphi_0^2}{2\varphi_1}\right).
\label{GeneralBoundaryAction}
\end{equation}
One can check that, on-shell, it satisfies the following relation
\begin{equation}
\delta I_\partial+B=i\kappa\int d\tau \,\delta\left(\frac{1}{\varphi_1}\right)C~,
\end{equation}
where the function
\be
C\equiv \varphi_1^2 T-\varphi_1 \varphi_0 P +\varphi_1 \varphi_0' -\varphi_1'\varphi_0-\frac{\varphi_0^2}{2}~,
\ee
is actually a constant on-shell. We conclude that this boundary action is practically the right one, the only thing we have to do is to impose an integrability condition. We impose
\begin{equation}
\varphi_1=\frac{i\bar{\phi}_r}{f'}\quad \text{with}\quad f(\tau+\beta)=f(\tau)+\beta,
\label{condition}
\end{equation}
so that the integrated variation vanishes and $I_\partial$ becomes the right boundary action. The constant $\bar{\phi}_r$ controls the renormalized boundary value of the dilaton \cite{Maldacena:2016upp}. We define also the parameter $\gamma=\kappa \bar{\phi}_r$. The integrability condition is responsible for the appearance of a diffeomorphism $f$ of the boundary circle.

We impose an additional constraint which consists in fixing the zero mode of $\varphi_0/\varphi_1$
\begin{equation}
\frac{1}{\beta}\int d\tau\, \le({\varphi_0 \ov\varphi_1}\ri)=-\bar{\mu},
\label{condition2bis}
\end{equation}
where $\bar{\mu}$ is another constant held fixed in the phase space. To implement this condition, we define $g$ satisfying $g(\tau+\beta)=g(\tau)$ such that  
\begin{equation}
\varphi_0=i\bar{\phi}_r\left(\frac{g'}{f'}-\bar\mu\right).
\label{condition2}
\end{equation}
The interpretation of $\bar{\mu}$ will become clear when we study the partition function. It is this condition that will give rise to the new symmetry. It will also allow for an interpretation of the solution space in terms of coadjoint orbits (a similar condition was considered in \cite{Grumiller:2017qao}). In terms of these new variables, the boundary action is
\begin{equation}\label{IbdyEuc}
I_\partial=\gamma\int \frac{d\tau}{f'} \left(T-g'P+g''-\frac{1}{2}g'^2\right)+ \g \bar\mu\int d\tau\left(P-\frac{f''}{f'}+g'\right)+\r{cste}~.
\end{equation}
The last term is a constant that can always be absorbed by a shift
\begin{equation}
I_\partial\to I_\partial-\r{cste}\,\frac{i\bar{\phi}_r}{\beta}\int d\tau \le({1\ov \varphi_1}\ri),
\end{equation}
which maintains a well-defined variational problem. One should note also that in the term proportional to $\bar{\mu}$, only $P$ contributes with the conditions we have on $f$ and $g$. To obtain the boundary action in the form \eqref{intro:bdyAction} given in the introduction, one should perform a redefinition of the fields described in Sec.\ \ref{Sec:AdSParticle}.

We have imposed additional conditions on the solution space, we therefore need to check if they are affecting the asymptotic symmetry group. The first condition does not lead to a change of the symmetry algebra but the second one requires that $\eta=\sigma'$, with $\sigma$ satisfying the same condition as $g$. The new transformations are
\begin{equation}
\begin{split}
&\delta_\xi f=\varepsilon f',\\
&\delta_\xi g=\sigma + \varepsilon g'.
\end{split}
\label{varfvarg}
\end{equation}
We see that $\ve$ is an infinitesimal reparametrization, while $\s$ acts by shifting $g$. 
% We recognize the action of an infinitesimal diffeomorphism $\varepsilon$ on $f$ and $g$ which transform like functions, while $g$ transforms like a phase under $\sigma$. 
We will now study this symmetry algebra in more details. This will be the occasion to review some mathematical results which are useful to understand the solution space and the boundary action.

\subsection{From BMS$_2$ to warped Virasoro}\label{From BMS2 to warped Virasoro}

As we said earlier, the set of all the vectors $\xi_{(\varepsilon,\eta)}$ forms an infinite-dimensional algebra called BMS$_2$ and the associated bracket is 
\begin{equation}
[(\varepsilon_1, \eta_1),(\varepsilon_2, \eta_2)]=\left(\varepsilon_1\varepsilon_2'-\varepsilon_2\varepsilon_1',(\varepsilon_1\eta_2-\epsilon_2\eta_1)'\right). 
\end{equation}
This algebra corresponds the finite spacetime coordinate transformations
\begin{equation}
\begin{split}
& \tau'=\mathcal{F}(\tau),\\
& r'=\frac{1}{\mathcal{F}'}\left(r+i\mathcal{H}(\tau)\right),
\end{split}
\end{equation}
where $\cF$ is a reparametrization of the circle while $\cH$ is an arbitrary periodic function. They are very similar to the BMS$_3$ transformations  at null infinity in three dimensions, see \cite{Barnich:2010eb}. A major difference is that the Euclidean time here plays the role of the celestial angle there. This has important consequences on the interpretation of the boundary theory. We will use the same terminology as the one used for BMS$_3$ to describe the transformations, $\mathcal{F}$ will be called a boost while $\mathcal{H}$ will be called a supertranslation.

The conditions that we have imposed on the phase space give a constraint on the supertranslation, so that the finite transformations become
\begin{equation}
\begin{split}
& \tau'=\mathcal{F}(\tau),\\
& r'=\frac{1}{\mathcal{F}'}\left(r+i\mathcal{G'}(\tau)\right),
\end{split}
\label{CoordWitt}
\end{equation}
where $\cG$ are arbitrary periodic functions. The corresponding algebra is spanned by the vectors
\begin{equation}
\xi=\varepsilon(\tau)\partial_\tau-(\varepsilon'(\tau)r-i\sigma'(\tau))\partial_r,
\label{KillingEuclid}
\end{equation}
which satisfy the algebra
\begin{equation}
[(\varepsilon_1, \sigma_1),(\varepsilon_2, \sigma_2)]=\left(\varepsilon_1\varepsilon_2'-\varepsilon_2\varepsilon_1',\varepsilon_1\sigma_2'-\varepsilon_2\sigma_1'\right). 
\label{AlgebraLaw}
\end{equation}
If we define the modes $L_n=\left(\frac{i\beta}{2\pi}e^{2\pi i n t/\beta},0\right)$ and $J_n=\left(0, \frac{\beta}{2\pi}e^{2\pi i n t/\beta}\right)$, it becomes
\bea
[L_n,L_m]\=(n-m)L_{n+m}~,\-
 [L_n,J_m]\=-m J_{m+n}~,\-
 [J_m,J_n]\=0,
\eea
which is known as the warped Witt algebra \cite{Detournay:2012pc}. This algebra can be centrally extended
\bea
[\mathcal{L}_n,\mathcal{L}_m] \= (n-m) \mathcal{L}_{n+m} +{c\ov 12}n(n^2-1) \d_{n+m,0}~,\-
[\mathcal{L}_n,\mathcal{J}_m] \=  - m\mathcal{J}_{m+n} - i \la n(n-1) \d_{n+m,0}~,\-
[\mathcal{J}_n,\mathcal{J}_m] \= {k\ov 2} n \d_{n+m,0}~.
\label{Centrally extended algebra}
\eea 
The corresponding group is the semidirect product of the diffeomorphisms of the circle with the smooth functions on the circle: $\mathrm{Diff}(S^1)\ltimes C^\infty(S^1)$. It has the same structure as BMS$_2$ but the group law induced by the coordinate changes \eqref{CoordWitt} is different
\begin{equation}
(\mathcal{F}_1,\mathcal{G}_1 )(\mathcal{F}_2,\mathcal{G}_2)=(\mathcal{F}_2\circ \mathcal{F}_1, \mathcal{G}_1+ \mathcal{G}_2\circ  \mathcal{F}_1).
\end{equation}
The coadjoint representation of this group will play an important role in what follows, and this is why we want to study it here. We will describe only the minimum in order to be self-contained, for a complete mathematical description see \cite{lectures_on_the_orbit_method}. An element of the Lie algebra will be denoted $v$ and we have 
\begin{equation}
v=\epsilon(\tau)\partial_\tau+\sigma(\tau)+a^i e_i.
\end{equation}
The constants $a^i$'s account for possible central extensions. The dual algebra, which is the space of forms on the Lie algebra is then spanned by covectors 
\begin{equation}
b=T(\tau)d\tau^2+P(\tau)d\tau+c_ie^i,
\end{equation}
where $P$ and $T$ are any function on the circle. The $c_i$'s are constants and the $e^i$'s satisfy $e^i(e_j)=\delta^i_j$. The action of $b$ on $v$ is then given by the bracket 
\begin{equation}
\langle b, v \rangle=\int_{S^1} b(v)=a^ic_i+\int_{0}^\beta d\tau(T(\tau)\epsilon(\tau)+P(\tau)\sigma(\tau)).
\end{equation}
Having defined the action of covectors on vectors, we can define the action of a group element on the covector $b$, which is the coadjoint representation
\begin{equation}
\langle \r{Ad}^*_{(\mathcal{F},\mathcal{G})}b,v\rangle=\langle b, \r{Ad}^{-1}_{(\mathcal{F},\mathcal{G})}v\rangle.
\end{equation}
The action of a group element on $b$ induces a transformation of the two functions $P$ and $T$ that are interpreted as currents. Exactly like the holomorphic and anti-holomorphic components of the energy-momentum tensor in a 2d CFT which transform in the coadjoint representation of the Virasoro algebra. The coadjoint representation of the warped Virasoro group is described in \cite{Afshar:2015wjm} and corresponds to the transformations
\begin{equation}
\begin{split}
&\widetilde{P}(\mathcal{F}(\tau))=\frac{1}{\mathcal{F}'(\tau)}\left[P(\tau)+\lambda \frac{\mathcal{F}''(\tau)}{\mathcal{F}'(\tau)}-\frac{k}{2}\mathcal{G}'(\tau)\right],\\
&\widetilde{T}(\mathcal{F}(\tau))=\frac{1}{\mathcal{F}'(\tau)^2}\left[T(\tau)+\frac{c}{12}\{\mathcal{F}(\tau),\tau\}-P(\tau)\mathcal{G}'(\tau)-\lambda \mathcal{G}''(\tau)+\frac{k}{4}\mathcal{G}'(\tau)^2\right].
\end{split}
\label{Coadjoint}
\end{equation}
The three constants $c$, $k$ and $\la$ are all the possible central extensions of the warped Virasoro group. We have defined the Schwarzian derivative 
\begin{equation}
\{\mathcal{F}(\tau),\tau\}=\frac{\mathcal{F}'''}{\mathcal{F}'}-\frac{3}{2}\left(\frac{\mathcal{F}''}{\mathcal{F}'}\right)^2.
\end{equation}
The functions $T$ and $P$ being periodic, we can define the generators $\mathcal{L}_n$ and $\mathcal{J}_n$ to be the modes of $T$ and $P$ on the circle. The centrally extended algebra \eqref{Centrally extended algebra} is then recovered for the bracket $[\mathcal{Q}_{(\epsilon_1,\sigma_1)},\mathcal{Q}_{(\epsilon_2,\sigma_2)}]=-\delta_{(\epsilon_1,\sigma_1)}\mathcal{Q}_{(\epsilon_2,\sigma_2)}$.

These transformations for $P$ and $T$ follow directly from group theory. We can also view the group as acting on the spacetime coordinates like in \eqref{CoordWitt}. This action on the Euclidean Bondi metric also induces finite transformations of the functions $P$ and $T$ appearing in the $uu$-component. Acting on the bulk metric with \eqref{CoordWitt}, we find that they correspond exactly to the coadjoint transormation \eqref{Coadjoint} for the central charges
\begin{equation}\label{AdS2charges}
c=0,\quad k=-2\quad  \text{and}\quad \lambda=-1.
\end{equation}
The transformations \eqref{TransfoPandT}, with $\eta=\sigma'$, are the infinitesimal version of these transformations. The fact that the solution space transforms in the coadjoint representation of the asymptotic symmetry group is mysterious but not rare. It also happens for the usual boundary conditions for AdS$_2$ in Fefferman-Graham gauge as we will see later. This was also shown for a version of the CGHS model in \cite{Afshar:2019axx}. Another important example is the case of 3d gravity with the Brown-Henneaux boundary conditions, for which the solution space is parametrized by the energy-momentum tensor of a 2d CFT.

But there is even more to say about the relation between the coadjoint representation of the warped Virasoro group and 2d gravity in AdS-Bondi gauge. We have found that the bulk action reduces to the boundary term
\begin{equation} 
I_\partial=\gamma\int \frac{d\tau}{f'} \left(T-g'P+g''-\frac{1}{2}g'^2\right)+ \gamma\bar{\mu}\int d\tau\left(P-\frac{f''}{f'}+g'\right),
\label{Boundary Action}
\end{equation}
when evaluated on the Bondi AdS solution space, and when the integrability conditions are taken into account. Interestingly we recognize also the coadjoint action of a group element $(f,g)$ on $T$ in the first integrand and on $P$ in the second one, such that the boundary action becomes
\begin{equation}\label{IasT}
I_\partial=\gamma \int d\tilde{\tau}\,\widetilde{T}(\tilde{\tau})+ \gamma\bar{\mu} \int d\tilde{\tau}\,\widetilde{P}(\tilde{\tau}),
\end{equation}
where we have defined $\tilde{\tau}=f(\tau)$. One should bear in mind that this time, the group element does not come from a coordinate change, but was really defined by the dilaton through the integrability conditions. The solution space of Bondi-AdS is therefore parametrized by a group element $(f,g)$ of the warped Virasoro group and a vector in the dual Lie algebra defining two currents $P$ and $T$. The action is simply given by the coadjoint action of the group element on $T$ and $P$. The transformations \eqref{varfvarg} are nothing but an infinitesimal version of the group law. Indeed, taking
% \begin{equation}
% \begin{split}
% & \mathcal{F}_1(\tau)=\tau+\varepsilon(\tau),\quad \mathcal{G}_1(\tau)=\tau+\sigma(\tau),\\
% & \mathcal{F}_2(\tau)=f(\tau)\quad\text{and}\quad \mathcal{G}_2(\tau)=g(\tau).
% \end{split}
% \end{equation}
\begin{align}
\cF_1(\tau)  & = \tau+\ve(\tau) &  \cF_2(\tau) & = f(\tau ) \\
\cG_1(\tau) & = \s(\tau)  & \cG_2(\tau) & = g(\tau)
\end{align}
gives the linearized version of the group law
\begin{equation}
(\mathcal{F}_1,\mathcal{G}_1 )(\mathcal{F}_2,\mathcal{G}_2)=(f+\varepsilon f',g+\sigma+\varepsilon g').
\label{Infgrouplaw}
\end{equation}
We will make use of this property later when studying the symmetry breaking. 

\subsection{Gravitational charges}
\label{Gravitational charges}

In gauge theories, charges associated with asymptotic symmetries can be constructed using the covariant phase space formalism \cite{Wald:1993nt, Iyer:1994ys, Wald:1999wa, compere_advanced_2018}. This gives a way to define surface charges associated to diffeomorphisms that are not pure gauge because of the presence of a boundary. The diffeomorphisms that we study here are given in \eqref{CoordWitt}.

At first one constructs the field variation of a charge (which corresponds to a one-form in the field configuration space) in the following way. Starting with the symplectic potential 
\begin{equation}
\bold{\Theta}=\Theta^\mu \sqrt{g}\,\varepsilon_{\mu\nu}\,dx^\nu=i\kappa \left(\varphi_0\delta P -\varphi_1\delta T\right)d\tau,
\end{equation}
the field-exterior derivative defines the symplectic form
\begin{equation}
\pmb{\omega}=\delta\bold\Theta=i\kappa(\delta\tilde{\varphi}_0\wedge\delta P-\delta\varphi_1\wedge\delta T)d\tau,
\end{equation}
where we have defined $\tilde{\varphi}_0=i\bar{\phi}_r g'/f'$. The sympetic form $\pmb{\w}$ is a $2$-form in the field space. Now the fundamental  theorem of the covariant phase space formalism tells us that when $\phi$ and $\delta \phi$ are on-shell, there exists a function $k_\xi$ such that 
\begin{equation}
\pmb{\omega}(\delta\phi, \delta_\xi\phi)=dk_\xi(\delta\phi),
\end{equation}
where $d$ is the spacetime exterior derivative. The function $k_\xi$ is a one-form in the field space. It can always be decomposed in the following way
\begin{equation}
k_\xi=\delta \mathcal{Q_\xi}+\Xi_\xi,
\end{equation}
and we say that $k_\xi$ is integrable when $\Xi_\xi$ vanishes. Usually, when we have integrability, $k_\xi$ is integrated over a codimension-two surface in order to define a charge. In two dimensions, this would just be a point so we leave it as it is. For 
\begin{equation}
\xi=\varepsilon\partial_\tau-(\varepsilon' r -i\sigma')\partial_r,
\end{equation}
we find that $k_\xi$ depends only on $\tau$ and is given by
\begin{equation}
k_\xi=-\frac{i}{2}\kappa \left(2\varepsilon T \delta\varphi_1+\varepsilon\varphi_1\delta T-\varepsilon P\delta\tilde{\varphi}_0+\varepsilon \delta\tilde{\varphi}_0'-\varepsilon'\delta\tilde{\varphi}_0-\sigma'\delta\varphi_1\right).
\end{equation}
Which can be decomposed into an integrable and non-integrable part in the following way
\begin{equation}
\begin{split}
\mathcal{Q}_\xi &=-\frac{i}{2}\kappa \left(2\varepsilon T \varphi_1-\varepsilon P\tilde{\varphi}_0+\varepsilon\tilde{\varphi}_0'-\varepsilon'\tilde{\varphi}_0-\sigma'\varphi_1\right),\\
\Xi_\xi &=-\frac{i}{2}\kappa\,\varepsilon(\tilde{\varphi}_0\delta P-\varphi_1 \delta T).
\end{split}
\label{GravitationalCharges}
\end{equation}
The usual representation theorem states that when the charges are integrable, the bracket
\begin{equation}
\{\mathcal{Q}_\xi,\mathcal{Q}_\chi\}\equiv\delta_\chi \mathcal{Q_\xi}
\end{equation}
defines a representation (possibly centrally extended) of the asymptotic symmetry group. Here, $\delta_\chi \cQ_\xi$ means that we take the field variation of the charge $\mathcal{Q}_\xi$, and replace $\delta\phi$ by $\delta_\chi \phi$. When the charges are not integrable,  one can define the modified bracket 
\begin{equation}
\{\mathcal{Q}_\xi,\mathcal{Q}_\chi\}^*\equiv\delta_\chi \mathcal{Q_\xi}+\Xi_\chi(\delta_\xi\phi).
\label{modifiedbracket}
\end{equation}
This bracket was introduced in \cite{Barnich:2011mi} where it was used to define a centrally extended representation of the BMS algebra in 4d asymptotically flat gravity. It was also used in the context of near-horizon symmetries in \cite{Donnay:2016ejv,Donnay:2019jiz}. One can show that in our case it also defines a representation
\begin{equation}
\{\mathcal{Q}_{(\varepsilon_1,\sigma_1)},\mathcal{Q}_{(\varepsilon_2,\sigma_2)}\}^*=\mathcal{Q}_{[(\varepsilon_1, \sigma_1),(\varepsilon_2, \sigma_2)]}+\frac{\gamma \bar{\mu}}{2}(\varepsilon_1\sigma'_2-\varepsilon_2\sigma_1'),
\label{ChargeAlgebra}
\end{equation}
where the bracket on the right hand side coincides with \eqref{AlgebraLaw}. The central extension is actually trivial since it can be absorbed in the following redefinition of the charge
\begin{equation}
\widetilde{\mathcal{Q}}_\xi=\mathcal{Q}_\xi+\frac{\gamma\bar{\mu}}{2} \sigma,
\end{equation} 
which becomes in terms of $f$ and $g$,
\begin{equation}
\widetilde{\mathcal{Q}}_\xi=\frac{\gamma\bar{\mu}}{2}\sigma+\frac{\gamma}{2f'}\left(2\varepsilon T-\varepsilon P g'-\varepsilon' g' -\sigma'-\varepsilon\frac{g'f''}{f'}+\varepsilon g''\right).
\end{equation}
From the representation theorem, we can  deduce the time evolution of the charge. Using  the vector $(\varepsilon=1, \sigma=0)$, the associated variations are all time derivatives $\delta_{(1,0)}\phi=\phi'$. We replace one of the two vectors by $(1,0)$ in \eqref{ChargeAlgebra} to obtain
\begin{equation}
\frac{d}{d\tau}\widetilde{\mathcal{Q}}_{(\epsilon,\sigma)}=\delta_{(1,0)}\widetilde{\mathcal{Q}}_{(\epsilon,\sigma)}+\widetilde{\mathcal{Q}}_{(\varepsilon',\sigma')}=-\Xi_{(1,0)}(\delta_{(\varepsilon,\sigma)}\phi).
\end{equation}
The non conservation of the charge is sourced by the non integrable part. We conclude that if we find a proper restriction of the phase space on which $\Xi_\xi$ vanishes, the associated symmetries will be true symmetries.

We would like to implement such a restriction on the solution space. An obvious choice is to require that the two currents are constant
\begin{equation}
P(\tau)=P_0\quad\text{and}\quad T(\tau)=T_0,
\end{equation}
held fixed in the solution space. The infinite-dimensional symmetry algebra is then broken to the subalgebra that leaves $P_0$ and $T_0$ invariant. Eq. \eqref{TransfoPandT} becomes
\begin{equation}
\begin{split}
\epsilon'P_0+\epsilon''-\sigma'=0,\\
2\epsilon' T_0+\sigma' P_0-\sigma''=0.\\
\end{split}
\end{equation}
Solving for $\varepsilon$ and $\sigma$, we find
\begin{equation}
\begin{split}
\varepsilon(\tau) &=\lambda_1+\lambda_2\, e^{ i\tau\sqrt{2s_0} }+\lambda_3 \,e^{-i\tau\sqrt{2s_0}},\\
\sigma(\tau) &=\la_2\left( P_0+\sqrt{-2s_0}\right) e^{i\tau\sqrt{2s_0}}+ \la_3 \left( P_0-\sqrt{-2s_0}\right) e^{-i\tau\sqrt{2s_0}}+\lambda_4,
\end{split}
\label{Solepsilon}
\end{equation}
where we have defined $-s_0=T_0+\frac{1}{2}P_0^2$. For the transformations associated to $\lambda_2$ and $\lambda_3$ to be well-defined (not considering winding), we need to have
\begin{equation}
s_0=- T_0-\frac{1}{2}P_0^2= \frac{2\pi^2}{\beta^2}.
\label{HypDisk}
\end{equation}
Then, the transformations associated to $\la_1,\la_2$ and $\la_3$ generate an $\r{SL}(2,\mathbb{R})$ symmetry. Moreover there is an extra $\r{U}(1)$ symmetry corresponding to shifting of $g$ by a constant, which is $\lambda_4$ here. Therefore, requiring $P$ and $T$ to be constant and to satisfy the relation \eqref{HypDisk} realizes the symmetry breaking 
\begin{equation}
\r{Diff}(S^1)\ltimes C^{\infty}(S^1)\to \r{SL}(2,\mathbb{R})\times \r{U}(1).
\end{equation}
We have found that using this boundary condition, which leads to integrable charges, the asymptotic symmetry group is bigger than the vacuum symmetry group $\r{SL}(2,\R)$.\footnote{The same phenomenon happens for example in AdS$_3$, with the Brown-Henneaux boundary condition. The charges are also integrable and the symmetry group is the infinite dimensional 2d conformal group while the vacuum symmetry is only the global part.} Morevover, the gravitational charge associated to the $\mathrm{U}(1)$ symmetry is 
\begin{equation}
\widetilde{\mathcal{Q}}_{(\varepsilon=0, \sigma=1)}=\frac{\gamma\bar{\mu}}{2}.
\label{U(1)charge}
\end{equation}
This additional symmetry will play an important role in the study of the Euclidean path integral. Moreover, one can show that with this condition, the gravitational charges agree with the Noether charges of the boundary action (ignoring the term proportional to $\bar{\mu}$ which does not contribute to the dynamics).

One should note that the global $\r{U}(1)$ transformation $(\varepsilon=0, \sigma=1)$ has actually no real effect on the initial fields as it leaves $\Phi$ and $g_{\mu\nu}$ invariant. This however does not mean that it is not a true symmetry. In fact, such symmetries arise frequently and are known as reducibility parameters. They act trivially on the fields on-shell but have a non trivial charge (see the generalized Noether's theorem in \cite{Ruzziconi:2019pzd}).\footnote{One can think of the gauge transformation $A\rightarrow A+d\lambda$ in a free $\r{U}(1)$ gauge theory. The transformation $\lambda=1$ leaves $A$ invariant and still, the corresponding charge is non trivial since it is the electric charge \cite{Henneaux:2018gfi}. }

In what follows we will also consider geometries for which $s_0\neq \frac{2\pi^2}{\beta^2}$ (in particular the trumpet geometry). In that case, the transformations associated to $\lambda_2$ and $\lambda_3$ are not defined so the symmetry breaking pattern is
\begin{equation}
\r{Diff}(S^1)\ltimes C^{\infty}(S^1)\to \r{U}(1)\times \r{U}(1),
\end{equation}
where the unbroken symmetries correspond to the zero modes of $\varepsilon$ and $\sigma$.

\subsection{Particle interpretation }\label{Sec:AdSParticle}

We consider again the boundary action \eqref{Boundary Action} and make the change of variable $\tilde{\tau}=f(\tau)$ together with the field redefinition $\tilde{f}=f^{-1}$ and $\tilde{g}=-g\circ f^{-1}$. The boundary action becomes
\be
I [f,g] = \int d\tau\,\cL[f,g]~,
\ee
with the following Lagrangian 
\bea\label{Effaction}
\cL[f,g]\= \gamma\le(T f'^2 -{1\ov 2}g'^2 + P f'g' + {g' f''\ov f'}-  g'' \ri)+ \g\bar\mu \le( P f' + {f''\ov f'} - g' \ri)~.
\eea
This can be interpreted as describing the dynamics of fluctuations $(f,g)\in \r{Diff}(S^1)\ltimes C^\infty(S^1)$ around the background geometry
\begin{equation}\label{backgeo}
ds^2=2\le(\frac{r^2}{2}-iP(\tau) r+T(\tau) \ri)d\tau^2+2id\tau dr~.
\end{equation}
As for the usual Schwarzian action, this boundary action can also be understood as the motion of a boundary particle \cite{Maldacena:2016upp, Maldacena:2017axo}. In this picture, the full dynamics is captured by the motion of a  particle near the boundary of rigid AdS$_2$. The simplest way to obtain this is to consider a "vacuum" boundary particle whose trajectory is at a constant $r=r_0$. Applying the diffeomorphism
\be
\tau\ra f(\tau),\qq r\ra {r +i g'(\tau)\ov f'(\tau)}~,
\ee
we obtain a new trajectory depending on $(f,g)$. The worldline action of the corresponding boundary particle on the geometry \eqref{backgeo} takes the form
\be
I_\r{particle} = \int ds \,\cL_\r{particle}=\frac{\gamma}{2} \int ds \,\dot{x}^2 ~.
\ee
We consider a particle whose trajectory is the image of the vacuum one
\begin{equation}
\tau=f(s),\quad r={r_0 +i g'(s)\ov f'(s)},
\end{equation}
and compute the corresponding Lagrangian, we obtain
\be
\cL_\r{particle}  + {1\ov 2}\gamma\, r_0^2 = \cL[f,g]~.
\ee
Up to a constant term, the particle Lagrangian is the same as the one appearing in the gravitational boundary action \eqref{Effaction}. Intriguingly, the match is achieved with the identification $\bar\mu = r_0$. When studying the Euclidean path integral in Sec.\ \ref{Sec:EucPath}, we will see that $\bar\mu$ is interpreted as the chemical potential associated with the $\r{U}(1)$ charge. The identification with the boundary particle suggests that this $\r{U}(1)$ charge is related to the radial direction of AdS$_2$. This particle usually lies at $r_0\to+\infty$, \emph{i.e.} close to the asymptotic boundary of AdS$_2$. It is interesting to note that we do not need to take such a limit here to  match the boundary action with the particle action. This suggests that there might be a relation with the finite cutoff versions of JT gravity discussed recently \cite{Iliesiu:2020zld,Stanford:2020qhm}.

\section{Relation to the Schwarzian}
\label{Relation to the Schwarzian action}

In this section, we show that the boundary action derived in the previous section can be seen as a generalization of the usual Schwarzian action. We start by writing a diffeomorphism that maps the solution space of Bondi gauge to the one FG gauge with the usual Dirichlet condition. With an additional boundary condition in Bondi AdS, we recover the Schwarzian action starting from the one derived in Bondi gauge \eqref{Boundary Action}. We also show how the asymptotic symmetries reduce to one copy of the Virasoro group and how to recover the FG gravitational charges from the Bondi ones given in \eqref{GravitationalCharges}. The computation of the partition function in Sec.\ \ref{Sec:EucPath} will show that the Schwarzian theory is actually a subsector of the Bondi theory, corresponding to setting the $\r{U}(1)$ charge to zero.

\subsection{From Bondi to Fefferman-Graham}
\label{From Bondi to Fefferman-Graham}

We start by writing the bulk metric in FG gauge. The metric takes the simple form
\begin{equation}\label{FGAdS2}
ds^2=r^2V(t,r)^2dt^2+\frac{dr^2}{r^2}.
\end{equation}
The most general solution to the equation $R=-2$ is 
\begin{equation}
V(t,r)=h(t)-\frac{s(t)}{2r^2}.
\end{equation}
The usual boundary condition is $h(t)=1$. It is the equivalent of the Brown-Henneaux boundary condition in two dimensions, \emph{i.e.} requiring the boundary metric to be flat. Looser boundary conditions were considered in \cite{Grumiller:2017qao}, where $h$ was allowed to fluctuate. When asking $h(t)=1$, the asymptotic Killing vectors are
\begin{equation}
\xi=\left(\varepsilon(t)-\frac{\varepsilon''(t)}{2r^2-s(t)}\right)\partial_t-\varepsilon'(t) r\partial_r,
\end{equation}
where $\varepsilon$ is any function of the time direction. The asymptotic symmetry group is therefore isomorphic to $\mathrm{Diff}(S^1)$. The effect of the Killing vector on the metric is to modify the function $s$ in the following way
\begin{equation}
\delta_\xi s=\varepsilon s'+2s\varepsilon'+\varepsilon'''.
\end{equation} 
Again, one can interpret this transformation as the coadjoint action of an element of the Lie algebra of $\mathrm{Diff}(S^1)$ on a covector. This means that the function $s$ transforms exactly like the holomorphic component of a 2d CFT energy-momentum tensor. The finite version of this transformation is 
\begin{equation}
\tilde{s}(\mathcal{F}(\tau))=\frac{1}{\mathcal{F}'(\tau)^2}\left[s(\tau)-\frac{c}{12}\{\mathcal{F}(\tau),\tau\}\right],
\label{coadjointVirasoro}
\end{equation}
where we have $c=12$.

An analysis similar to the one realized in Sec.\ \ref{sec:boundaryaction} was done in \cite{Grumiller:2017qao} for the Schwarzian action. They derive the boundary term by demanding a well-defined variational problem. We are going to take another route here, which consists in recovering the Schwarzian action from the boundary action in Bondi gauge.

We would like to find a diffeomorphism that maps the metric in Bondi gauge \eqref{backgeo} to the metric in FG gauge (with $h(t)=1$).\footnote{See \cite{Poole:2018koa, Compere:2019bua, Compere:2020lrt} for a similar construction in higher dimensions.} We start by going in tortoise coordinates
\begin{equation}
\tau=t^*-i r^*, \quad r=-\cot r^*,
\end{equation}
followed by 
\begin{equation}
t^*=t+\sum_{n=1}^\infty v_n(t)r^{-n},\quad r^*=\sum_{n=1}^\infty w_n(t)r^{-n}.
\end{equation}
We can solve order by order in $r^{-1}$ to find the coefficients\footnote{The same notation has been used for the radial coordinate in Bondi and FG gauge.}. The first ones are 
\begin{equation}
v_1(t)=0, \quad w_1(t)=1,\quad v_2(t)=-P(t),\quad w_2(t)=iP(t), \quad ...
\end{equation}
The resulting metric is indeed in FG gauge with a function $s$ written in terms of $P$ and $T$:
\begin{equation}
s(t)=-\left(T(t)+\frac{1}{2}P(t)^2-P'(t)\right).
\label{FunctionS}
\end{equation}
% We conclude that if we want to describe the solution space of the FG gauge, we should consider the Bondi gauge with an additional boundary condition. 
Compared to the Schwarzian action, the boundary action in Bondi gauge has an additional mode $g$ and a new physical symmetry. To recover the Schwarzian action, we need to kill this additional mode by imposing a further boundary condition in Bondi gauge. We choose to impose $P(t)=P_\star(t)$ where $P_\star$ is fixed in the solution space. This translates into a condition on the asymptotic Killing vectors. The first equation of \eqref{TransfoPandT} gives
\begin{equation}
\sigma'=\varepsilon' P_\star+\varepsilon P_\star'+\varepsilon'',
\label{SymCond}
\end{equation}
so that the Bondi asymptotic Killing \eqref{KillingEuclid} is now parametrized by $\varepsilon$ only. We can deduce the transformation of $s$ from the second equation of \eqref{TransfoPandT} and \eqref{FunctionS}
\begin{equation}
\delta_\xi s=\varepsilon s'+2s\varepsilon'+\varepsilon'''.
\label{TransfoS}
\end{equation}
This is exactly the transformation we have obtained from the asymptotic Killing vectors in FG gauge. 

Now, we would like to impose this further boundary condition at the level of the boundary action and write it in terms of the function $s$. We recall that we have two remaining equations of motion \eqref{Dilaton} for the dilaton components $\varphi_0$ and $\varphi_1$. In terms of the group element $(f,g)$ they are
\begin{equation}
\begin{split}
\frac{1}{f'}\left(P_\star-\frac{f''}{f'}+g'\right)&=\bar{\mu},\\
\left[\frac{1}{f'^2}\left(T-P_\star\,g'+g''-\frac{g'^2}{2}\right)\right]' &=0.
\end{split}
\label{Dilatonequationfandg}
\end{equation} 
Solving the first equation for $g'$ and using the definition of $s$ in \eqref{FunctionS} (together with the two conditions \eqref{condition} and \eqref{condition2} on the solution space), we can rewrite the boundary term \eqref{GeneralBoundaryAction} in terms of $f$ and $s$ as
\begin{equation}
I_\partial=-\gamma\int \frac{d\tau}{f'}\left(s-\{f,\tau\}\right)=-\gamma\int d\tilde{\tau}\,\tilde{s}(\tilde{\tau}).
\label{SchwarzianAction}
\end{equation}
The integrand is exactly the coadjoint action \eqref{coadjointVirasoro} of $f$ on the current $s$.  We observe the same structure as the one discussed in the previous section. The only difference is that the group has changed. Asking the current $P$ to be held fixed in the solution space removes the shift symmetry so that the symmetry group reduces to $\r{Diff}(S^1)$. Interestingly, the integrand of the boundary action (after integrating one of the equations of motion) remains written in terms of the coadjoint action of the central extension of $\r{Diff}(S^1)$, \emph{i.e.} the Virasoro group.

\subsection{Gravitational charges and the Schwarzian}

Having found the mapping from Bondi to FG, we are also able compute the gravitational charges. To do so, we consider the charges found in Bondi gauge \eqref{GravitationalCharges}, and impose the condition $P=P_\star$ with the corresponding condition on the symmetries \eqref{SymCond}. After integrating out the first equation of \eqref{Dilatonequationfandg}, we obtain (up to a constant)
\begin{equation}
\begin{split}
\widetilde{Q}_\varepsilon &=-\frac{\gamma}{2f'}\left(2\varepsilon s -\varepsilon\left(\frac{f'''}{f'}-2\frac{f''^2}{f'^2}\right)+ \varepsilon'\frac{f''}{f'}+\varepsilon''\right),\\
\Xi_\varepsilon &=\frac{\gamma\,\varepsilon}{2f'}\,\delta s.
\end{split}
\label{GravitationalChargesFG}
\end{equation}
Again, one can show that under the modified bracket \eqref{modifiedbracket}, these non-integrable charges belong to a representation the $\r{Diff}(S^1)$ algebra. A consistent condition to have integrability is to impose $s(\tau)=s_0$, a constant held fixed in the solution space. This condition must be preserved by the asymptotic symmetries. From \eqref{TransfoS} we obtain
\begin{equation}
2s_0\,\varepsilon'+\varepsilon'''=0,
\label{StabS0}
\end{equation}
which is solved by 
\begin{equation}
\varepsilon(\tau) =\lambda_1+\lambda_2\, e^{i\tau\sqrt{2s_0}}+\lambda_3\, e^{-i\tau\sqrt{2s_0}}.
\label{SL}
\end{equation}
We recover the $\mathrm{SL}(2,\mathbb{R})$ symmetry of AdS$_2$ for $s_0=2\pi^2/\beta^2$, but we do not have the extra $\mathrm{U}(1)$ symmetry anymore. With the above choice for $s_0$, we define $f^{-1}=\tilde{f}$ and make the change of variable $\tilde{\tau}=f(\tau)$ to obtain the action
\begin{equation}
I_\partial=-\gamma\,\int_0^\beta d\tilde{\tau}\left(\frac{2\pi^2}{\beta^2}\tilde{f}^{\,'2}+ \{\tilde{f}, \tilde{\tau}\}\right)=-\gamma\,\int_0^\beta d\tilde{\tau}\,\le\{\r{tan}\le(\tfrac{\pi}{\beta}\tilde{f}\ri),\tilde{\tau}\ri\}.
\end{equation}
This is the usual Euclidean Schwarzian action. The finite version of the symmetry \eqref{SL} corresponds to the $\mathrm{SL}(2,\,\mathbb{R})$ symmetry of the Schwarzian derivative
\begin{equation}
F\to \frac{a F+b}{c F+d},
\end{equation}
where $F=\r{tan}\le(\frac{\pi}{\beta}\tilde{f}\ri)$. Thus, requiring $s$ to be constant realizes the symmetry breaking from $\mathrm{Diff}(S^1)$ to $\r{SL}(2,\mathbb{R})$. Moreover, one can show that the charges \eqref{GravitationalChargesFG} give three charges, one for each generator of \eqref{SL}, which precisely match with the Noether charges of the Schwarzian action.

\section{Euclidean path integral}\label{Sec:EucPath}

We would like to compute the Euclidean path integral of the boundary action of Bondi AdS. We start with the action given in \eqref{Effaction} which we reproduce below
\be\label{EucAction}
I[f,g]= \gamma\int_{S^1} d\tau\le(T f'^2 -{1\ov 2}g'^2 + P f' g' + { g'  f'' \ov f'}  \ri)+  \g \bar\mu\int_{S^1}d\tau\,P~.
\ee
In the following we will take constant values $P=P_0$ and $ T=T_0$. The on-shell action is obtained for $f(\tau)=\tau$ and $g(\tau)=0$ and reads
\be\label{onshelleucaction}
I_\r{on-shell} = \g \le( \b T_0 +  \mu P_0\ri)~,
\ee
where we have defined $\mu \equiv \b \bar\mu$ which will be interpreted as the chemical potential associated to the $\r{U}(1)$ symmetry. 
The path integral is defined as
\begin{equation}
Z=\int_\cM \mathcal{D}f\mathcal{D}g \,e^{-I}.
\end{equation}
The fields $f$ and $g$ belong to the warped Virasoro group and the action has a symmetry that corresponds to the stabilizer of $P_0$ and $T_0$ under the coadjoint action of the group. Therefore we should integrate over the manifold
\begin{equation}
\cM = (\mathrm{Diff}(S^1)\ltimes C^\infty(S^1))/\mathrm{Stab}(P_0,T_0).
\end{equation}
This manifold is generically infinite-dimensional and is isomorphic to the coadjoint orbit of the covector $(P_0, T_0)$. It can be endowed with a canonical symplectic form which provide a measure for the path integral. 

We will start by computing the partition function on geometries that correspond to the hyperbolic disk. These geometries need to satisfy
\begin{equation}\label{relations0P0T0}
s_0=-\le(T_0+\frac{1}{2}P_0^2\ri)=\frac{2\pi^2}{\beta^2}.
\end{equation}
we recall that $s_0$ is the variable appearing in the  geometry in FG gauge. The value of $P_0$ is fixed on-shell in terms of the constant $\bar\mu$ which appeared in the boundary action. This follows from dividing the first equation in \eqref{Dilaton} by $\vphi_1$ and integrating over the circle, then using the boundary condition \eqref{condition2bis} it leads to
\be\label{P0andmu}
P_0 = \bar\mu= {\mu\ov \b}~.
\ee
The relation \eqref{relations0P0T0} then fixes $T_0$ and we get for the disk
\begin{equation}\label{DiskPT}
\mathrm{Disk:}\quad P_0= {\mu\ov \b}, \quad T_0=-\frac{2\pi^2}{\beta^2}-\frac{\mu^2}{2\b^2}~.
\end{equation}
We see that we have a one-parameter family of configurations, labeled by $\mu$. Note that a particular configuration can be generated by acting with the diffeomorphism \eqref{CoordWitt} with
\begin{equation}
\mathcal{F}(\tau)=\tau,\quad \mathcal{G}(\tau)={\mu\ov \b}\tau+\r{cste},
\end{equation}
on the Euclidean metric with $P_0=0$ and $T_0=-2\pi^2/\beta^2$. The function $\mathcal{G}$ is not periodic, therefore the couple $(\mathcal{F}, \mathcal{G})$ does not belong to the group. This means that $\mu$ parametrizes a family of inequivalent coadjoint orbits. Each of them define good phase spaces but they are still physically inequivalent according to the study of gravitational charges in Sec.\ \ref{Gravitational charges}. 

The disk partition function is then computed as a path integral over $f,g$ for the action defined with the above values of $P_0$ and $T_0$. As a result, it depends on $\b$ and $\mu$
\begin{equation}
Z(\b,\mu)=\int \mathcal{D}f\mathcal{D}g\,e^{-I}~.
\end{equation}
The parameter $\b$ appears because we are fixing the circle to have renormalized length $\b$. The parameter $\mu$ appears as an arbitrary parameter labeling a family of possible boundary actions. We will interpret below the parameter $\mu$ as the chemical potential associated to the $\mathrm{U}(1)$ symmetry. 

\subsection{Cardy thermodynamics} 

Before delving into the computation of the exact partition function. It is instructive to consider the thermodynamics in the saddle-point approximation. This derivation will be similar to the derivation of the Cardy formula in 2d CFT. In fact, we point out in Sec.\ \ref{Sec:warped CFTs} that there is a precise match between the entropy of JT in Bondi gauge and the entropy of warped CFTs.

\ms

We assume that the free energy is approximated by the on-shell action
\be
F \equiv - \log Z =  I_{\r{on-shell}} ~,
\ee
which, as will be shown later, is achieved when $\beta$ is small. For a thermal state, the on-shell action is computed using the values for $P$ and $T$ given in Eq. \eqref{DiskPT} and the classical solution corresponds to $f(\tau)=\tau$ and $g(\tau)=0$. This gives the free energy
\be
F = -{2\pi^2\g\ov \b}+{\g\mu^2\ov 2\b} ~.
\ee
The entropy is obtained using a Legendre transform
\be\label{entropyJTsaddle}
S =-\le(1-\b {\p\ov\p\b}-\mu{\p\ov \p\mu}\ri) F= {4\pi^2 \g\ov \b}~.
\ee
The saddle-point value of the energy and the charge are given by
% \footnote{We use the following definition of the density of states
% $Z(\b,\mu) = \int_0^\infty dE \int_{-\infty}^{\infty}dQ\, \rho_0(E,Q)e^{-\beta E +i \mu Q}$. }
\bea
E\= {\p F\ov \p\b}= {2\pi^2\g\ov \b^2} - {\g\mu^2\ov 2\b^2} ,\-
Q\= i {\p F\ov\p\mu}={i \g\mu\ov \b}~.
\label{saddlevalue}
\eea
Inverting these relations gives the  temperature and chemical potential at the saddle-point
\bea
\b = {2\pi \g\ov \sqrt{2 \g E - Q^2}},\qq \mu = -{2\pi i Q\ov \sqrt{2 \g E-Q^2}}~.
\eea
This allows to rewrite the entropy \eqref{entropyJTsaddle} as 
\be\label{entropyJTsaddleEQ}
S = 2\pi \sqrt{2 \g E-Q^2}~.
\ee
The saddle-point approximation is valid for large energy and charge. After computing the full partition function, we will see that we reproduce this entropy by taking the Cardy limit of the exact density of states \eqref{densityofstates}.  In Sec.\ \ref{Sec:warped CFTs}, we match this entropy with the Cardy formula of a warped CFT. In Sec.\ \ref{Complex SYK}, we also match it with the entropy of the complex SYK model.

\subsection{Partition function}
\label{Partition function computation}

We now consider the computation of the exact partition function.\footnote{We are not including higher genus configurations in this computation. We will come back to them in the next subsection.} 
The Kirillov-Kostant-Souriau symplectic form associated to coadjoint orbit of the warped Virasoro group was obtained in \cite{Afshar:2019tvp}, it is given by
\begin{equation}
\omega=-\alpha \int_{S^1}\left(\frac{df'\wedge df''}{f'^2}-\frac{4\pi^2}{\beta^2}df\wedge df'-d\tilde{g}\wedge d\tilde{g}'\right),
\end{equation} 
where we have defined $\tilde{g}=g-i\frac{\mu}{\beta}f-\log f'$. Here, $\a$ is a normalization constant for the symplectic form.  As in \cite{Stanford:2017thb}, we can use the Duistermaat-Heckman theorem that states that if the action generates a $\mathrm{U}(1)$ symmetry on the integration manifold then the partition function is one-loop exact. Indeed one can check that we have 
\begin{equation}
\omega(.\, ,\delta \phi=\phi')\propto d I,
\end{equation}
which means that $I$ generates a time translation symmetry on the coadjoint orbit.\footnote{This relation implies $\mathcal{L}_{\delta\phi=\phi'}\omega=(d\circ i_{\delta\phi=\phi'})\omega+(i_{\delta\phi=\phi'}\circ d)\omega=0$.} To organize the path integral, it is useful to write it as a classical contribution and a one-loop contribution
\be
Z(\b,\mu) = Z_{\text{1-loop}} \,e^{-I_\r{on-shell}} ~.
\ee
The classical contribution is given in \eqref{onshelleucaction} and gives with the values \eqref{DiskPT} 
\be
I_\r{on-shell} = {\g \mu^2\ov 2\b} - {2\pi^2\g\ov \b}~.
\ee
The one-loop part is obtained by integrating over the fluctuations $\ve(\tau)$ and $\s(\tau)$. The quadratic action is given by
\be\label{quadratic action}
I_\r{quad} =\g \int_{S^1} d\tau\le( T_0 \ve'^2 -{1\ov 2} \s'^2 + P_0 \ve'\s' + \s'\ve''\ri)~.
\ee 
We will start with generic values of $P_0$ and $T_0$ since some of these results will be useful when computing the partition function on the trumpet. The boundary conditions impose that $\ve$ and $\s$ are periodic with period $\b$. Hence, they can be decomposed into modes
\be\label{epsilonsigmadec}
\ve(\tau)= {\b\ov 2\pi}\sum_{n\in \Z} \ve_n e^{-{2\pi\ov \b}i n \tau},\qq \s(\tau) ={\b\ov 2\pi} \sum_{n\in\Z} \s_n e^{-{2\pi\ov \b} i n \tau}~.
\ee
Since $\ve(\tau)$ is real and $\s(\tau)$ is pure imaginary, we have the relations
\be\label{realcondepsilonsigma}
\ve_{-n} = \ve_n^\ast,\qq \s_{-n} = -\s_n^\ast~.
\ee
Using the decomposition, we can write the one-loop integral as
 \be
Z_\text{1-loop}=\int  \prod_{n\in \Z } d\ve_{n} d\s_n\,\r{Pf}(\w)\, e^{-I_\r{quad}}~.
 \ee
We decompose the symplectic form using the modes
\bea
\w \= \alpha {2\b^2\ov \pi i} \sum_{n\geq 1} T_0 n\, d\ve_n \wg d\ve_{-n}- \alpha{\b^2\ov \pi i  } \sum_{n\geq 1} n \,d\s_n \wg d\s_{-n}\-
&& +\alpha{\b^2\ov \pi i } \sum_{n\geq 1}\le[\le(-{2\pi i\ov \b} n^2+ P_0 n\ri) d\ve_n\wg d\s_{-n} +\le(-{2\pi i\ov \b} n^2- P_0 n\ri) d\ve_{-n} \wg d\s_n \ri]~.
\eea
We note that the zero mode $\ve_0$ and $\s_0$ do not appear and are hence degenerate directions of the symplectic form. This corresponds to the $\mathrm{U}(1)\times\mathrm{U}(1)$ symmetry of the coadjoint orbit. As they are degenerate directions, they must be removed in the computation of the partition function. We decompose $\omega$ into blocks
\begin{equation}
\omega=\begin{pmatrix}
\omega_{\varepsilon\varepsilon}&&\omega_{\varepsilon\sigma}\\
\omega_{\varepsilon\sigma}^T&&\omega_{\sigma\sigma}
\end{pmatrix}.
\end{equation}
The Pfaffian takes the form
\bea
\r{Pf}(\w)  \= \r{Pf}(\w_{\ve\ve} + \w_{\ve\s}\w_{\s\s}^{-1} \w_{\ve\s}^T ) \r{Pf}(\w_{\s\s})~.
\eea
We can represent each block of $\w$ as a  $2M\times 2M$ indexed by $(n,m)$ where $-M \leq n \leq M$, $-M \leq
m \leq M$ and $n,m\neq 0$, and we take $M\to+\infty$ at the end. With this parametrization, the components of $\omega$ take the form
\bea
(\w_{\ve\ve})_{nm}\= \alpha{\b^2\ov \pi i} T_0  n\,\d_{n+m}~, \-
(\w_{\ve\s})_{nm} \=\alpha {\b^2\ov 2\pi i }\le(-{2\pi i\ov \b} n^2+P_0 n\ri) \d_{n+m}~, \-
(\w_{\s\s})_{nm} \= \alpha{\b^2\ov 2\pi i} n \,\d_{n+m}~. 
\eea
Making use of the identity
\be
\r{Pf}\bpm 0 & A \\ -A^T & 0 \epm  = (-1)^{M(M-1)\ov 2}\r{det}\,A~,
\ee
we obtain
\be
\r{Pf}(\w_{\s\s})=\prod_{n= 1}^M\le( -\alpha{ \b^2\ov 2\pi i} n \ri)~.
\ee
We also compute
\be
(\w_{\ve\ve} + \w_{\ve\s}\w_{\s\s}^{-1}\w_{\ve\s}^T)_{nm} = {\b^2 n\ov  \pi i}\le({2\pi^2 \ov \b^2}n^2 + {1\ov 2}P_0^2 + T_0\ri)\delta_{n+m}~,
\ee
which gives
\be
\r{Pf}(\w_{\ve\ve} + \w_{\ve\s}\w_{\s\s}^{-1}\w_{\ve\s}^T ) = \prod_{n= 1}^M \le(\alpha{\b^2 n\ov \pi i}\le({2\pi^2 \ov \b^2}n^2 + {1\ov 2}P_0^2 + T_0 \ri)\ri)~.
\ee
Putting everything together gives 
\be\label{Pfaffianfinal}
\r{Pf}(\w )= \prod_{m=1}^M \le(\alpha{\b^2m\ov \pi i}\le({2\pi^2 \ov \b^2}m^2 + {1\ov 2}P^2_0 + T_0 \ri)\ri)\prod_{n=1}^M\le( -\alpha{\b^2\ov 2\pi i} n \ri)~.
\ee
We can write the Pfaffian as
\be
\r{Pf}(\w) = \prod_{n=1}^M \r{Pf}_n(\w) ,\qq \r{Pf}_n(\w) =\alpha^2 \b^2 n^2 \le(n^2 + {\b^2\ov 2\pi^2} \le({1\ov 2}P_0^2 + T_0\ri) \ri)
\ee
We also decompose the quadratic action \eqref{quadratic action} into modes
\be\label{QuadAction}
I_\r{quad} =\g \b \sum_{n\in \Z} \le(T_0 n^2\ve_n\ve_{-n}-{1\ov 2} n^2\s_n\s_{-n} + P_0  n^2 \ve_n\s_{-n}- {2\pi i\ov \b} n^3 \ve_n\s_{-n}\ri)~.
\ee
We restrict the sums to $n\geq 1$ using the constraint \eqref{realcondepsilonsigma} and decompose the modes into their real part and imaginary part
\be
\ve_n = \ve_n^{(R)} + i \ve_n^{(I)},\qq 
\s_n = \s_n^{(R)} + i \s_n^{(I)}~.
\ee
This allows us to write
\be
I_\r{quad} = \sum_{n\geq 1} I_\r{quad}^{(n)}~,
\label{QuadActionmodes}
\ee
where
\bea\nt
I_\r{quad}^{(n)} \= 2 \g \b \le( T_0n^2 |\ve_n|^2+{1\ov 2} n^2|\s_n|^2 + {1\ov 2} P_0  n^2 (\ve_n^\ast \s_n -\ve_n\s_{n}^\ast)+ {\pi i\ov \b} n^3 (\ve_n\s_{n}^\ast+\ve_n^\ast \s_n)\ri) \\ 
\= 2\g \b  \le( T_0 n^2 \le( (\ve_n^{(R)})^2 + (\ve_n^{(I)})^2\ri)+{1\ov 2} n^2\le( (\s_n^{(R)})^2 + (\s_n^{(I)})^2)\ri)\ri.  \-
&&\hspace{1cm} \le. + i  P_0  n^2 \le(\ve_n^{(R)} \s_n^{(I)} -\ve_n^{(I)}\s_{n}^{(R)}\ri)+{2\pi i \ov \b} n^3  \le(\ve_n^{(R)} \s_n^{(R)} +\ve_n^{(I)}\s_{n}^{(I)}\ri)\ri) ~.
\eea
We can perform the Gaussian integral and we obtain\footnote{Here, we have
\be
d^2 \ve_n = d\ve_n^{(R)}d\ve_n^{(I)}= {i\ov 2} d\ve_n d\ve_{-n}~,\qq d^2 \s_n = d\s_n^{(R)}d\s_n^{(I)}=- {i\ov 2} d\s_n d\s_{-n}~
\ee
}
\be\label{generalIntmoden}
\r{Pf}_n(\w) \int d^2\ve_n d^2\s_n  e^{-I_\r{quad}^{(n)}} = {\alpha^2\b^2\ov  n^2 \g^2}~.
\ee
The analysis above was done for arbitrary values of $P_0$ and $T_0$ and will be useful later, when we consider the trumpet. 

We now evaluate the Pfaffian for the values of $P_0$ and $T_0$ corresponding to the disk. From the relation 
\be
{1\ov 2}P_0^2 + T_0 = -s_0= -{2\pi^2\ov \b^2}~,
\ee
we obtain
\be
\r{Pf}^\r{disk}(\w )= \prod_{m=2}^M \le(-2\alpha i \pi m\le(m^2-1 \ri)\ri)\prod_{n=1}^M\le( -{ \alpha\b^2\ov 2\pi i} n \ri)~,
\ee
where we have removed the $m=1$ contribution which would be degenerate. This is because the  coadjoint orbit has an  enhanced symmetry $\mathrm{SL}(2,\mathbb{R})\times \mathrm{U}(1)$, corresponding to the modes $\varepsilon_{\pm 1}$. The Pfaffian for the disk can be written as
\be
\r{Pf}^\r{disk}(\w ) =\r{Pf}_{\sigma_{\pm 1}}(\w)\prod_{n=2}^M\ \r{Pf}_n(\w),\qq \r{Pf}_{\sigma_{\pm 1}}(\w)  \equiv -{\alpha \b^2\ov 2\pi i}~.
\ee
The one-loop path integral can be decomposed as
\begin{equation}
Z_{\text{1-loop}}=Z_{\sigma_{\pm 1}} Z_{\vert n \vert\geq 2},
\end{equation}
where we  compute
\bea\label{quadintngeq2}
Z_{|n|\geq 2} &=&  \prod_{n\geq 2}\r{Pf}_n(\w)\int d^2\ve_n d^2\s_n\, \, e^{-I_\r{quad}^{(n)}}  = {2 \pi\g^3\ov  \alpha^3\b^3}~,\-
 Z_{\sigma_{\pm 1}}  &=& \r{Pf}_{\s_{\pm 1}}(\w)\int d \s_1 d\s_{-1}\, e^{- I_{\r{quad}}^{(1)} }  = { \alpha\b \ov \g}~,
\eea
using zeta regularization for the infinite product
\be\label{infiniteproductzeta}
\prod_{n\geq 1}{x\ov n} = \exp\le[ -\p_s\le. \sum_{n\geq 1} \le( {x\ov n}\ri)^{-s}\ri|_{s=0}\ri] = \exp\le( \z(0)\log x  - \z'(0)\ri) = \sqrt{2 \pi\ov x}~.
\ee
Finally combining everything, we obtain the disk partition function
\be\label{Zbetamu}
Z(\b,\mu) =  {2\pi\g^2\ov \alpha^2\b^2}\exp\le( {2\pi^2\g \ov \b} - {\g\mu^2\ov 2\b}\ri)~.
\ee
Interpreting $\mu$ as a chemical potential associated to our new $\mathrm{U}(1)$ symmetry, we have a density of states $\rho_0(E,Q)$ defined from the formula
\begin{equation}
Z(\b,\mu) = \int_0^\infty dE \int_{-\infty}^{\infty}dQ\, \rho_0(E,Q)e^{-\beta E +i \mu Q}.
\end{equation}
From \eqref{Zbetamu}, we obtain
\begin{equation}\label{densityofstates}
\rho_0(E,Q)=\frac{\gamma}{\alpha^2\pi}\sinh\left(2\pi\sqrt{2\gamma E-Q^2}\right).
\end{equation}
This density of states is the same as the Schwarzian density of states but with a shifted ground state energy. We see that the Schwarzian theory corresponds to the $Q=0$ sector of the Bondi theory. Another way to see it is to consider the mixed ensemble partition function obtained using an inverse Fourier transform
\begin{equation}
Z(\beta, Q)= \sqrt{2\pi}\,\frac{\gamma^{3/2}}{\alpha^2\beta^{3/2}}\exp\left(\frac{2\pi^2\gamma}{\beta}-\frac{Q^2\beta}{2\gamma}\right).
\label{DiskQ}
\end{equation}
We see that for $Q=0$ we recover the Schwarzian partition function. For  $Q=0$, the Fourier transform is just an integration over $\mu$, which can be interpreted gravitationally as a sum over inequivalent fluctuations of the same geometry: the hyperbolic disk. The dynamics of these fluctuations is governed by an action whose coupling constants depend on $\mu$. From a more mathematical perspective, the integration over $\mu$ can also be seen as a sum over non-equivalent coadjoint orbits.

\subsection{Genus expansion and matrix integrals}

In the previous section, we computed the Euclidean partition function of the boundary theory. This can also be seen as the gravitational path integral on the disk geometry. In this section, we consider the full gravitational path integral, including the sum over all possible Euclidean geometries with fixed boundary conditions. For JT gravity in Fefferman-Graham gauge, it  was shown in \cite{Saad:2019lba} that the gravitational path integral was given by correlations functions of a double scaled matrix model. We will see that our new boundary conditions leads to a simple generalization of this result. 

\sss{Gravitational genus expansion}

The full Euclidean JT action  
\begin{equation}
I =I_\r{top}+I_\r{JT},\\
\end{equation}
We have included the pure gravity term
\begin{equation}
I_\r{top}=-\kappa \,\Phi_0 \left[\frac{1}{2}\int_{\mathcal{M}}\sqrt{g}R+\int_{\partial\mathcal{M}}K\right],
\end{equation}
where $\Phi_0$ is a constant which sets the extremal entropy $\mathcal{S}_0=2\pi\kappa\Phi_0$.\footnote{The terminology for $\mathcal{S}_0$ comes from the fact that JT gravity describes the the near-extremal thermodynamics of higher-dimensional black holes. One should think of $\Phi_0$ as much greater than $\Phi$ and of the action $I$ as the linear approximation in $\Phi$, see \cite{Maldacena:2016upp}.} This term is topological in two dimensions. The action $I_{\r{JT}}$ contains a boundary term for the variational problem to be well posed. It ensures that the solutions to the equations of motion are indeed saddle point of the action. When considering the Bondi phase space it will simply reduce to one copy of our boundary action for each circle.

Consider the gravitational path integral on geometries with multiple boundaries. On each boundary $i=1,...,n$ we fix the length $\beta_i$ and the chemical potential $\mu_i$. The gravitational path integral becomes  schematically
\begin{equation}
Z_n(\{\beta_i\}, \{\mu_i\})=\sum_{g=0}^\infty \frac{Z_{g,n}(\{\beta_i\}, \{\mu_i\})}{\left(e^{\mathcal{S}_0}\right)^{2g+n-2}},
\end{equation}
since we have $I_\r{top}=-\mathcal{S}_0\chi=-\mathcal{S}_0(2-2g-n)$. Now each $Z_{g,n}$ corresponds to the path integral on a fixed topology. The integration over the dilaton is crucial since it imposes the constraint $R=-2$. The bulk term of $I_{\r{JT}}$ vanishes in the exponential so that path integral becomes
\begin{equation}
Z_{g,n}(\{\beta_i\}, \{\mu_i\})= \int d(\mathrm{bulk\,\, moduli})\int \mathcal{D}\phi\,e^{-I_\partial}.
\end{equation}
The first integral corresponds to the sum over inequivalent hyperbolic Riemann surfaces with $g$ handles and $n$ boundaries. The second integration corresponds to a sum over the fluctuations associated to the boundary action, where of each boundary the values of $\beta$ and $\mu$ are fixed. We can equivalently fix the the charge $Q_i$ at each boundary using Fourier transform
\begin{equation}
Z_{g,n}(\{\beta_i\}, \{Q_i\})=\frac{1}{(2\pi)^n}\int d\mu_1...d\mu_n\,Z_{g,n}(\{\beta_i\}, \{\mu_i\})\,e^{-i\mu_1 Q_1}...\,e^{-i\mu_n Q_n}.
\end{equation}

\begin{figure}
  \centering
  \includegraphics[width=11cm]{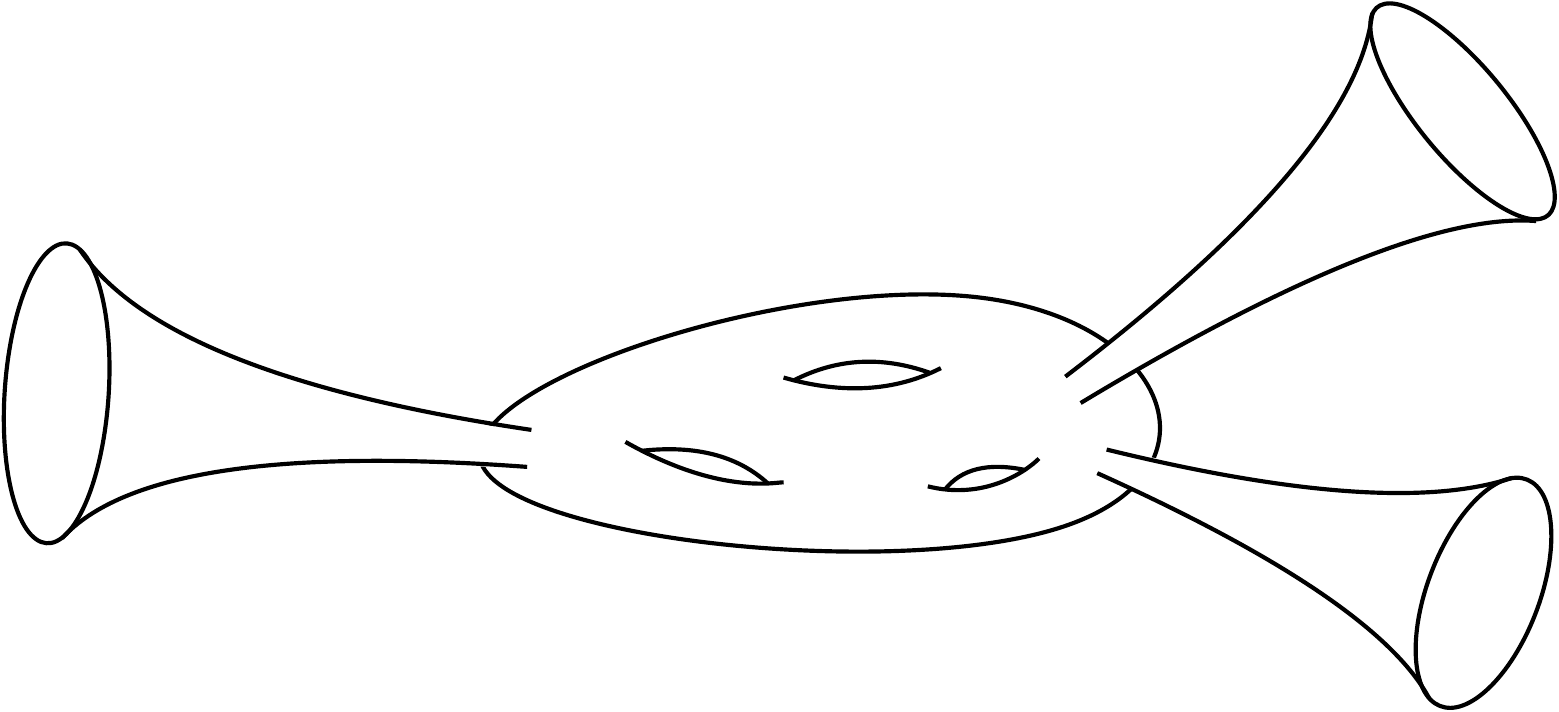}
    \put(0,5){{$(\b_2,Q_2)$}}
  \put(-360,62){{$(\b_1,Q_1)$}}
  \put(-10,140){{$(\b_3,Q_3)$}}
  \caption{Cartoon of the geometry that computes $Z_{g,n}(\{\b_i\},\{Q_i\})$. At the end of each trumpet $i$, we specify the boundary conditions $\b_i$ and $Q_i$, which corresponds to the insertion of $\Tr\,e^{-\b_i \le( H + {Q_i^2/ (2\g)} \ri)}$ in the matrix ensemble.}\label{Fig:CartoonTrumpets}
\end{figure}

\begin{figure}
  \centering
  \includegraphics[width=5cm]{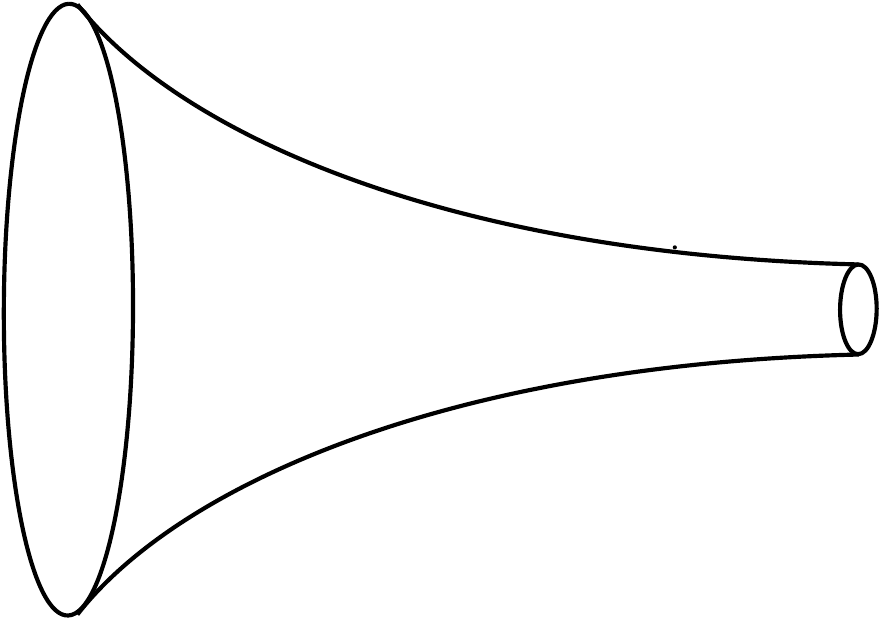}
  \put(-182,45){{$(\b,Q)$}}
      \put(8,47){{$b$}}
  \caption{The trumpet geometry: we specify the boundary conditions $\b$ and $Q$ at the end of the asymptotic boundary. The small end is labeled by the length $b$ of the geodesic boundary.}\label{Fig:Trumpet}
\end{figure}

We will need to consider two types of boundary fluctuations: on the disk and on the trumpet. This is because any hyperbolic Riemann surface with $n$ asymptotic boundaries can be constructed by gluing $n$ trumpets to a genus $g$ surface with $n$ geodesics boundaries. A typical example of such a geometry is depicted in Fig.\ \ref{Fig:CartoonTrumpets}. The case $g=0$ is special since there is only the disk. The disk has no modulus, while for the trumpet there is the length $b$ of the geodesic at the small end, see Fig.\ \ref{Fig:Trumpet}. We obtain 
\begin{equation}
\begin{split}
& Z_{0,1}(\beta,Q) =Z^{\mathrm{disk}}(\beta,Q),\\
& Z_{0,2}(\beta_1, \beta_2,Q_1,Q_2) =\alpha\int_0^{\infty} bdb \,Z^{\mathrm{trumpet}}(\beta_1,Q_1,b)Z^{\mathrm{trumpet}}(\beta_2,Q_2,b),\\
& Z_{g,n}(\{\beta_i\}, \{Q_i\})=\alpha^n\int_0^{\infty}b_1db_1...\int_0^\infty b_ndb_n V^\alpha_{g,n}(b_1,...,b_n)\prod_{i=1}^n Z^{\mathrm{trumpet}}(\beta_i,Q_i,b_i).
\end{split}
\end{equation}
The constant $\alpha$ multiplies the symplectic form\footnote{There is a constant $\alpha$ in front of the Weil-Petersson form, \emph{i.e.} the form which induces a measure on the moduli space and allows to define its volume. This constant $\alpha$ is also in front of the symplectic form on the coadjoint orbit.}. It will ultimately be absorbed in a redefinition of $\mathcal{S}_0$. The function $V^\alpha_{g,n}(b_1,\dots,b_n)$ is the volume of the moduli space of  hyperbolic Riemann surfaces with prescribed geodesic boundaries. The only new feature compared to the computation in \cite{Saad:2019lba} is that our definitions for the disk and the trumpet contributions are different since we have this additional boundary chemical potential $\mu$. The geometries do not change, they are characterized by the value of the current $s_0$ when we write them in FG gauge. For the disk it is $s_0=\frac{2\pi^2}{\beta^2}$, while for the trumpet, which is some sort of hyperbolic disk with a conical defect, it is $s_0=-\frac{b^2}{2\beta^2}$. To obtain these values, one can compute the on-shell value of the Schwarzian action on these geometries.

\subsubsection{Trumpet geometry}

We would like now to compute the trumpet contribution $Z^{\mathrm{trumpet}}$. From the relation 
\be
T_0+\frac{1}{2}P_0^2=-s_0={b^2\ov 2\beta^2}~,
\ee 
and the relation \eqref{P0andmu} between $P_0$ and $\mu$, we obtain the values
\begin{equation}
\begin{split}
\mathrm{Trumpet: } \quad P_0=\frac{\mu}{\beta}, \quad T_0=\frac{b^2}{2\beta^2}-\frac{\mu^2}{2\beta^2}.
\end{split}
\end{equation}
The on-shell action \eqref{onshelleucaction} evaluates to
\be
I_\r{on-shell} = {\g b^2\ov 2\b }+ {\g\mu^2\ov 2\b}~.
\ee
In this case, the Pfaffian \eqref{Pfaffianfinal} takes the form 
\be\label{PfaffianTrumpet}
\r{Pf}^\r{trumpet}(\w )= \prod_{m=1}^M\alpha\frac{ m}{2\pi i}(b^2+4\pi^2m^2)\prod_{n=1}^M\alpha\le( -{\b^2\ov 2\pi i} n \ri)= \prod_{n=1}^M  {\mathrm{Pf}}_n(\w)~.
\ee
The computation is very similar to the disk, which was considered in the previous section. The difference is that  we must not remove the two modes $\varepsilon_{\pm 1}$ because they are not degenerate directions of the symplectic form anymore. 
% The path integral becomes 
% \begin{equation}
% Z^{\mathrm{trumpet}}(\beta, Q, b)=\frac{1}{2\pi}\int_{-\infty}^{+\infty} \frac{d\mu}{2\pi}\sqrt{\alpha} \int \mathcal{D}f\mathcal{D}g\, e^{-I_\partial[\mu]}e^{-i\mu Q},
% \end{equation}
% Again, we integrate over all the inequivalent orbits that correspond to a trumpet geometry. 
The integration is now over the infinite-dimensional manifold
\begin{equation}
(\mathrm{Diff}(S^2)\times C^{\infty}(S^1))/(\mathrm{U}(1)\times \mathrm{U}(1)),
\end{equation}
where two $\mathrm{U}(1)$'s correspond to the zero modes $\varepsilon_0$ and $\sigma_0$. To compute the one-loop contribution, we can use the formula \eqref{generalIntmoden}, which is valid for a general values of $P_0$ and $T_0$. We obtain
\be
Z_\text{1-loop} =\prod_{n\geq 1} \r{Pf}_n(\w)\int d^2\ve_n d^2\s_n\, \, e^{-I_\r{quad}^{(n)}}=\frac{2\pi\gamma}{\beta \a }~,
\ee
where we have used the formula \eqref{infiniteproductzeta} for the infinite product. Combining the one-loop contribution and the on-shell action, we obtain the trumpet contribution
\be
 Z^{\mathrm{trumpet}}(\beta,\mu,  b) =\frac{2\pi\gamma}{\beta\a}\exp\left(-\frac{\gamma b^2 }{2\beta}-\frac{\gamma \mu^2 }{2\beta}\right).
 \ee

\subsubsection{Matrix integrals}

The genus zero contribution $Z^{\mathrm{disk}}$ was computed in the previous section, see Eq. \eqref{DiskQ}. To have the same normalization as \cite{Saad:2019lba}, we make the following redefinition of the disk partition function at fixed $Q$
\begin{equation}
Z^{\mathrm{disk}}(\beta,Q)=\frac{1}{2\pi}\int_{-\infty}^{+\infty}\frac{d\mu}{2\pi}\sqrt{\alpha}\,e^{-i\mu Q}Z^\r{disk}(\beta,\mu).
\end{equation}
This just corresponds to a redefinition of the chemical potential and the charge
\begin{equation}
\mu \to \frac{\sqrt{\alpha}}{2\pi}\mu, \qquad Q\to \frac{2\pi}{\sqrt{\alpha}}Q.
\end{equation}
We obtain 
\begin{equation}
Z^{\mathrm{disk}}(\beta,Q)=\frac{1}{\alpha^{3/2}}\frac{\gamma^{3/2}}{(2\pi)^{1/2}\beta^{3/2}}\exp\left(\frac{2\pi^2\gamma}{\beta}-\frac{Q^2\beta}{2\gamma}\right).
\end{equation}
We see that for $Q=0$, this coincides with the partition function of the Schwarzian action on the disk. We redefine in same way the trumpet contribution to obtain
\begin{equation}
Z^{\mathrm{trumpet}}(\beta, Q, b)=\frac{1}{\alpha^{1/2}}\frac{\gamma^{1/2}}{(2\pi)^{1/2}\beta^{1/2}}\exp\left(-\frac{\gamma b^2}{2\beta}-\frac{Q^2\beta}{2\gamma}\right).
\end{equation}
For $Q=0$, this matches with the Schwarzian partition function on the trumpet \cite{Saad:2019lba}.

We now have all the ingredients to give the final result. Before, we should make a comment this normalization constant $\alpha$. The overall $\a$ dependence of $Z_{g,n}$ is
\begin{equation}
\alpha^n \cdot \alpha^{3g+n-3}\cdot \alpha^{-n/2}=\alpha^{-3\chi/2}~.
\end{equation}
Since the power of $\a$ is proportional to the Euler characteristic $\chi$, we can absorb $\a$ in a redefinition of the extremal entropy $\mathcal{S}_0$ and set $\a=1$.

We denote with the superscript $^{\mathrm{SSS}}$ the quantities of Saad-Shenker-Stanford \cite{Saad:2019lba}. Here, they correspond to setting to zero all the $\r{U}(1)$ charges. Let's consider the example of the double trumpet, obtained by gluing two trumpets along their geodesic boundaries (or small ends). Its contribution is computed by integrating over the modulus $b$.\footnote{The measure is $bdb$ because we also have the freedom to twist one trumpet with respect to the other.} We find
\begin{equation}
\begin{split}
Z_{0,2}(\beta_1,\beta_2,Q_1,Q_2) &=\int_0^\infty bdb\,Z^{\mathrm{trumpet}}(\beta_1, Q_1, b)Z^{\mathrm{trumpet}}(\beta_2, Q_2, b)\\
& = Z^{\mathrm{SSS}}_{0,2}(\beta_1,\beta_2)\,e^{-\beta Q_1^2/(2\gamma)}e^{-\beta Q_2^2/(2\gamma)},
\end{split}
\end{equation}
where $Z^{\mathrm{SSS}}_{0,2}(\beta_1,\beta_2)=\frac{\sqrt{\beta_1\beta_2}}{2\pi(\beta_1+\beta_2)}$. The difference with \cite{Saad:2019lba} is that each boundary receives a contribution from the corresponding $\r{U}(1)$ charge. This contribution corresponds to a multiplicative factor at each boundary. Indeed, for a generic number of boundaries, we have
\begin{equation}
Z_n(\{\beta_i\},\{Q_i\})=Z^{\mathrm{SSS}}_n(\{\beta_i\})\prod_{i=1}^n e^{-\beta Q_i^2/(2\gamma)}.
\label{Result}
\end{equation}
It was shown in  \cite{Saad:2019lba} that the quantity $Z^{\mathrm{SSS}}_n(\{\beta_i\})$ is equal to
\begin{equation}
\langle \mathrm{Tr}\,e^{-\beta_1 H}...\,\mathrm{Tr}\,e^{-\beta_n H}\rangle_c,
\end{equation}
where the average is taken over a double-scaled random matrix ensemble. 
% The matrix $H$ is an $L\times L$ Hermitian matrix drawn in an ensemble, which in the double-scl The $L\to\infty$ maps to a genus expansion. 
This implies that the insertion of a boundary of length $\b$ in the path integral corresponds to an insertion of $\Tr\,e^{-\b H}$ in the correlation function. 
% (the connectedness of the correlation being dual to the connectedness of the spacetime). 

The formula \eqref{Result} tells us that, for our theory, each boundary of length $\b$ and charge $Q$ corresponds to the insertion of 
\be
\Tr\,e^{-\b (H+Q^2/(2\g))}
\ee
in the same matrix ensemble. In other words, we can compute the complete  Euclidean path integral with $n$ boundaries, and boundary conditions $(\b_i,Q_i)$, using the formula
\begin{equation}
Z_n(\{\b_i\},\{Q_i\}) = \le\langle \mathrm{Tr}\,e^{-\beta_1 \le(H+Q_1^2/(2\g)\ri)}\dots \,\mathrm{Tr}\,e^{-\beta_n \le(H+Q_n^2/(2\gamma)\ri)}\ri\rangle_c~,
\end{equation}
where the average is taken in the same matrix ensemble. Note that the effect of the $Q_i$'s is to shift the energy at each boundary. We emphasize that the charges $Q_i$ are not matrices but scalars here.

A similar structure appears when boundary global symmetries are added to the matrix ensemble of JT gravity \cite{Kapec:2019ecr, Iliesiu:2020qvm}. This can be realized by adding a BF theory in the bulk. In this setting, each boundary is labeled by an irreducible representation $r_i$ of the corresponding group. The result is that there is also a factorization 
\begin{equation}\label{BFgenusexpansion}
Z_n(\{\beta_i\},\{r_i\})\propto \delta_{r_1,...r_n}Z^{\mathrm{SSS}}_n(\{\beta_i\})\prod_{i=1}^n e^{-\beta c_2(r_i)/2}.
\end{equation}
The similarity is that the gauge symmetry in the bulk also shifts the ground state energy, by the Casimir $c_2(r)$ of the representation, which is indeed proportional to $Q^2$ for us. The difference is that in our model,  we can have different charges at different boundaries of a connected geometry: there is no constraint enforcing equality of the charges. This stems from the fact that in the trumpet, there is no charge at the small end of the trumpet, which was possible in \cite{Kapec:2019ecr, Iliesiu:2020qvm} due to the presence of bulk gauge fields. Our model only contains pure gravity and all the dynamics reside at the boundary.

\section{Complex SYK}\label{Complex SYK}

In this section, we show that JT gravity in Bondi gauge reproduces the low-energy effective action of the complex  SYK model. This implies that the complex SYK model contains a subsector describing JT gravity in Bondi gauge, in the same way that the Majorana SYK model contains JT gravity in FG gauge as a subsector.  A similar match between a flat space version of our boundary action and the complex SYK model was given in \cite{Afshar:2019axx}. 

\subsection{The complex SYK model}

The complex SYK model \cite{Davison_2017, Bulycheva:2017uqj, Gu:2019jub} is a generalization of the SYK model \cite{Polchinski:2016xgd, Maldacena:2016hyu}, obtained by replacing the Majorana fermions with complex fermions. This model is of special interest because it is maximally chaotic and solvable at large $N$, but closer to condensed-matter systems than the original SYK model. For example, it was recently used to investigate strange metals \cite{Guo:2020aog}.  

\ms

The complex SYK model is a quantum mechanical model involving a large number $N$ of complex fermions with a random interaction. The Hamiltonian is
\be
H =  \sum_{\{i_a\}}  J_{i_1\dots i_q} \cA\le\{\psi_{i_1}^\dg \dots \psi_{i_{q/2}}^\dg \psi_{i_{q/2+1}} \dots\psi_{i_q}\ri\}~,
\ee
where $\cA\{\dots\}$ denotes the antisymmetrized product of operators. The couplings $J_{i_1\dots i_q}$ are independent random complex variables with zero mean and variance
\be
\ol{ |J_{i_1\dots i_q} |^2} = J^2 {(q/2)!(q/2-1)!\ov N^{q-1}}~.
\ee 
where $q$ is an even integer greater than four. There is a global $U(1)$ charge
\be
\widehat{Q} = \sum_{i=1}^n \cA\le\{ \psi_i^\dg\psi_i\ri\}~.
\ee
We also define the specific charge $\cQ = {1\ov N}\ln \widehat{Q}\rn$ which is related to the UV asymmetry of the Green function
\be
G(\tau_1-\tau_2) = -\ln T\{\psi(\tau_1)\psi^\dg(\tau_2)\}\rn,\qq G(0_+) = -{1\ov 2} +\cQ,\qq G(0^-) = {1\ov 2}+\cQ~.
\ee
In the IR, the spectral asymmetry parameter $\cE$ characterizes the long-time behavior of the zero temperature Green function 
\be
G_{T=0}(\pm \tau) = \mp e^{\pm \pi \cE} \tau ^{-2/q}\qq \text{for}\qq \tau\gg J^{-1}~.
\ee
Let's denote by $S(N,N\cQ, T)$ the entropy of the complex SYK model at fixed $N$, fixed $\cQ$ and temperature $T$. In the large $N$ limit, the complex SYK model has a zero temperature entropy
\be
\lim_{T\to 0}\lim_{N\to+\infty}{S(N, N\cQ,T)\ov N} = \cS(\cQ)~,
\ee
Here, $\cS(\cQ)$ is a universal function, in the sense that it is fully determined by the structure of the low-energy theory. At small but non-zero temperature $T=\b^{-1}$, the complex SYK model is described by an effective low-energy action given by \cite{Davison_2017}
\be\label{IcSYK}
I_\r{cSYK} = {N K\ov 2} \int_{S^1} d\tau\le(\tilde{g}'+{2\pi i\cE\ov \b}f' \ri)^2 -{N\g_\r{SYK}\ov 4\pi^2} \int_{S^1} d\tau\,\le\{\r{tan}(\tfrac\pi\b f), \tau\ri\}~,
\ee
where $f(\tau)$ is the Schwarzian mode that reparametrizes time and $g(\tau)$ is a $\mathrm{U}(1)$ mode which is periodic in the absence of winding. This action is the analog of the Schwarzian action for the Majorana SYK model. In comparison, the complex SYK model has an additional $\mathrm{U}(1)$ mode. The parameters $\g_\r{SYK}$ and $K$ are defined from thermodynamical properties: $\g_\r{SYK}$ characterizes the linear response in temperature of the entropy
\be
{S(N,N\cQ,T)\ov N} = \cS(\cQ) + \g_\r{SYK}T + O(T^2)~.
\ee
While $K$ is the zero temperature compressibility defined as 
\be
K = \le( {\p \cQ\ov \p \mu_\r{SYK}}\ri)_{T=0}~,
\ee
where $\mu_\r{SYK}$ is the chemical potential associated with the $\mathrm{U}(1)$ charge. The spectral asymmetry parameter is also related to the zero temperature entropy according to
\be
{d \cS(\cQ)\ov  d \cQ} = 2\pi\cE~.
\ee

\subsection{Matching with boundary action}

\pg{Symmetries.} A connection between the complex SYK model and warped CFTs was explained in \cite{Chaturvedi:2018uov}. It was observed that the complex SYK model has an underlying warped Virasoro symmetry which is spontaneously and explicitly broken down to its global part
\be\label{SYKsymbreak}
\r{Diff}(S^1)\ltimes C^\infty(S^1)\lra \mathrm{SL}(2,\R)\times \mathrm{U}(1)~.
\ee
This is the direct analog of the spontaneous and explicit breaking of $\r{Diff}(S^1)\ra \mathrm{SL}(2,\R)$ in the Majorana SYK model. The symmetry breaking pattern \eqref{SYKsymbreak} of the complex SYK model is the same as the symmetry breaking which controls the version of JT gravity described in this paper. This is the first hint of a relation with our Bondi-AdS version of JT gravity and the complex SYK model.

\pg{Action.} In the rest of this section, we will show that the effective action precisely matches with our boundary action. A similar matching between the action of complex SYK and the CGHS model was achieved in \cite{Afshar:2019axx} in the context of flat holography. This matching was involving a special scaling limit, which should correspond to the flat limit of our theory. Note that a different holographic interpretation of complex SYK, with a 2d gauge field, was proposed in \cite{Gaikwad:2018dfc}.
\ms

We start with our boundary action for AdS-Bondi JT gravity 
\be
I[f,g]= \g \int_{S^1} d\tau \le(T_0 f'^2 -{1\ov 2\l_\r{AdS}^2}g'^2 + P_0 f' g' + { g'f''\ov f'} -  g'' \ri)+\r{cste}~,
\label{Euclaction2}
\ee
where we have restored the AdS$_2$ radius $\l_\r{AdS}$. We recall that $\gamma$ is written in terms of the renormalized value of the dilaton and the Newton constant as follows
\be
\g = {\bar\phi_r\ov  8 \pi G_N^{(2)}}~.
\ee
We consider this theory at a temperature $\b$ and chemical potential $\mu$. This corresponds to the choice 
\be
P_0 =  { \mu\ov \b}  ,\qq T_0=- {2\pi^2\l_\r{AdS}^2 \ov \b^2} - {\mu^2\l_\r{AdS}^2\ov 2\b^2}~,
\ee
where we have reinstated the factors of the AdS$_2$ radius $\l_\r{AdS}$.  We now redefine $g$ according to
\be\label{SYKredefg}
 g(\tau ) = \l_\r{AdS}^2 \le( i\la_0 \tilde{g}(\tau)   +\log f'(\tau)+ {\mu_0\ov \b} f(\tau) \ri)~,
\ee
where $\la_0$ and $\mu_0$ are two arbitrary parameters. Note that this redefinition changes the periodicity of $g$ since we have
\be
\tilde{g}(\tau+\b) = \tilde{g}(\tau) + {i \mu_0\ov \la_0}~.
\ee
We can take $\mu_0 = 0$ to obtain a function $\tilde{g}$ that remains periodic. The action can then be written as
\be
I={\g \l_\r{AdS}^2 \la_0^2\ov 2 } \int_{S^1} d\tau\le(\tilde{g}'-{i(\mu_0-\mu)\ov \la_0 \b}f' \ri)^2 - \g \l_\r{AdS}^2\int_{S^1} d\tau\,\{\r{tan}(\tfrac\pi\b f), \tau\}~.
\ee
This is precisely the action of the complex SYK model \eqref{IcSYK} with parameters
\be\label{SYKparameters}
K = {\g \l_\r{AdS}^2 \la_0^2\ov  N},\qq \g_\r{SYK} = {4\pi^2 \g \l_\r{AdS}^2 \ov N}, \qq \cE = {\mu-\mu_0\ov 2\pi \la_0} ~.
\ee
We can also report the gravitational parameters in terms of the SYK parameter
\be
\g \l_\r{AdS}^2=  {\bar\phi_r \l_\r{AdS}^2 \ov  8 \pi G_N^{(2)}}= {\g_\r{SYK}N\ov 4\pi^2} ,\qq \mu = \mu_0 + 2\pi  \la_0\cE,\qq \la_0^2= {4\pi^2 K\ov  \g_\r{SYK}}~.
\ee
We see that $N$ controls the gravitational coupling and the AdS$_2$ radius.  We note that the value $\mu_0=\mu$ corresponds to $\cE=0$ (this value for $\mu_0$ is special because it diagonalizes the symplectic form, see $(3.9)$ of \cite{Afshar:2019tvp}). We have also reported the value of $\la_0$, which is the other arbitrary parameter in the matching.

We have shown that the AdS-Bondi JT gravity described in this paper reproduces the low-energy effective action of the complex SYK model. This relation is on the same footing as the relation between the standard JT gravity and the Majorana SYK model. 

\pg{Thermodynamics.} It is instructive to also match the large $N$ entropy of the SYK model. Following \cite{Chaturvedi:2018uov}, the entropy of the complex SYK model can be written
\be
S_\r{SYK}= 2\pi \sqrt{ {N\g_\r{SYK}\ov 2\pi^2} \le( E-{Q_\r{SYK}^2\ov 2 NK}\ri)}~,
\ee
where we have defined $Q_{\r{SYK}}=\langle \widehat{Q}\rangle$. In comparison, the entropy of JT gravity computed in \eqref{densityofstates} leads at large $2 \l_\r{AdS}^2 \g E -Q^2$ to 
\be
S_\r{JT} = 2\pi \sqrt{2 \l_\r{AdS}^2 \g E -Q^2}.
\ee
With the choice of parameters \eqref{SYKparameters}, we get a precise match provided the relation
\be
Q_\r{SYK} = \la_0 Q~.
\ee
The parameter $\la_0$, which appeared as an arbitrary parameter in the matching of the actions, has the interpretation of a relative rescaling of the $\r{U}(1)$ charges between gravity and the SYK model.

It is also interesting to see that the leading logarithmic correction to the entropy can also be matched. In the grand canonical ensemble, the expression \eqref{Zbetamu} for the partition function shows that we have the correction
\be
\d S_\r{JT} = -2 \log \b~.
\ee
This matches with the logarithmic correction to the corresponding entropy of the complex SYK model, as reported in \cite{Gu:2019jub}. 

We should make a comment on this matching of the entropy. In Sec.\ \ref{Partition function computation}, the partition function  \eqref{Zbetamu} of  JT gravity in Bondi gauge is not exactly the same as the partition function of the complex SYK model considered in \cite{Chaturvedi:2018uov}. This is because our boundary action \eqref{Effaction} has an additional term, involving the coadjoint transformation of $P$, which does not contribute to the dynamics but gives an additional constant in the partition function. The effect of this constant is to change the sign of the $\mu^2$ term in the exponential, and hence corresponds to the replacement $\mu\ra i\mu$. This change can be absorbed in a redefinition of the relation between the partition function and the density of states, leading to the same formula for the density of states and therefore the same entropy.

\section{Relation to warped CFTs}\label{Sec:warped CFTs}

The near-horizon geometry of any extremal black hole has an $\r{SL}(2,\R)\times \r{U}(1)$ symmetry. This was one of the motivation to introduce the notion of warped CFTs in \cite{Detournay:2012pc}. Warped CFTs are theories which are analogous to 2d CFTs but whose symmetry algebra is the  warped Virasoro algebra,  which consists in one copy of the Virasoro algebra together with a $\r{U}(1)$ Kac-Moody algebra. The corresponding global symmetry is $\r{SL}(2,\R)\times \r{U}(1)$. In this section we describe a relation between AdS$_2$ gravity in Bondi gauge and warped CFTs (see \cite{Castro:2014ima} for an early discussion on the subject). We will also comment on some intriguing observations related to the Kerr/CFT correspondence.

\ms
For a warped CFT whose coordinates are $t$ and $\phi$, the warped conformal symmetry corresponds to the following change of coordinates
\begin{equation}
t\rightarrow t+\mathcal{G}(\phi), \quad \phi \rightarrow \mathcal{F}(\phi).
\end{equation}
In our setup, the coordinate $\phi$ becomes the Euclidean boundary time of the AdS$_2$ spacetime while the coordinate $t$ has no immediate gravitational interpretation. In this context, the quantities $T(\phi)$ and $P(\phi)$ are the associated conserved currents; they transform like in Eq. \eqref{Coadjoint} with the replacement $\tau\to \phi$. The modes of $T$ and $P$, denoted respectively $\mathcal{L}_n$ and $\mathcal{J}_n$, satisfy the algebra \eqref{Centrally extended algebra} for the bracket $[\mathcal{Q}_{(\epsilon_1,\sigma_1)},\mathcal{Q}_{(\epsilon_2,\sigma_2)}]=-\delta_{(\epsilon_1,\sigma_1)}\mathcal{Q}_{(\epsilon_2,\sigma_2)}$. The central charge $\la$, called the twist parameter, can always be absorbed in a redefinition of the generator $\mathcal{L}_n$ \cite{Afshar:2019axx}
\be
\cL_n \ra\cL_n + {2 i\la\ov  k} n \cJ_n~,
\ee
together with a shift of the zero modes $\cL_0$ and $\cJ_0$ by suitably chosen constants. This gives the same algebra with $\la=0$  and shifts the central charge $c^{*} = c -{24 \k^2 /k}$, while leaving $k$ unchanged. For the algebra with the central charges taking the values corresponding to AdS$_2$, as reported in \eqref{AdS2charges}, the new central charges after the shift are
\be
c^{*}=12,\qq k^{*} = -2 ,\qq \la^{*}=0 ~.
\label{Newcentralcharges}
\ee
We would like to compute the vacuum values of $P$ and $T$ on the cylinder. The mapping between the plane and the cylinder is realized by the coordinate change \cite{Detournay:2012pc}
\begin{equation}
t\to t-\frac{\alpha}{\beta}\,\phi, \quad \phi\to e^{2\pi i \phi /\beta}~,
\end{equation}
which leads to the thermal identification 
\begin{equation}
(t,\phi)\sim (t+\alpha,\phi+\beta)~.
\end{equation}
Setting the vacuum values on the plane to zero and using the transformations \eqref{Coadjoint}, we obtain the vacuum values on the cylinder 
\be
P^{\r{vac}}_0 =  -\frac{2\pi i \lambda}{\beta}+\frac{\alpha k}{2} ,\qq T^{\r{vac}}_0=-\frac{\pi c}{6\beta^2}-\frac{2\pi \lambda\alpha}{\beta}+\frac{\alpha^2k}{4}.
\ee
We define the effective chemical potential $\alpha^{*}=\alpha-\frac{2\lambda}{k}(2\pi i /\beta)$. This redefinition of the chemical potential allows to rewrite the vacuum values of $P$ and $T$ in terms of the new central charges $c^{*}$, $\lambda^{*}$ and $k^{*}$ as 
\be
P^{\r{vac}}_0 = \frac{\alpha^* k^*}{2} ,\qq T^{\r{vac}}_0=-\frac{\pi c^*}{6\beta^2}+\frac{\alpha^{*2}k^*}{4}.
\ee
We see that the twist parameter has been absorbed in the redefinitions of the central charges and the chemical potential. With our values for the new central charges \eqref{Newcentralcharges} and the identification $\alpha^*\rightarrow \mu$ the vacuum values become
\be
P^{\r{vac}}_0 =  { \mu\ov \b}  ,\qq T^{\r{vac}}_0=- {2\pi^2 \ov \b^2} - {\mu^2\ov 2\b^2}~.
\ee
They correspond exactly to the values of $T$ and $P$ we have been using to compute the gravitational path integral at fixed temperature and chemical potential in Sec.\ \ref{Sec:EucPath}. In the AdS$_2$ bulk, the map from the plane to the cylinder is realized by the following diffeomorphism 
\begin{equation}
\begin{split}
& \tau\to e^{2\pi i \tau /\beta},\\
& r\to \frac{\beta}{2\pi i}\,e^{-2\pi i \tau /\beta}\left(r-{ i \mu\ov \b} -{2\pi\ov\b}\right).
\end{split}
\end{equation}
The connection between our boundary action for AdS$_2$ and warped CFTs follows from the results of \cite{Chaturvedi:2018uov}. There, it was shown that the partition function of the complex SYK model matches with a particular limit of the vacuum character of a warped CFT. This is similar to the relation between the Schwarzian partition function and the vacuum character of a CFT$_2$ \cite{Mertens:2017mtv}. 

This also leads to a matching of the  leading thermodynamics as explained in \cite{Chaturvedi:2018uov}. We find that the thermodynamics of JT gravity in Bondi gauge matches with that of a warped CFT in the Cardy regime. The entropy of a warped CFT can be written as 
\be
S =S_L + S_R~,
\ee
where in the Cardy regime, we have
\be
S_L = {4\pi i \ov k}P_0 P_0^\r{vac},\qq S_R = 2\pi\sqrt{ {c\ov 6}\le(L_0 - {P_0^2\ov k} \ri)}~.
\ee
This was derived in \cite{Detournay:2012pc} and here, we follow the conventions of \cite{Chaturvedi:2018uov}. Using our values for the central charges \eqref{Newcentralcharges}, and writing the quantum numbers as
\be
L_0 = \g E,\qq P_0 = iQ~,
\ee
where $\g = \k \bar\phi_r$ is the parameter controlling the deviation from extremality, we obtain
\be
S_L = {2\pi  \mu Q\ov \b},\qq S_R = 2\pi \sqrt{2 \g E - Q^2}~.
\ee
The right moving entropy $S_R$ matches with the entropy of AdS-Bondi JT gravity  given in \eqref{entropyJTsaddleEQ}. As in \cite{Chaturvedi:2018uov}, we interpret $S_L$ as contributing to the ground state entropy. It can also be noted that the one-loop correction to the entropy in a warped CFT gives also a logarithmic correction
\be
\d S = -2 \log\b~,
\ee
which agrees with  our Euclidean path integral computation.

\ms

We would like to make a final comment on the central charges in Eq. \eqref{Newcentralcharges} and a potential application to the Kerr/CFT correspondence. In the gravitational context, it is natural to rescale $P$ and $T$ by $\k^{-1} = 8\pi G^{(2)}_N$ to obtain physical  central charges multiplied by $\k$
\be
c^*= 12 \k,\qq k^*= -2 \k~.
\ee 
For the AdS$_2$ spacetime appearing in the extreme Kerr black hole, we have $\k= J$\footnote{This is because the Newton constant for the AdS$_2$ factor in the near-horizon region of an extreme Kerr black hole is given by $A_{\mathcal{H}}/G^{(4)}=1/G^{(2)}$, where $A_{\mathcal{H}}$ is the area of the horizon, given by $A_{\mathcal{H}}=8\pi J$ \cite{Bardeen:1999px}. With the convention $G^{(4)}=1$ we obtain $\kappa=J$.} so our central charge reproduces the central charge $c=12 J$ \cite{Guica:2008mu}. A warped Virasoro symmetry has also been described for Kerr in \cite{Aggarwal:2019iay} and we reproduce the central charges $c=12 J$ and  $k=-2J$ that were obtained there. A major shortcoming of the Kerr/CFT approach is a lack of knowledge of the classical phase space on which the symmetry algebra is acting. We hope that the near-AdS$_2$ realization of these symmetries will shed light on this issue.

\section{Near-extremal black holes}\label{Sec:Near-extremal}

Extremal black holes have a universal AdS$_2$ factor in their near-horizon geometries. This makes near-AdS$_2$ holography a nice framework to understand near-extremal black hole dynamics. In this section, we will demonstrate the relevance of our boundary conditions in this context. We will show that they are sensitive to deformations of the extremal black hole beyond the usual addition of mass at fixed charges, using the example of the Reissner-Nordström black hole. We will also describe the gravitational perturbation that takes an extreme Kerr black hole away from extremality and how its dynamics is related to JT gravity in Bondi gauge. In particular we will give an interpretation of the additional $\r{U}(1)$ symmetry as the axial symmetry of the rotating black hole. In this section, we will use the Lorentzian conventions for the variables $P,T,\vphi_0,\vphi_1$. The dictionary  with the Euclidean variables is given in \eqref{Dictionary}.

\subsection{Deformations of Reissner-Nordström}\label{sec:RN}

We start with the near-extreme Reissner-Nordström black hole. It is  known to give JT gravity after a Kaluza-Klein reduction on the sphere \cite{Sarosi:2017ykf,Brown:2018bms}. The 4d geometry is given by
\be
ds^2 = -{ (r - r_+)(r-r_-)\ov r^2} dt^2 + {r^2 dr^2\ov (r-r_+)(r-r_-)} +r^2 d\Om^2~.
\ee
The inner and outer horizons are at
\be
r_\pm = M \pm \sqrt{M^2-Q_{\r{RN}}^2}~,
\ee
where $M$ is the mass and $Q_{\r{RN}}$ is the electric charge of the black hole. In this section, we use units in which the 4d Newton constant is set to $G^{(4)}_N = 1$. The black hole is extremal when the two horizons coincide
\be
r_+=r_- = M_0~,
\ee
where $M_0=Q_{\r{RN}}$ is the extremal mass. The near-extreme black hole is obtained with a deformation
\be\label{deformationhorizons}
r_+ = M_0 + \la \d r_+ + O(\la^2), \qq r_- = M_0 - \la \d r_-+ O(\la^2)~,
\ee
where $\la$ is a small parameter, and $\d r_\pm$ are two constants, which can be translated into deformations of the black hole mass and charge. The near-horizon geometry is then obtained by replacing 
\be
r\ra M_0+\la r,\qq t \ra 2 M_0^2 {t\ov \la}~,
\ee
and taking the limit $\la \to 0$, where we used the  same parameter $\la$ as in \eqref{deformationhorizons}. In the limit $\la\to0$, we obtain the AdS$_2\times S^2$ metric
\be
ds^2 =ds^2 (\r{AdS}_2)+  ds^2(S^2)~,
\ee
where the metric of the sphere is $ds^2(S^2)= M_0^2 d\Om^2$. In FG gauge, we can write the AdS$_2$ metric as
\be\label{RNAdS2FG}
ds^2(\r{AdS}_2)= M_0^2\le(- r^2 \le(1 - {(\d r_++\d r_-)^2\ov 16 r^2} \ri) dt^2 + {dr^2\ov r^2}\ri)~.
\ee
This is the usual AdS$_2$ geometry \eqref{FGAdS2} and we can read\footnote{We consider here the Lorentzian current $s$ which is related to its Euclidean counterpart according to $s^{\r{L}}=-s^{\r{E}}$.}
\be\label{s0RNFG}
s_0= - {1\ov 8}{(\d r_+ +\d r_-)^2}=-\frac{2\pi^2}{\beta}~,
\ee
which determines the AdS$_2$ inverse temperature $\b$. The above equation shows that the formulation in FG gauge is only sensitive to the sum of the deformations $\d r_+ + \d r_-$. 

\ms

In contrast, our Bondi-AdS formulation of JT gravity will be sensitive to the independent values of $\d r_+$ and $\d r_-$. For this reason, it can be seen as a finer version of near-AdS$_2$ holography which differentiates between a larger set of deformations. The additional information will be related to the $\r{U}(1)$ symmetry of our AdS$_2$ boundary action. 

To take the near-horizon limit to AdS$_2$-Bondi, we consider the Reissner-Nordström black hole in Eddington-Finkelstein coordinates
\be
ds^2 = - { (r-r_+)(r-r_-)\ov r^2} du^2 - 2 du dr + r^2d\Om^2~.
\ee
We make the deformation \eqref{deformationhorizons} and take the near-horizon limit  using
\be
r \ra M_0+\la r,\qq u\ra M_0^2 {u\ov \la}~.
\ee
This leads, after rescaling some of the coordinates, to the AdS$_2\times S^2$ geometry with
\be
ds^2(\r{AdS}_2) = 2  \le( -{r^2\ov 2 M_0^2} + P_0 r+ T_0 \ri)du^2 - 2 du dr 
\ee
where
\bea
P_0 = {\d r_+-\d r_-\ov 2 M_0},\qq T_0 ={\d r_+ \d r_-\ov 2}~.
\eea
We note that these are precisely the geometries that are captured by the Bondi-AdS version of JT gravity. We see that the we are indeed sensitive to the independent values of $\d r_+$ and $\d r_-$. As a consistency check, we can verify that
\be
{1\ov 2}\l_\r{AdS}^2 P_0^2 + T_0 = -s_0~,
\ee
as given in \eqref{s0RNFG}, and with $\l_\r{AdS} =M_0$. The usual formulation of JT gravity, in FG gauge, is only sensitive to the combination that gives $s_0$, while our formulation distinguishes between different values of $P_0$ and $T_0$.

\pg{Thermodynamics.}

The simplest deformation of an extremal black hole is to add some mass at fixed charge
\be
M = M_0 + \la^2  \d M+ O(\la^3),\qq  Q_{\r{RN}} \text{ fixed}~,
\ee
this gives the deformation  \eqref{deformationhorizons} with
\be
\d r_+ = \d r_- = \sqrt{2  \d M M_0}~,
\ee
The entropy of the near-extremal black hole is then 
\be
S = S_0+ 2 \pi \lambda M_0\sqrt{ 2 M_0 \delta M} + O(\lambda^2)~,
\ee
where $S_0 =\pi M_0^2$ is the extremal entropy. The second piece is the entropy added by the small addition of mass. This entropy can be reproduced using the Schwarzian theory \cite{Maldacena:2016upp}, whose entropy is given by
\be
S_\r{Schwarzian} = 2\pi \sqrt{2 \g E}~.
\ee
This matches with the perturbation of the near-extremal black hole entropy for $E=\lambda\delta M$ and $\g = \k\bar\phi_r  =\lambda M_0^3$. 

\ms

It is natural to consider a more general deformation of the two horizons, \emph{i.e.} to take $\d r_+\neq \d r_-$. For example, this can be achieved by a deformation of the mass and charge given by
\bea
M \= M_0 +{1\ov 2}(\d r_+-\d r_-)\la + {1\ov 8M_0} (\d r_+ +\d r_-)^2\la^2+ O(\la^3)~,\\\label{RNdefQ}
Q_{\r{RN}}\=  M_0 + {1\ov 2}(\d r_+-\d r_-)\la +O(\la^3)~.
\eea
This corresponds to the following deformation of the extremal black hole: we first increase both the mass and charge by the same $O(\la)$ amount so that the black hole remains extremal and we then increase  the mass by an $O(\la^2)$ amount. This regime is necessary to get the effect that we want: a deformed geometry with $\d r_+\neq \d r_-$.

In this more general case, the near-extremal entropy and temperature are 
\bea
S \= \pi M_0^2 + 2\pi \la M_0 \d r_+ + O(\la^2)~,\-
T_H\= {(\d r_+ +\d r_-)\la\ov 4\pi M_0^2} + O(\la^2)~. 
\eea
The near-extremal entropy can be written as
\be
S = S_0 + \d S_0+ {4\pi^2 M_0^3\ov \b_H} + O(\la^2)~,
\ee
where the first term is a correction to the extremal entropy taking the form
\be
\d S_0 = \pi M_0 (\d r_+ - \d r_-)\la + O(\la^2)~,
\ee
and the second term is linear in the Hawking temperature $\b_H^{-1}$. We expect that JT gravity captures only the term linear in temperature. Indeed, the entropy of JT gravity takes the form
\be
S_\r{JT} = {4\pi^2 \l_\r{AdS}^2 \g\ov \b}= \pi M_0 (\d r_++\d r_-) \lambda ~,
\ee 
where $\b^{-1}$ is the AdS$_2$ temperature which is related to the Hawking temperature  $\b_H^{-1}$ according to
\be
{1\ov \b_H}  = {\la \l_\r{AdS}^2 \ov \b} ~.
\ee
We can also obtain the values of the charges $E$ and $Q$ introduced in Sec.\ \ref{Sec:EucPath}. In the saddle-point approximation, they are given by
\bea
E \={2\pi^2 \g \l_\r{AdS}^2\ov \b^2} - {\g \l_\r{AdS}^2\mu^2\ov 2\b^2}= {\d r_+ \d r_-\ov 2}\lambda~ ,\-
Q \={ i \g \mu \l_\r{AdS}^2\ov \b} =  { i M_0\ov 2}(\d r_- -\d r_+)\lambda~.
\eea
We note that the additional charge $Q$ allows us to probe deformations with $\d r_+\neq \d r_-$. Therefore the version of JT gravity in Bondi gauge is a finer probe of near-extremal black holes deformations. As a consistency check, we can  further verify that the entropy
\be
S_\r{JT} = 2\pi \sqrt{ 2 \l_\r{AdS}^2 \g E - Q^2} = \pi M_0 (\d r_+ + \d r_-)\lambda~,
\ee
is the correct linear response in temperature in the near-extremal entropy. It is also interesting to note that our $\r{U}(1)$ charge $Q$ corresponds here to the change in the electric $\r{U}(1)$ charge $Q_\r{RN}$ of the black hole, as can be seen from \eqref{RNdefQ}.

\subsection{Breaking away from extreme Kerr}\label{Sec:breakingKerr}

The focus of this work is on JT gravity. Nevertheless,  the ideas developed here should be applicable in  more general near-AdS$_2$ spacetimes, even in cases where it is not clear if JT gravity is a good description, \emph{e.g.} when it cannot be obtained by Kaluza-Klein reduction. One such case of interest is the near-horizon geometry of the extreme Kerr black hole (NHEK) \cite{Bardeen:1999px}.

The near-AdS$_2$ physics of near-extreme Kerr was realized in \cite{Castro:2019crn} as a linearized  perturbation of its near-horizon geometry, where one of the mode of the perturbation was shown to be the Schwarzian mode. In this section, we perform an identical analysis in Eddington-Finkelstein coordinates, which allows us to obtain the AdS$_2$ factor in Bondi gauge.  We  consider a consistent perturbation of the NHEK where one of the mode will satisfy the dilaton equations of motion \eqref{Dilaton}. We show how the infinite-dimensional asympotic symmetry algebra of AdS$_2$ is embedded in the the near-horizon geometry of extreme Kerr. We also see that our additional $\r{U}(1)$ charge gets interpreted as the angular momentum of the 4d geometry.

\ms

The NHEK shares many properties with the AdS$_2\times S^2$ geometry, although the angular dependence makes it more complicated. Starting from extreme Kerr in Eddington-Finkelstein coordinates, the near-horizon limit is obtained by performing the change of coordinates
\begin{equation}
u\to \frac{2M_0^2}{\lambda}u, \quad r\to M_0+r\lambda, \quad \phi\to \phi +\frac{M_0}{\lambda}u~.
\label{NHlimit}
\end{equation}
The limit $\la\to 0$ gives the NHEK geometry
\be
ds^2 = M_0^2 (1+\r{cos}^2\t) \le(-r^2 du^2 -2 du dr + d\t^2 \ri) + {4 M_0^2 \,\r{sin}^2\t\ov 1+\r{cos}^2\t} (d\phi+r du)^2~.
\label{NHEK}
\ee
This geometry has an $\r{SL}(2,\R)\times \r{U}(1)$ isometry group. It contains an AdS$_2$ factor in Bondi coordinates. We can consider a more general AdS$_2$ background as given in \eqref{Bondi} and \eqref{Vonshell}. To do so, one can act on the $(u,r)$-coordinates with the Lorentzian version of the diffeomorphism \eqref{CoordWitt}, together with a change in the azimuthal coordinate $\phi$:
% \begin{equation}
% \begin{split}
% & u \to\mathcal{F}(u),\\
% & r\to\frac{1}{\mathcal{F}'}\left(r+\mathcal{G'}(u)\right),\\
% & \phi \to \phi - \mathcal{G}(u).
% \end{split}
% \label{CoordWitt}
% \end{equation}
\be
u \to\mathcal{F}(u),\qq  r\to\frac{1}{\mathcal{F}'}\left(r+\mathcal{G'}(u)\right),\qq
 \phi \to \phi - \mathcal{G}(u)~.
\ee
We notice that the azimuthal coordinate is sensitive to the zero mode of $\mathcal{G}$. This already shows that our $\r{U}(1)$ symmetry is related to the change of angular momentum. This diffeomorphism leads to the geometry
\bea
ds^2 \= M_0^2 (1+\r{cos}^2\t) \le( 2\le(-{r^2\ov 2}+ P(u)r +T(u) \ri) du^2 -2 du dr + d\t^2 \ri) \-
&& \hspace{0.3cm} + {4 M_0^2 \,\r{sin}^2\t\ov 1+\r{cos}^2\t} (d\phi+r du)^2~,
\eea
with general values for $P(u)$ and $T(u)$, written in terms of $\mathcal{F}$ and $\mathcal{G}$. To be able to distinguish between these geometries and obtain non-trivial dynamics, we need to deform the geometry away from extremality. This is done by considering a linearized gravitational perturbation which brings us to the near-AdS$_2$ regime. We consider a similar ansatz than the one used in \cite{Castro:2019crn} for the perturbation:
\bea\nt
ds^2 \= M_0^2 (1+\r{cos}^2\t+\la \chi) \le( 2\le(-{r^2\ov 2}+ P(u)r +T(u) + \la \psi \ri) du^2 -2 du dr + d\t^2 \ri) \\
&& \hspace{0.3cm} + {4 M_0^2\,\r{sin}^2\t\,(1+\la \Phi) \ov 1+\r{cos}^2\t+\la \chi} (d\phi+r du + \la A)^2 + O(\la^2)~,
\label{deformed geometry}
\eea
where a linearization in $\lambda$ is implied. The deformation is parametrized by three functions $\psi$, $\chi$ and $\Phi$ that depend on $u$ and $r$, and a gauge field
\be
A = A_u(u,r,\t) du + A_r (u,r,\t) dr~.
\ee
We now solve the 4d linearized Einstein equation
\be
R_{ab} = 0~.
\ee
The coordinates are denoted $x^a=(x^\mu, \theta,\phi)=(u,r, \theta,\phi)$. Let us explain some of the steps to obtain the solution. Firstly, from $R_{rr}=-{1\ov 2}\p_r^2\Phi = 0$, we obtain
\be
\Phi(u,r) = \tilde\vphi_0(u) + r\vphi_1(u)~,
\ee
where $\tilde\vphi_0$ and $\vphi_1$ are arbitrary functions of $u$. The equations $R_{r\t}= R_{\t\phi} = R_{t\t}=0$ allow us to determine the $\t$-dependence in $A_u$ and $A_r$. Then, the equation $R_{\t\t} = 0$ gives the equation
\be
\Box_2\chi = 2 \chi~,
\label{Boxchi}
\ee
where $\Box_2$ is the Laplacian for the AdS$_2$ metric
\be\label{AdS2backKerr}
ds^2(\r{AdS}_2) =  g_{\mu\nu} dx^\mu dx^\nu = 2 \le(-{r^2\ov 2} + P(u)r + T(u) \ri) du^2 - 2 du dr~.
\ee
We see that $\chi$ is an AdS$_2$ scalar with conformal dimension $\D=2$ as was obtained in \cite{Castro:2019crn}. From $R_{r\phi} = R_{\phi\phi}=0$, we can determine $A_u$ and $A_r$. Then, $R_{tr} = 0$ allows us to determine $\psi$. Finally, from $R_{tt}=R_{t\phi}=0$, we obtain the JT equations of motion. To see this, we first define $\tilde\vphi_0(u) = c_0 + \vphi_0(u)$ which gives
\be
\Phi(u,r) = c_0+ \Phi_\r{JT}(u,r),\qq \Phi_\r{JT}(u,r) = \vphi_0(u) + \vphi_1(u) r,
\ee 
where $c_0$ arises here as an integration constant. Then, $R_{tt}=R_{t\phi}=0$ gives
\be
\begin{split}
&\varphi_1'+P\varphi_1+\varphi_0=0~,\\
&\varphi_0''-P\varphi_0'+\varphi_1T'+2T\varphi_1'=0~.
\label{DilatonNHEK}
\end{split}
\ee
 We see that $\Phi_\r{JT}$ satisfies the equations of motion of the JT dilaton on the background \eqref{AdS2backKerr}, as reported in \eqref{Dilaton}. The additional constant $c_0$ was also observed in \cite{Castro:2019crn}. We report the solution for the gauge field 
\be\label{Kerrgaugefield}
A(u,r,\t) =- {c_0\ov 2}  r du+  \ve_\mn \p^\mu \Psi(u,r,\t) dx^\nu + \p_\mu \a(u,r) dx^\mu ~,
\ee
where $\a(u,r)$ is an arbitrary function arising from the integration and we have defined
\be
\Psi(u,r,\t) \equiv {1\ov 2\,\r{sin}^2\t} (\Phi(u,r)-\chi(u,r))  - {1+\r{cos}^2\t\ov 8} \Phi(u,r)~.
\ee 
We also have a relation that determines $\psi(u,r)$
\be
\p_r^2 \psi(u,r) = 3 \Phi_\r{JT}(u,r)~.
\ee
which is solved by 
\be
\psi(u,r) = \psi_0(u) + r\psi_1(u) + {1\ov 2}r^2 \le( (r -3 P(u)) \vphi_1(u) - 3\vphi_1'(u) \ri)~.
\ee
We can then verify that all the components of the Einstein equation are satisfied. The solution space is now parametrized by $\a(u,r),\psi_0(u), \psi_1(u), \varphi_0(u), \varphi_1(u), \chi(u,r)$ and $\Phi_0$, and their dynamics is controlled by the equations \eqref{Boxchi} and \eqref{DilatonNHEK}. The functions $\a(u,r),\psi_0(u), \psi_1(u)$ are expected to be pure diffeomorphisms in the sense that they can be generated by acting on the unperturbed background with Lie derivatives. This is shown for a similar function $\alpha$ in Appendix B of \cite{Castro:2019crn}.

The constant $c_0$ corresponds to the perturbation of the angular momentum induced by the deformation. Indeed, note that the deformed geometry \eqref{deformed geometry} possesses also an axial $\r{U}(1)$ symmetry, materialized by the Killing $\partial_\phi$. In the AdS$_2$ factor, this symmetry is associated with a 2d  gauge field which appears in the circle fibration and is given by
\be
A_\r{total} = r du  +  \la A  + O(\la^2)~.
\ee
This is an electric field in AdS$_2$ with leading charge normalized to unity. From the expression of $A$ in \eqref{Kerrgaugefield}, we see that  the effect of $c_0$ is to change this charge at linear order in $\la$, which corresponds to a change of the 4d angular momentum.

The constant $c_0$ is also related to the $\r{U}(1)$ gravitational charge introduced in Sec.\ \ref{Gravitational charges}. To see this, we compute the value of $\Phi$ that corresponds to the linearized perturbation to \eqref{NHEK} that one obtain when considering the next to leading order of the near-horizon limit \eqref{NHlimit}. This "vacuum" value of is $\Phi_{\r{vac}}=\r{cste}\cdot r$. If we now parametrize $\varphi_0$ and $\varphi_1$ in terms of $f$ and $g$ as in \eqref{condition} and \eqref{condition2} and ask their vacuum values, \emph{i.e.} $f(u)=u$ and $g(u)=0$, to correspond to the vacuum value of $\Phi$, we obtain the relation
\begin{equation}
c_0+i\bar{\phi}_r\bar{\mu}=0,
\end{equation}
between the constant $c_0$ and the chemical potential. This relation translate into a relation between the 2d gravitational $\r{U}(1)$ charge \eqref{U(1)charge} and the constant $c_0$
\begin{equation}
\widetilde{\mathcal{Q}}_{\r{U(1)}}=\frac{\kappa}{2}c_0.
\end{equation}
Hence, the $\r{U}(1)$ symmetry discussed in Sec.\ \ref{Gravitational charges} is realized geometrically in the near-extreme Kerr black hole: it corresponds to the axial symmetry of the deformation. The corresponding charge computes the change in angular momentum of the 4d solution. Although the Killing $\partial_\phi$ does not act geometrically on AdS$_2$, it has a non-trivial effect on the near-AdS$_2$ physics. We emphasize again that this effect would be invisible in the Schwarzian theory but is captured in the AdS-Bondi boundary action introduced in this paper.

\section{Flat holography in two dimensions}\label{sec:Flat}

In this section, we study a flat space version of our boundary action. This action was derived in \cite{Afshar:2019axx} and shown to correspond to a version of the CGHS model. After a brief study of its solution space and its gravitational charges, we show that it can also be interpreted as a particle moving in a 2d flat spacetime. We compute the exact partition function of the boundary theory and the full gravitational path integral. The latter is greatly simplified by the fact that there are only two flat surfaces with boundaries: the disk and the cylinder. We find a non-vanishing contribution of the cylinder indicating that this theory is holographically dual to an ensemble average.

\subsection{The CGHS model and its boundary action}\label{sec:CGHSandboundaryaction}

The CGHS model \cite{Callan:1992rs} is a theory of two-dimensional gravity in asymptotically flat spacetime. Following \cite{Cangemi:1992bj}, the authors of  \cite{Afshar:2019axx} define a modified version of the CGHS model, dubbed $\wh{\text{CGHS}}$, whose action is
\be\label{CGHShat}
I = {\k\ov 2} \int d^2 x\sqrt{g}\le( \Phi R - 2 \Psi + 2 \Psi \ve^\mn \p_\mu A_\nu\ri) + I_\partial~.
\ee
The corresponding equations of motion are 
\begin{align}
R & = 0, \\
 \ve^\mn \p_\mu A_\nu & = 1,\\
\n_\mu \n_\nu \Phi -g_\mn \Box\Phi & = g_\mn \Psi, \\ \Psi & = \r{const}.
\label{EOMFLAT}
\end{align}
In App.\ \ref{App: Gauge-theoretic formulation} we comment on the AdS version of this model and show that under appropriate gauge-fixing conditions, its dynamics reproduces the one described in Sec.\ \ref{sec:BondiJT}. The flat version of the metric \eqref{backgeo} is
\begin{equation}\label{backgeoflat}
ds^2=2\le(-iP(\tau) r+T(\tau) \ri)d\tau^2+2id\tau dr~.
\end{equation}
where $P$ and $T$ are arbitrary functions. It is the most general flat metric which satisfies $R=0$ in Bondi gauge. It is readily obtained as the $\ell_{\r{AdS}}\to \infty$ limit of its AdS counterpart.
% The solution can be written as
% \begin{align}
% ds^2 & = 2\le(P(u) r+T(u) \ri)du^2-2dudr, & A & = r du +  d\la~,\\
% \Phi & =\varphi_0(u)+\varphi_1(u)r~,
% \end{align}
% and we have the equations
% \begin{equation}
% \begin{split}
% &\varphi_1'+P\varphi_1=  \Psi~,\\
% &\varphi_0''-P\varphi_0'+\varphi_1T'+2T\varphi_1'=0.
% \end{split}
% \end{equation}
% As explained below, the thermal solution corresponds to the choice $P =- {2\pi/ \b}, T = 0$. The first equation then fixes the value of $\Psi$ to be

\ms

We will see that the theory $\wh{\text{CGHS}}$ has thermal solutions corresponding to the Rindler spacetime, which we describe explicitly below. In \cite{Afshar:2019axx} it is noted that the on-shell value of $\Psi$ controls the temperature of the Rindler solution according to
\be
\Psi=-i\bar{\phi}_r P = {2\pi\bar{\phi}_r\ov \b}~.
\label{relationtempflat}
\ee
The above model is closely related to another version of the CGHS model described by the action
\be\label{badCGHSaction}
I= {\k\ov 2} \int d^2 x\sqrt{-g}\le(\Phi R -2 \Lambda\ri)~.
\ee
where $\Lambda$ is a "cosmological constant"\footnote{Not to be confused with the $1/\ell^2_{\r{AdS}}$ in the JT action.} that is fixed from the start. This theory has the same equations of motion as the theory \eqref{CGHShat}  with $\Psi=\Lambda$. From the above discussion, we see that this theory has only a thermal solution with temperature $\b = 2\pi\bar{\phi}_r/ \Lambda$. In particular, flat JT gravity (corresponding to $\Lambda=0$) has only the zero temperature solution.\footnote{This relation is also discussed, in a different context, in Appendix C of \cite{Stanford:2020qhm}.} Since we would like to obtain a non-trivial Euclidean path integral, the theory should contain solutions with different temperatures. This is why we consider the theory \eqref{CGHShat}.

\ms
After a choice of appropriate boundary conditions, the gauge field does not lead to additional degrees of freedom. We refer to \cite{Afshar:2019axx} for a detailed analysis of the solution space, including the derivation of the boundary action 
\be\label{EucFlatAction}
I[f,g]= \gamma\int_{S^1} d\tau\le(T f'^2  + P f' g' + { g'  f'' \ov f'} -g'' \ri)~.
\ee
This action can also be obtained as the $\ell_{\r{AdS}}\to \infty$ limit of our Bondi AdS action \eqref{IbdyEuc}.

\ms

The asymptotic symmetry algebra is the same than in AdS, it is spanned by the vectors 
\begin{equation}
\xi=\varepsilon(\tau)\partial_\tau-(\varepsilon'(\tau)r-i\sigma'(\tau))\partial_r~,
\label{KillingEuclidFlat}
\end{equation}
where $\varepsilon$ and $\sigma$ are periodic functions. When expanded in modes, they satisfy the warped Witt algebra. The corresponding variations of $P$ and $T$  read
\begin{equation}
\begin{split}
\delta_\xi P &=\varepsilon P'+\varepsilon'P+\varepsilon'',\\
\delta_\xi T &=\varepsilon T'+2\varepsilon' T+\sigma' P-\sigma''.\\
\end{split}
\label{TransfoPandTflat}
\end{equation}
They transform in the coadjoint representation of the warped Virasoro group with central charges 
\begin{equation}
k=0,\quad  \lambda=-1\quad  \text{and} \quad c=0~,
\end{equation}
where we refer to Sec.\ \ref{From BMS2 to warped Virasoro} for more details. Following a similar procedure to the one in Sec.\ \ref{Gravitational charges}, we derive the gravitational charges. The computations are given in App.\ \ref{Flat gravitational charges}. In terms of $f$ and $g$, the gravitational charges read
\begin{equation}
\begin{split}
\mathcal{Q}_\xi &=\frac{i\kappa }{2}\Psi\sigma+\frac{\gamma}{2f'}\left(2\varepsilon T-\varepsilon P g'-\varepsilon' g' -\sigma'-\varepsilon\frac{g'f''}{f'}+\varepsilon g''\right),\\
\Xi_\xi &=\frac{\gamma}{2f'} \varepsilon \left(g'\delta P-\delta{T}\right).
\end{split}
\end{equation}
We recall that $\Psi$ is just a constant on-shell. These charges define a centerless representation of the asymptotic symmetry algebra under the modified bracket \eqref{modifiedbracket}. The expression of the non-integrable part indicates that a proper phase space is achieved when $ P$ and $T$ are constants, held fixed in the solution space, which we denote $P_0$ and $T_0$. This condition realizes the symmetry breaking from the infinite-dimensional algebra of asymptotic symmetries to the subset that preserves these values of $P$ and $T$, \emph{i.e.} satisfying 
\begin{equation}
\begin{split}
&\varepsilon'P_0+\varepsilon''=0~,\\
&2\varepsilon' T_0+\sigma' P_0-\sigma''=0~.\\
\end{split}
\end{equation}
This is solved by 
\begin{equation}
\begin{split}
&\varepsilon=\lambda_1+\lambda_2\;e^{-P_0\tau},\\
&\sigma=\lambda_4+\lambda_3\, e^{P_0\tau}-\lambda_2\frac{T_0}{P_0}\, e^{-P_0\tau}.
\end{split}
\label{flatstabilizer}
\end{equation}
The generators $\lambda_2$ and $\lambda_3$ are well defined only if $P_0=\pm 2\pi i/\beta$ (we do not consider winding here). The corresponding algebra is $\r{ISO}(2)\times \r{U}(1)$, a central extension of the 2d Poincaré group. Defining the corresponding generators $K_i\,$, for $i=1,...,4$, the only non-vanishing commutators are
\begin{equation}
\begin{split}
&[K_1, K_+]=-i K_-,\\
&[K_1, K_-]=-i K_+,\\
&[K_+, K_-]=2 K_4,
\end{split}
\end{equation}
where $K_\pm=K_2\pm K_3$. Therefore, we see that asking $P_0=\pm 2\pi i/\beta$ and $T=T_0$ realizes the symmetry breaking 
\be
\r{Diff}(S^1)\ltimes C^\infty(S^1) \ra \r{ISO}(2)\times\r{U}(1)~.
\ee
In that case, the charges become integrable and conserved. For the same values of $P$ and $T$, the boundary action \eqref{EucFlatAction} is invariant under the same symmetry. The centrally extended Poincaré algebra acts on the fields according to 
\begin{equation}
\begin{split}
&\delta_\xi f=\varepsilon f',\\
&\delta_\xi g=\sigma + \varepsilon f',
\end{split}
\label{varfvargflat}
\end{equation}
where $\varepsilon$ and $\sigma$ are given by Eq. \eqref{flatstabilizer}. Moreover, the gravitational charges are then identified with the Noether charges of the boundary action.

With the boundary conditions considered in \cite{Afshar:2019axx}, one of the components of the dilaton equation in \eqref{EOMFLAT} becomes
\begin{equation}
\frac{\bar{\phi}_r}{f'}\left(P-\frac{f''}{f'}\right)=i\Psi.
\end{equation}
Integrating this equation and considering the thermal solution, we obtain the relation \eqref{relationtempflat} between the temperature and the value of $\Psi$.\footnote{ In \cite{Afshar:2019axx}, the constant $\bar{\phi}_r$ is set to one and a different choice of convention is made for $P$, leading to the relation $P=2\pi/\beta$. } The latter is not fixed by the equations of motion, allowing us to consider solutions with different temperatures.

We should make a comment on the fact that asking a maximum of symmetry does not  uniquely fix the solution. Indeed we still have the freedom to choose the value of $T_0$. A similar freedom is present in AdS-Bondi, where only the combination $s_0=-T_0-\frac{1}{2}P_0^2$ was fixed by the temperature, see the end of Sec.\ \ref{Gravitational charges}. Again this freedom is materialized by transformations that do not belong to the group, indeed, all values for $T_0$ are obtained by acting on the couple $(P_0=2\pi i/\beta, T_0=0)$ with the transformation
\begin{equation}
r\to r+\frac{\beta }{2\pi}T_0,
\end{equation}
which corresponds to $\mathcal{G}(\tau)={\b\ov 2\pi i } T_0\tau$ and $\mathcal{F}(\tau)=\tau$, which does not belong to the group because $\mathcal{G}$ is not periodic.

\subsection{A particle moving in flat space}
\label{FlatSolutions}

\begin{figure}
  \centering
  \includegraphics[width=7cm]{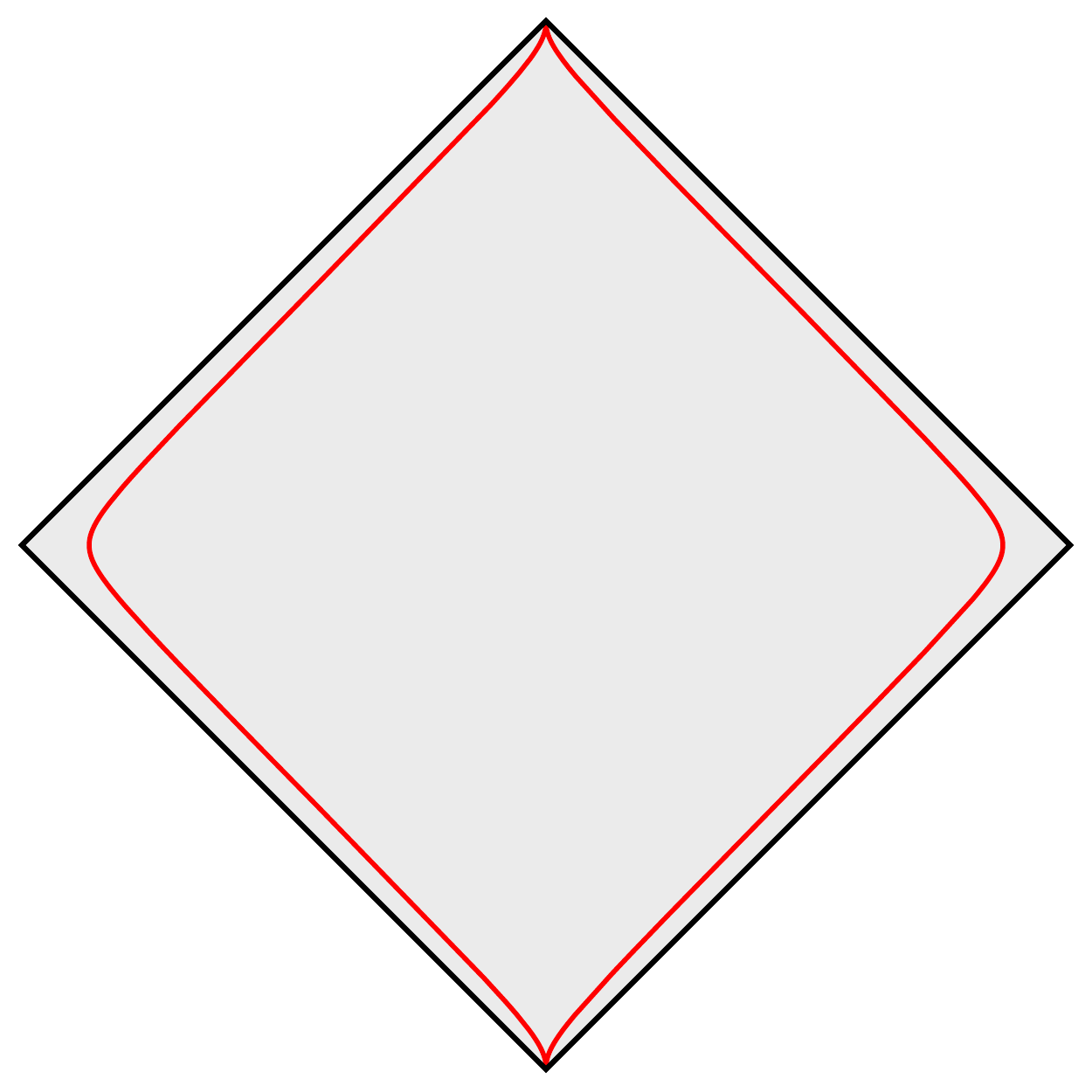}
    \includegraphics[width=7cm]{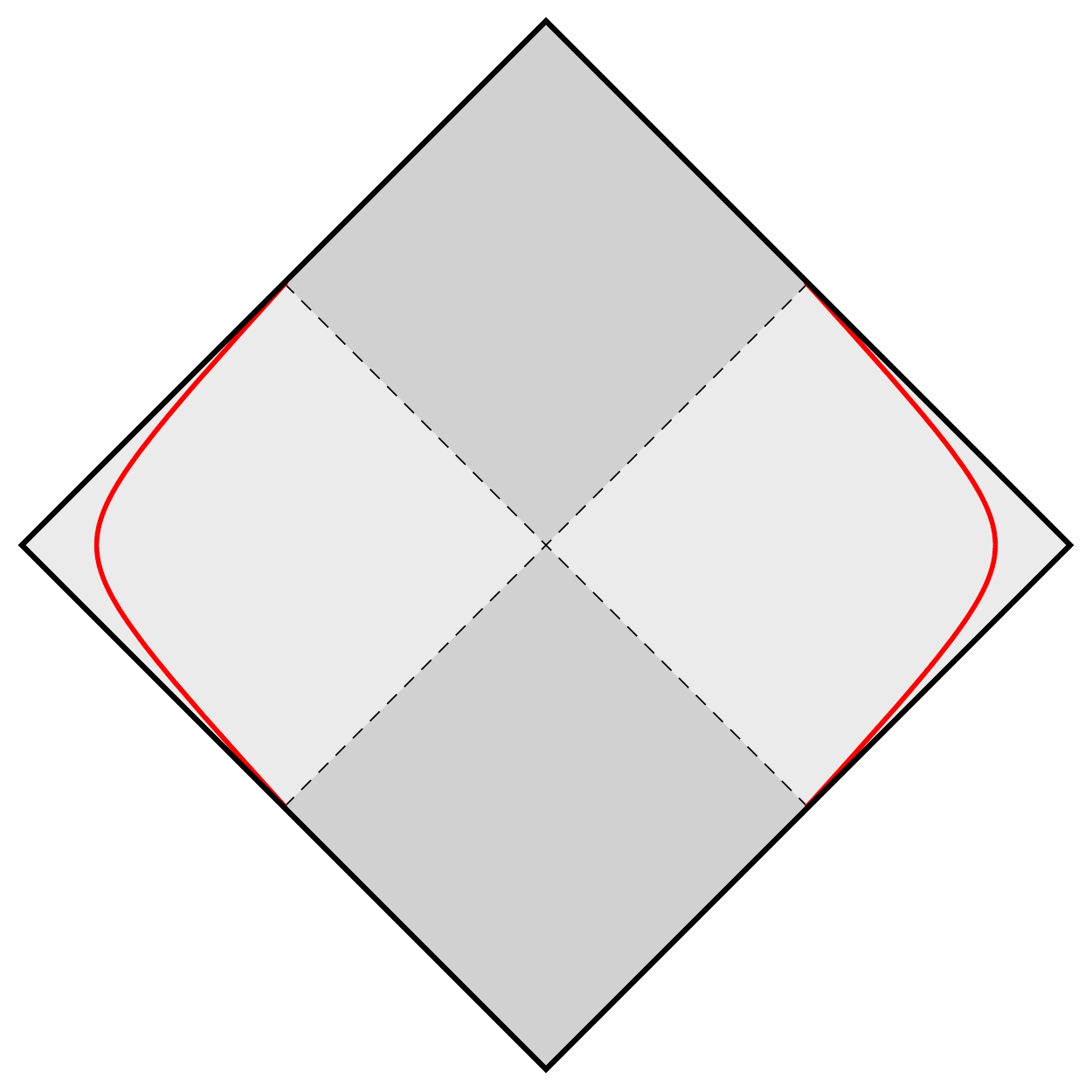}

  \caption{Flat space boundary particles in Lorentzian signature. On the right, we have the thermal state which is analogous to a Schwarzschild black hole. On the left, we have the vacuum Minkowski state which is similar to the global AdS$_2$ solution of JT gravity and should be interpreted as an eternal traversable wormhole \cite{Maldacena:2018lmt}.  }\label{FlatBparticles}
\end{figure}

 In this section, we will show that the boundary action \eqref{EucFlatAction} can be interpreted as a particle moving in flat space. Let's consider a particle moving on the background geometry \eqref{backgeoflat} for $P(\tau)=P_0$ and $T(\tau)=T_0$. We consider  a "vacuum" trajectory which corresponds to a particle lying at $r=r_0 $. Applying a warped Virasoro transformation $(f,g)$, we obtain a new particle with trajectory
\begin{equation}
\tau=f(s),\quad r={r_0 +i g'(s)\ov f'(s)},
\end{equation}
The worldline action of the particle is defined as
\be
I_\r{particle} = {\g\ov 2}\int ds \,\dot{x}^2~.
\ee
The computation shows that this matches with the boundary action \eqref{EucFlatAction} up to a constant term
\be
I_\r{particle} = I [f,g] + \r{const}~.
\ee
As a result, the dynamics of the $\wh{\text{CGHS}}$ model is captured by the motion of this particle. We depict the boundary particles of classical solutions in Fig.\ \ref{FlatBparticles}.

\paragraph{Minkowski particle.}

The vacuum of the theory is 2d Minkowski space 
\be
ds^2=  -du^2 -2 dudr~,
\ee
where $u = - i \tau$ is the Lorentzian Bondi coordinate. This corresponds to the geometry \eqref{backgeoflat} with
\be
P=0,\qq T = {1\ov 2}~.
\ee
The corresponding boundary particle lies at a constant value $r=r_0$. In fact, this solution actually has two boundary particles because there is another asymptotic boundary at $r=-r_0$. This geometry is the analog of the global AdS$_2$ solution of JT gravity studied in \cite{Maldacena:2018lmt} which has the interpretation of an eternal traversable wormhole, see Fig.\ \ref{FlatBparticles}.

\ms

\paragraph{Rindler particle.}
There is also a thermal solution which corresponds to the Rindler spacetime. This is a 2d analog of the Schwarzschild black hole. To describe this configuration, we can start with the Rindler metric with inverse temperature $\b$
\be
ds^2= -{4\pi^2\ov \b^2}x^2 dt^2 + dx^2~,
\ee
where we have $t \sim t+i \b$. We can write this in Bondi gauge with the change of variables
\be
x= \sqrt{\b r\ov  \pi },\qq t= u+{ \b \ov 4\pi}\log r~,
\ee
which leads to the metric
\be
ds^2 = -{4\pi r\ov \b} du^2 -  2 du dr~.
\ee
The Euclidean metric is obtained by $u = - i\tau$. This corresponds to the geometry \eqref{backgeoflat} with
\be
P= {2\pi i \ov \b},\qq T = 0~.
\ee
Note that this geometry can be generated by the diffeomorphism \eqref{CoordWitt} with the choice
\be\label{fgRind}
\cF(\tau) =  {\b\ov 2\pi i} e^{ 2\pi  i \tau/\b},\qq \cG(\tau)= -{\b^2\ov 16\pi^2} e^{4\pi  i \tau/\b}~.
\ee
We depict the boundary particles corresponding to these two states in Fig.\ \ref{FlatBparticles}.

\subsection{Euclidean path integral}

We consider the Euclidean path integral of the theory \eqref{CGHShat}. The integration over $\Phi$ fixes the 2d metric to be flat while the integration over $\Psi$ imposes the constraint $\ast \,d A = 1$ on the gauge field. Assuming that there exists a solution $A_0$ to the constraint, the general solution is $A= A_0 + d\la$ where $\la$ is an arbitrary scalar field. The integration over $A$ becomes an integration over $\la$, which needs to be quotiented by the trivial gauge transformations (\ie the ones that vanish at infinity). Hence, the only degree of freedom in the gauge field is the boundary value of $\la$, which is actually fixed by the boundary condition considered in \cite{Afshar:2019axx}. What remains is the integral over $f$ and $g$ with the boundary action \eqref{EucFlatAction}, which we consider below.

In this section, we study the only two flat surfaces with asymptotic boundaries, the disk and the cylinder, which completely determine the gravitational path integral. $\wh{\r{CGHS}}$ is defined for Lorentzian metrics in Bondi gauge with a boundary at $r=+\infty$. Going to Euclidean signature, one can show that this statement implies that the distance between any bulk point and the boundary has to be infinite in the Euclidean geometry. This prevents the inclusion of Euclidean geometries with boundaries at finite distance, like a disk pierced with many holes. Such geometries would appear in a ``finite-cutoff'' version of $\wh{\r{CGHS}}$, analogous to \cite{Iliesiu:2020zld, Stanford:2020qhm}, which would be interesting to investigate. 

 % We compute the disk partition function, the cylinder geometry connecting two boundaries and we consider the genus expansion. We will see that the path integral is much simpler than in the hyperbolic case because of the simplicity of flat Riemann surfaces. Indeed, there are only two topologically distinct regular Riemann surfaces with a flat metric and an asymptotic boundary: the plane and the cylinder.

\subsubsection{Partition function}

The partition function is obtained by computing the path integral of the action \eqref{EucFlatAction}  on the disk geometry. The result for the disk partition function can be deduced from a computation done in \cite{Afshar:2019tvp}. We revisit it here, allowing general values of $P_0$ and $T_0$ in the Pfaffian, which will be necessary for the computation of the cylinder contribution in the next section.

This computation is very similar to the corresponding AdS computation in Sec.\ \ref{Partition function computation}.  As in the AdS case, the disk partition function is one-loop exact because of the Duistermaat-Heckman theorem. 

\ms

To compute the one-loop contribution, we decompose 
\be
f(\tau) = \tau+\ve(\tau),\qq g(\tau) = \s(\tau)~,
\ee
where $\ve$ and $\s$ are taken to be infinitesimal. The boundary conditions impose that $\ve$ and $\s$ are periodic with period $\b$. Hence, they can be decomposed as
\be\label{epsilonsigmadecFlat}
\ve(\tau)= {\b\ov 2\pi}\sum_{n\in \Z} \ve_n e^{-{2\pi\ov \b}i n \tau},\qq \s(\tau) ={\b\ov 2\pi} \sum_{n\in\Z} \s_n e^{-{2\pi\ov \b} i n \tau}~.
\ee
Since $\ve(\tau)$ is real and $\s(\tau)$ is pure imaginary, we have the relations
\be
\ve_{-n} = \ve_n^\ast,\qq \s_{-n} = -\s_n^\ast~.
\ee
The symplectic form on the phase space takes the form
\bea
\w \= \alpha{2\b^2\ov \pi i} \sum_{n\geq 1} T_0 n\, d\ve_n \wg d\ve_{-n}\-
&& + \alpha {\b^2\ov \pi i } \sum_{n\geq 1}\le[\le(-{2\pi i\ov \b} n^2+ P_0 n\ri) d\ve_n\wg d\s_{-n} +\le(-{2\pi i\ov \b} n^2- P_0 n\ri) d\ve_{-n} \wg d\s_n \ri]~.
\eea
We can write $\w_{\ve\s}$ as a $2M\times 2M$ indexed by $(n,m)$ where $-M \leq n,m \leq M $ and $n,m\neq 0$.
\be
(\w_{\ve\s})_{nm} = \alpha {\b^2\ov 2\pi i }\le(-{2\pi i\ov \b} n^2+ P_0 n\ri) \d_{n+m}~.
\ee
Its Pfaffian is given by
\bea\label{PfaffianFlat}
\r{Pf}(\w) \= (-1)^{M-1} \r{Pf}(\w_{\ve\s})^2,\-
\= (-1)^{M-1} \prod_{n= 1}^M   \alpha\b \le(n^2-{\b\ov 2\pi i} P_0 n\ri)  \prod_{n= 1}^M \alpha \b\le(n^2+ {\b\ov 2\pi i} P_0 n \ri)~,\-
\=(-1)^{M-1} \prod_{n= 1}^M \r{Pf}_n(\omega) ~,
\eea
where we have defined 
\begin{equation}
\r{Pf}_n(\omega)=\alpha^2\beta^2n^2\left(n^2+\frac{\beta^2}{4\pi^2}P_0^2\right).
\end{equation}
The classical piece of the action is given by
\be\label{Ssaddleflat}
I_\r{on-shell} = \g \b T_0~,
\ee
and the quadratic action is given as
\bea
I_\r{quad} \=\sum_{n \geq 1} I_\r{quad}^{(n)} ~,
\eea
where
\bea
I_\r{quad}^{(n)}  \=2\b\gamma \le( T_0 n^2 \le( (\ve_n^{(R)})^2 + (\ve_n^{(I)})^2\ri)+ i  P_0  n^2 \le(\ve_n^{(R)} \s_n^{(I)} -\ve_n^{(I)}\s_{n}^{(R)}\ri)  \vphantom{{2\pi i\ov \b}}\ri. \-
&& \hspace{1cm}\le. +{2\pi i\ov \b} n^3  \le(\ve_n^{(R)} \s_n^{(R)} +\ve_n^{(I)}\s_{n}^{(I)}\ri)\ri)~.
\eea
We can perform the Gaussian integral for generic values of $P_0$ and $T_0$ and we obtain
\begin{equation}\label{generalPfQuadFlat}
 \r{Pf}_n(\w)  \int  d^2\ve_n d^2\s_n e^{-I_\r{quad}^{(n)}} = {\alpha^2\b^2\ov  n^2 \g^2 }~.
\end{equation}
We now consider the values of $P_0$ and $T_0$ that correspond to the Rindler spacetime
\begin{equation}
P_0=\frac{2\pi i }{\beta}, \quad T_0=0.
\end{equation}
To compute the  one-loop piece, we rewrite the Pfaffian as
\be
\begin{split}
\r{Pf}^{\r{disk}}(\w) &= (-1)^{M-1} \prod_{n= 2}^M  \alpha \b \le(n^2- n\ri) \prod_{n= 1}^M \alpha \b\le(n^2+   n\ri) ~,\\
&\equiv \r{Pf}_{\varepsilon_{-1}, \sigma_1} \prod_{n= 2}^M \r{Pf}_n(\omega),
\end{split}
\ee
where we have removed the degenerate directions $\ve_{1}$ and $\sigma_{-1}$ and defined $\r{Pf}_{\varepsilon_{-1}, \sigma_1}=2\alpha\beta$. The one-loop path integral can be decomposed as
\begin{equation}
Z_{\text{1-loop}}= Z_{\varepsilon_{-1}, \sigma_1} Z_{|n|\geq 2},
\end{equation}
where we compute
\begin{equation}
\begin{split}
&Z_{|n|\geq 2} = \prod_{n\geq 2} \r{Pf}_n(\w) \int d^2\ve_n d^2\s_n e^{-I_\r{quad}^{(n)}}  = {2 \pi\gamma^3\ov \alpha^3\b^3}~,\\
&Z_{\varepsilon_{-1}, \sigma_1}=\r{Pf}_{\varepsilon_{-1}, \sigma_1}\int  d\ve_{-1} d\s_1e^{- I_\r{quad}^{(1)}} = {\alpha\b\ov \gamma}~.
\end{split}
\end{equation}
This leads to
\be
Z_\text{1-loop}   = {2\pi \g^2\ov \b^2}~.
\ee
Since $T_0=0$, the classical piece \eqref{Ssaddleflat} vanishes here so the partition function takes the form
\be
Z^{\r{disk}}(\b) = {2\pi\g^2\ov\alpha^2 \b^2}  ~,
\ee
which matches the result of \cite{Afshar:2019tvp}.\footnote{The comparison should be made after normalizing the symplectic form in the same way, which corresponds here to the choice $\a=1/2$} An inverse Laplace transform gives the density of states 
\be\label{densityCGHS}
\rho(E) = {2\pi \g^2\ov \a^2}  E~.
\ee
Since there are no higher genus surfaces with a flat metric, this is actually the exact density of states.\footnote{$\rho(E)$ could still be corrected by nonperturbative effects in the genus expansion corresponding to doubly-nonperturbative effects in the gravitational coupling.} The fact that $\rho(E)$ is not a sum of delta functions implies that our model is not a standard quantum system. In fact, we will see that this theory should be interpreted as an ensemble average.

We note that \eqref{densityCGHS} is different from the density of states of the CGHS model that was proposed in a different context in \cite{Stanford:2020qhm}. They actually consider the theory \eqref{badCGHSaction} which  has a fixed temperature and is different from the theory \eqref{CGHShat} considered here.

\subsubsection{Cylinder}

\begin{figure}
  \centering
  \includegraphics[width=8cm]{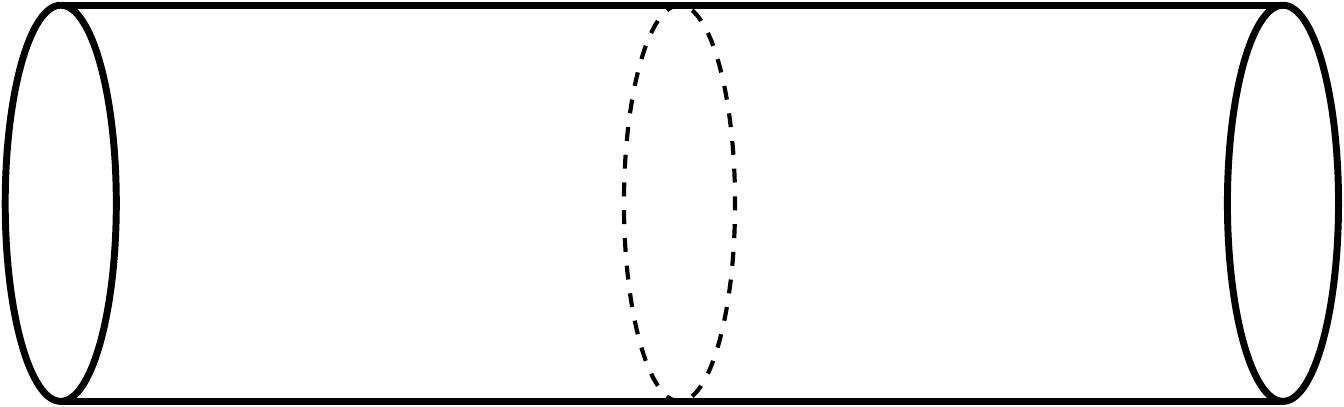}
  \put(-245,30){{$\b_1$}}
  \put(-115,75){{$b$}}
      \put(4,30){{$\b_2$}}
  \caption{The cylinder geometry. We specify two boundary conditions $\b_1$ and $\b_2$ at each end. The only modulus is the circumference $b$.}\label{Fig:Cylinder}
\end{figure}

The cylinder is the flat space analog of the double trumpet. It is described by the metric 
\be
ds^2 = dt^2 + d x^2~,\qq t\sim t +b~,
\ee
where $b>0$ is the circumference of the cylinder and is depicted in Fig.\ \ref{Fig:Cylinder}. 

To compute its contribution, we first compute the contribution of half-cylinder which is described by the same metric, where we only focus on the asymptotic boundary at $x=+\infty$. We can define a Bondi coordinate $\tau$ with 
\be
t= {b\ov \beta}\tau +{\beta\ov b} i x,\qq x = {\b\ov b}r ~,
\ee 
to get
\be
ds^2 = {b^2\ov \b^2} d\tau^2 + 2 i d\tau dr ~.
\ee 
This corresponds to the geometry \eqref{backgeoflat} with
\be
P_0 = 0,\qq T_0=  {b^2\ov 2 \b^2} ~.
\ee
The classical contribution gives
\be
I_\r{on-shell} = \g \b T = {\g b^2\ov 2 \b} 
\ee
The one-loop contribution is computed in the same way as in the previous section. The difference is that because $P_0=0$, the Pfaffian \eqref{PfaffianFlat} takes the form
\be
\r{Pf}^\text{half-cyl}(\w) = (-1)^{M-1} \prod_{n= 1}^M \alpha \b n^2  \prod_{n= 1}^M \alpha\b n^2~.
\ee
As a result, we should not remove the contribution of $\ve_{1}$ and $\s_{-1}$. This is analogous to what happens in the trumpet computation for AdS$_2$. From the formula \eqref{generalPfQuadFlat}, we obtain
\be
Z_\text{1-loop} =\r{Pf}^\text{half-cyl}(\w) \int \prod_{n\geq 1} d^2\ve_n d^2\s_n e^{-I_\r{quad}^{(n)}}  =   {2\pi \g\ov\alpha \b}~,
\ee 
where we have computed the infinite product  using \eqref{infiniteproductzeta}. This gives
\be
Z^\text{half-cyl}(\b,b) = {2\pi \g\ov\alpha \b} \exp\le(-{\g b^2\ov 2\b} \ri) ~.
\ee
We note that this is the same contribution as the AdS trumpet for $\mu=0$. 

\ms

We can now compute the formula for the cylinder using
\be
Z^\text{cyl}(\b_1,\b_2) =  \int_0^{+\infty} b db \, Z^\text{half-cyl}(\b_1; b) Z^\text{half-cyl}(\b_2; b)~,
\ee
where the factor $b$ in the measure $b db $ is necessary because of the freedom to twist one of the half-cylinder relative to the other when gluing. In AdS, this factor follows from the Weil-Peterson measure \cite{Saad:2019lba}. This gives
\be\label{Zcylinder}
Z^\text{cyl}(\b_1,\b_2)  = {4\pi^2\g\ov \b_1+\b_2}~,
\ee
where we have set $\alpha=1$. It is interesting that we find a non-zero answer. This implies that if the $\wh{\text{CGHS}}$ model has a holographic dual, it has to be an ensemble of theories, as in the case of AdS JT gravity.  

Let us introduce the notation $\ln Z(\b_1)\dots Z(\b_n)\rn$ to represent the Euclidean path integral with $n$ asymptotic circles of lengths $\b_1,\dots,\b_n$. The fact that the cylinder does not vanish implies that
\be
\ln Z(\b_1)Z(\b_2) \rn \neq \ln Z(\b_1) \rn\ln Z(\b_2)\rn~.
\ee
This indeed shows that the path integral should be interpreted as an ensemble average. The answer \eqref{Zcylinder} is not the universal answer for double-scaled matrix ensembles so the dual of the $\wh{\text{CGHS}}$ model has to be something different.

\subsubsection{Genus expansion}

The Euclidean path integral that we consider involves flat Riemann surfaces with at least one asymptotic boundary. The only such surfaces are the plane (or disk) and the cylinder. Therefore, the connected contribution to the path integral with more than three boundaries identically vanishes:
\be
\ln Z(\b_1)Z(\b_2)\dots Z(\b_n)\rn_c = 0 \qq n\geq 3~.
\ee
This shows that the path integral with an arbitrary number of boundaries is completely determined using  Wick contractions involving the cylinder and the disk. This implies that the corresponding third-quantized theory is a Gaussian theory. Thus, the $\wh{\text{CGHS}}$ model constitutes an interesting example of a theory where the full Euclidean path integral can be done, while not being completely trivial and giving rise to an ensemble average.

\appendix

\section{A new counterterm for JT gravity}\label{app:NewCount}

In this paper we have worked with a spacetime whose boundary is fixed while the boundary value of the dilaton is allowed to fluctuate and defines the Schwarzian mode. Another approach, which is the one of \cite{Maldacena:2016upp}, is to define the boundary of the spacetime as the place where the dilaton is equal to a fixed constant (taken to infinity at the end) and it is the shape of this boundary that defines the Schwarzian mode. The equivalence between these two formulations is explained in \cite{Gaikwad:2018dfc}. We would like to show that our new boundary action can also be obtained using an approach similar to that of \cite{Maldacena:2016upp}.

 In the second formulation, the Schwarzian action is derived from the renormalized extrinsic curvature term
\be\label{app:Iextrinsic}
I_\r{Schw} = \k \int d u  \sqrt{-h} \, (\Phi K- \Phi) ~,
\ee
where $h$ is the boundary metric. We will show that this term does not give the new boundary action discussed in this paper, but that it  arises from a new counterterm
\be\label{app:Inew}
I_\r{new} = \k \int \,du \sqrt{-h} \, (\Phi K-n^\mu \p_\mu\Phi) ~.
\ee
Let's consider a general Lorentzian metric satisfying $R=-2$ in Bondi gauge
\be
ds^2 = 2\le( -{r^2\ov 2}+ P(u)r + T(u)\ri) du^2 -2 du dr~.
\ee
It can be obtained by acting on the Poincaré AdS$_2$ metric, corresponding to $P = T =0 $, with an element of the warped Virasoro group
\be
u\ra \cF(u),\qq r \ra {r+\cG'(u)\ov \cF'(u)}~,
\ee
which gives explicitly
\be\label{app:explicitPT}
P(u) = {\cF''(u)\ov \cF'(u)} -\cG'(u),\qq T(u) = -{1\ov 2} \cG'(u)^2+{\cG'(u) \cF''(u)\ov \cF'(u)}-\cG''(u) ~.
\ee
We refer to \eqref{Dictionary} for the dictionary between Lorentzian and Euclidean signature.

\pg{The Schwarzian action.} We take the dilaton to be asymptotically\footnote{The minus sign here is a consequence of the mapping between Bondi and FG gauge.} 
\be
\Phi \sim -\bar\phi_r r~,
\ee
so that the boundary is at $r=1/\ve$. We can  then compute the action \eqref{app:Iextrinsic} which gives
\be
I_\r{Schw}[\cF,\cG] = -{\g} \int du \,\le(T(u)  + {1\ov 2 } P(u)^2 - P'(u) \ri)~.
\ee
This is precisely the Schwarzian action since, using \eqref{app:explicitPT}, we have
\be
{-}T(u)  - {1\ov 2 } P(u)^2 + P'(u) = \{ \cF(u),u\}
\ee
We see that this action does not depend on $\cG(\tau)$ and matches with the Schwarzian action discussed in Sec.\ \ref{Relation to the Schwarzian action}. A different counterterm is needed to recover our boundary action.

\pg{Our boundary action.} We now take the dilaton to be
\be\label{bdyCondBondi}
\Phi = -\bar\phi_r \le(r+\bar\mu\ri)~,
\ee
so that the boundary is still at $r = {1/ \ve}$. Note that here, we also fix the subleading piece of the dilaton in terms of a constant $\bar\mu$. This actually fixes the dilaton everywhere in the bulk. We now consider the following boundary action
\be\label{app:Inew}
I_\r{bdy}^\r{new} = \k\int \,du \sqrt{-h} \, (\Phi K-n^\mu \p_\mu\Phi) ~.
\ee
It can be checked that this counterterm does not affect the well-posedness of the variational problem.  The integrand gives
\bea
\sqrt{-h}\,(\Phi K-n^\mu \p_\mu\Phi) = \bar\phi_r(\bar\mu+P(u)) \,r +\bar\phi_r (2 T(u) -\bar\mu P(u)- P'(u)) + O(r^{-1})~.
\eea
We see that this counterterm cancels the $ r^2$ divergence. The $ r$ divergence is actually also cancelled in the integral since we have
\be
\bar\mu+P(u) = 0~,
\ee
thanks to one of the dilaton equations of motion in the bulk. Finally, we obtain
\be\label{finalbondi}
I_\r{bdy}^\r{new}[\cF,\cG]= 2 \g \int du\,T(u) + \r{const}~.
\ee
This precisely matches our boundary action, as written in \eqref{IasT}, where the term involving  $\int du\,P(u)$ has been included in the constant. This shows that our boundary action arises from a new boundary condition  \eqref{bdyCondBondi} on the dilaton together with a new counterterm.

\section{Gauge-theoretic formulation}
\label{App: Gauge-theoretic formulation}

In this paper, we have been focusing on an AdS-Bondi version of JT gravity. It is worth mentioning that there is a closely related theory which gives an alternative formulation of the same physics. This is an AdS version of the $\wh{\text{CGHS}}$ model studied in Sec.\ \ref{sec:CGHSandboundaryaction} and described by the action
\begin{equation}
I[\Phi,g,\Psi,A]=\frac{\kappa}{2}\int dx^2 \sqrt{g}\,\left[\Phi\left(R+2\right)-2\Psi+2\Psi \varepsilon^{\mu\nu}\partial_\mu A_\nu\right] + I_\p~.
\label{CGHSADS}
\end{equation} 
This theory can be interpreted as JT gravity coupled to a BF theory via a scalar field $\Psi$. The equations of motion are 
\bea
0 \=R+2, \\
0 \=\nabla^\mu\nabla^\nu \Phi-g^{\mu\nu}\nabla_\rho\nabla^\rho \Phi+ g^{\mu\nu}\Phi- g^{\mu\nu}\Psi,\\
0 \= \varepsilon^{\mu\nu}\partial_\mu A_\nu-1,\label{EOMGAUGE}\\
0 \=\partial_\mu \Psi.
\eea
We note that the field $\tilde{\Phi} = \Phi-\Psi$ satisfies the JT equation of motion. The solution for the metric is AdS$_2$ therefore we can write it in Bondi gauge
\be
ds^2 = 2\le({r^2\ov 2}-i P(\tau)r +T(\tau) \ri) d\tau^2 -2i d\tau dr ~,
\ee
In axial gauge, a solution to the equation of motion \eqref{EOMGAUGE} is given by the Coulomb field
\begin{equation}
A= rd\tau.
\end{equation}
Following \cite{Afshar:2019axx}, we require this solution to be preserved in the phase space, under combined diffeomorphisms and gauge transformations. A similar condition in AdS$_2$ was considered in \cite{Hartman:2008dq}. This leads to the same asymptotic symmetry algebra than the one considered in Sec.\ \ref{sec:BondiJT}. Indeed asking the metric to be preserved gives
\begin{equation}
\xi=\varepsilon(\tau)\partial_\tau-(\varepsilon'r-\eta(\tau))\partial_r.
\end{equation}
while asking combined diffeomorphisms and gauge transformations to preserve the Coulomb solution gives
\begin{equation}
 \delta_{\varepsilon,\sigma} A=\mathcal{L}_\xi A + d\sigma=0\Rightarrow \eta=\sigma'.
\end{equation}
Moreover, with the following parametrization of the dilaton \\
\begin{equation}
\varphi_1=\frac{i\bar{\phi}_r}{f'},\qq \varphi_0=i\bar{\phi}_r\frac{g'}{f'},
\end{equation}
one can show that the boundary action corresponds exactly to the first terms in Eq. \eqref{IbdyEuc}, which contains all the boundary dynamics. Asking the dilaton equations to be the same than the one in Eq. \eqref{Dilatonequationfandg} when written in terms of $f$ and $g$, we obtain the relation 
\begin{equation}
\bar{\mu}=\frac{i\Psi}{\bar{\phi}_r}.
\end{equation}
With these conditions on the solution space, the dynamics of the theory \eqref{CGHSADS} matches exactly with the dynamics of pure AdS-JT gravity in Bondi gauge.

\section{Gravitational charges of the $\wh{\text{CGHS}}$ model}
\label{Flat gravitational charges}

We derive here the expression of the gravitational charges given in Sec.\ \ref{sec:CGHSandboundaryaction}, associated with the warped Witt asymptotic symmetry algebra of 2d Minkowski spacetime. This result follow from boundary conditions on the solution space of the $\widehat{\text{CGHS}}$.

\ms

We recall the action 
\be
I = {\k\ov 2} \int d^2 x\sqrt{g}\le( \Phi R - 2 \Psi + 2 \Psi \ve^\mn \p_\mu A_\nu\ri) + I_\partial~,
\ee
and its equations of motion
\begin{align}
R & = 0, \\
 \ve^\mn \p_\mu A_\nu & = 1,\label{gauge}\\
\n_\mu \n_\nu \Phi -g_\mn \Box\Phi & = g_\mn \Psi, \\ \Psi & = \r{const}.
\end{align}
The flat metric is given in Bondi gauge
\begin{equation}
ds^2=2(-iP(\tau)r+T(\tau))d\tau^2+2id\tau dr.
\end{equation}
The corresponding asymptotic symmetry algebra is spanned by the vectors
\begin{equation}\label{app:WarpedWitt}
\xi=\varepsilon(\tau)\partial_\tau-(\varepsilon'r-i\eta(\tau))\partial_r,
\end{equation}
and the corresponding transformations of $P$ and $T$ are 
\begin{equation}
\begin{split}
&\delta_\xi P =\varepsilon P'+\varepsilon' P +\varepsilon'',\\
&\delta_\xi T=\varepsilon T'+2\varepsilon' T+\eta P-\eta'.
\end{split}
\end{equation} 
In axial gauge, the equation of motion for the gauge field \eqref{gauge} is solved by the Coulomb field
\begin{equation}
A= rd\tau.
\end{equation}
Follow \cite{Afshar:2019axx}, we require this solution to be preserved in the phase space, under combinations of diffeomorphisms  and gauge transformations. This gives a relation between the supertranslation $\eta$ and the gauge parameter $\sigma$
\begin{equation}
\delta_{\xi, \sigma} A=0 \Rightarrow \eta=\sigma'.
\end{equation}
The total symmetry algebra is now warped Witt. One of the dilaton equations gives
\begin{equation}
\Phi(u,r)=ir\varphi_1(\tau)-\varphi_0(\tau),
\end{equation}
while the two other equations being
\begin{equation}
\begin{split}
&\varphi_1'+P\varphi_1=-\Psi,\\
&\varphi_0''-P\varphi_0'+\varphi_1T'+2T\varphi_1'=0.
\end{split}
\label{Dilatonflat}
\end{equation}
The solution space is now parametrized by the functions $P$ and $T$ in the metric, the functions $\varphi_0$ and $\varphi_1$ in the dilaton and the constant $\Psi$. From now on, on-shell means that the two equations \eqref{Dilatonflat} are satisfied (together with their linearized versions for the linear perturbation).

To derive the charges associated to the warped Witt symmetry \eqref{app:WarpedWitt}, we need to compute the presymplectic potential $\mathbf{\Theta}$. This is done by varying the bulk term in the action and extracting the boundary term, which leads to
\begin{equation}
\mathbf{\Theta}=i\kappa(\varphi_0\delta P-\varphi_1 \delta T)d\tau.
\end{equation}
The corresponding symplectic form is 
\begin{equation}
\pmb{\omega}=i\kappa (\delta \varphi_0\wedge \delta P-\delta \varphi_1\wedge \delta T)d\tau.
\end{equation}
The fundamental theorem of the covariant phase space formalism states that, when $\phi$ and $\delta\phi$ are on-shell, there exists a function $k_\xi$ such that
\begin{equation}
\pmb{\omega}(\delta \phi, \delta_\xi\phi)=dk_\xi(\delta \phi).
\end{equation}
We find
\begin{equation}
k_\xi=\delta \mathcal{Q}_\xi+\Xi_{\xi},
\end{equation}
where we have split $k_\xi$ into an integrable and a non-integrable part\footnote{One should keep in mind that $\Psi$ is a constant whose value is not fixed by the equations of motion, so $\delta\Psi$ is a non-trivial direction in the solution space.}
\begin{equation}
\begin{split}
\mathcal{Q}_\xi &=i\Psi \sigma-\frac{i}{2}\kappa \left(2\varepsilon T \varphi_1-\varepsilon P\varphi_0+\varepsilon\varphi_0'-\varepsilon'\varphi_0-\sigma'\varphi_1\right),\\
\Xi_\xi &=-\frac{i}{2}\kappa\,\varepsilon(\varphi_0\delta P-\varphi_1 \delta T).
\end{split}
\label{GravitationalChargesFLAT}
\end{equation}
One can show that these charges define a centerless representation of the warped Witt algebra under the modified Dirac bracket \eqref{modifiedbracket}.

Exactly like in our AdS analysis, the authors of \cite{Afshar:2019axx} impose conditions on $\varphi_0$ and $\varphi_1$ to have a well-defined variational poblem 
\begin{equation}
\varphi_1=\frac{i\bar{\phi}_r}{f'},\qq \varphi_0=i\bar{\phi}_r\frac{g'}{f'},
\end{equation}
where $f(\tau+\beta)=f(\tau)+\beta$ and $g(\tau+\beta)=g(\tau)$. These conditions allow them to derive the boundary action \eqref{EucFlatAction} and to interpret it in terms of a coadjoint action of the warped Witt algebra. Using this parametrization of $\varphi_0$ and $\varphi_1$, our gravitational charges are
\begin{equation}
\begin{split}
\mathcal{Q}_\xi &=\frac{i\kappa }{2}\Psi\sigma+\frac{\gamma}{2f'}\left(2\varepsilon T-\varepsilon P g'-\varepsilon' g' -\sigma'-\varepsilon\frac{g'f''}{f'}+\varepsilon g''\right),\\
\Xi_\xi &=\frac{\gamma}{2f'} \varepsilon \left(g'\delta P-\delta{T}\right).
\end{split}
\end{equation}

\section*{Acknowledgements}

We would like to thank Alejandra Castro, Luca Ciambelli, Ricardo Espindola, Seth Koren, Blagoje Oblak, Gabriel Verastegui and Brianna White, for insightful discussions. We woud like to give a special thanks to Romain Ruzziconi for his explanations on the subtleties of asymptotic symmetries. CM would like also to thank the Kavli Institute for the hospitality and support given while part of this work was completed. The work of CM was supported in part by the National Science Foundation under Grant No. NSF PHY-1748958, by the Heising-Simons Foundation and by the ANR-16-CE31-0004 contract Black-dS-String. The work of VG is supported by the Delta ITP consortium, a program of the NWO that is funded by the Dutch Ministry of Education, Culture and Science (OCW).

\bibliography{bibliography}

\end{document}

%% file: victorsmacros.tex
\usepackage[utf8x]{inputenc}
\usepackage[T1]{fontenc}
\usepackage{amsthm}
\usepackage{slashed}
\usepackage{mathtools}
\usepackage{float}
\usepackage{empheq}
\usepackage{bm}
\usepackage{framed}
\usepackage{comment}
\usepackage{xcolor}

\def\sss{\subsubsection}
\def\pg{\paragraph}

\def\ie{\emph{i.e.} }

\def\nt{\notag}
\def\ol{\overline}

\def\wh{\widehat}

\def\R{\mathbb{R}}
\def\Z{\mathbb{Z}}

\def\cA{\mathcal{A}}

\def\cE{\mathcal{E}}
\def\cF{\mathcal{F}}
\def\cG{\mathcal{G}}
\def\cH{\mathcal{H}}

\def\cJ{\mathcal{J}}

\def\cL{\mathcal{L}}
\def\cM {\mathcal{M}}

\def\cQ {\mathcal{Q}}

\def\cS{\mathcal{S}}

\def\dg {\dagger}
\def\p{\partial}

\def\ov{\over}

\def\rn{\rangle}
\def\ln{\langle}

\def\t{\theta}
\def\s{\sigma}

\def\ve{\varepsilon}
\def\vphi{\varphi}
\def\a{\alpha}
\def\b{\beta}
\def\d{\delta}
\def\k{\kappa}
\def\g {\gamma}
\def\la {\lambda}
\def\w {\omega}
\def\z{\zeta}

\def\l{\ell}
\def\mn{{\mu\nu}}

\def\n {\nabla}

\def\D{\Delta}

\def\Om {\Omega}

\def\ra{\rightarrow}
\def\lra{\longrightarrow}

\def\Tr{\mathrm{Tr}}

\def\r{\mathrm}

\def\_{\hspace{2cm}}
\def\-{\\\notag}
\def\={&=&}

\newcommand\be{\begin{equation}}
\newcommand\ee{\end{equation}}

\newcommand{\bea}{\begin{eqnarray}}
\newcommand{\eea}{\end{eqnarray}}

\newcommand{\bpm}{\begin{pmatrix}}
\newcommand{\epm}{\end{pmatrix}}

\newcommand{\bit}{\begin{itemize}}
\newcommand{\eit}{\end{itemize}}

\newcommand{\ben}{\begin{enumerate}}
\newcommand{\een}{\end{enumerate}}

\newcommand\bsp{\begin{split}}
\newcommand\esp{\end{split}}

\def\le{\left}
\def\ri{\right}

\def\ms{\medskip}

\def\wg{\wedge}
\def\l{\ell}

\def\qq{\qquad}

\def\sinh{\r{sinh}}